\documentclass[11pt,twoside]{article}

\usepackage{geometry} 
\usepackage{axodraw}
\usepackage{amsmath}
\usepackage{amssymb}
\usepackage{graphicx}
\usepackage{subfigure}

\newcommand{\beq}{\begin{equation}}
\newcommand{\eeq}{\end{equation}}
\newcommand{\bea}{\begin{eqnarray}}
\newcommand{\eea}{\end{eqnarray}}
\newcommand{\ba}{\begin{array}}
\newcommand{\ea}{\end{array}}
\newcommand{\bec}{\begin{center}}
\newcommand{\eec}{\end{center}}
\newcommand{\bei}{\begin{itemize}}
\newcommand{\eei}{\end{itemize}}

\newcommand{\GeV}{\,\mathrm{GeV}}

\newcommand{\eV}{\,\mathrm{eV}}
\newcommand{\SM}{$SU(3)_C \times \protect 
        \linebreak[0]SU(2)_L \times \protect\linebreak[0]U(1)_Y$}

\DeclareMathOperator{\tr}{Tr}

\newlength{\myem}
\newcommand{\sep}[1]{#1}
\newcounter{mysubequation}[equation]
\renewcommand{\themysubequation}{\alph{mysubequation}}
\newcommand{\mytag}{\stepcounter{mysubequation}%
\tag{\theequation\protect\sep{\themysubequation}}}
\newcommand{\globallabel}[1]{\refstepcounter{equation}\label{#1}}

\makeatletter
\renewcommand{\section}{\@startsection{section}{1}{0em}%
        {-3.5ex \@plus -1ex \@minus -.2ex}%
        {2.3ex \@plus.2ex}%
        {\normalfont\large\bfseries}}
\renewcommand{\subsection}{\@startsection{subsection}{2}{0em}%
        {-3.25ex\@plus -1ex \@minus -.2ex}%
        {1.5ex \@plus .2ex}%
        {\normalfont\bfseries}}
\renewcommand{\subsubsection}%
        {\@startsection{subsubsection}{3}{0em}%
        {-3.25ex\@plus -1ex \@minus -.2ex}%
        {1.5ex \@plus .2ex}%
        {\normalfont\itshape}}
\makeatother

\newcommand{\SISSA}{SISSA/ISAS and INFN, I--34013 Trieste, Italy}

\newcommand{\preprintnumber}{%
SACLAY-T09/158\\SISSA--64/2009/EP}
 
\newcommand{\titletext}{Flavour violation in supersymmetric SO(10) \\
unification with a type II seesaw mechanism}
\newcommand{\authortext}{\large Lorenzo Calibbi$^{\, a}$, Michele Frigerio$^{\, b}$,
St\'ephane Lavignac$^{\, b}$ \\ and Andrea Romanino$^{\, a}$
\bigskip\\\em\normalsize 
$\mbox{}^a$ \SISSA
\\[0.1\baselineskip] 
$\mbox{}^b$ Institut de Physique Th\'{e}orique, CEA-Saclay, 
F-91191 Gif-sur-Yvette Cedex, France~\footnote{Laboratoire de la Direction
des Sciences de la Mati\`ere du Commissariat \`a l'Energie Atomique
et Unit\'e de Recherche Associ\'ee au CNRS (URA 2306).}}
\newcommand{\abstracttext}{We study flavour violation in a
supersymmetric SO(10) implementation of the type II seesaw mechanism,
which provides a predictive realization of triplet leptogenesis.
The experimental upper bounds on lepton flavour
violating processes have a significant impact on the leptogenesis dynamics,
in particular they exclude the strong washout regime.
Requiring successful leptogenesis then constrains the otherwise
largely unknown overall size of flavour-violating observables,
thus yielding testable predictions.
In particular, the branching ratio for $\mu \to e \gamma$
lies within the reach of the MEG experiment if the superpartner
spectrum is accessible at the LHC,
and the supersymmetric contribution to $\varepsilon_K$
can account for a significant part of the experimental value.
We show that this scenario can be realized in a consistent SO(10)
model achieving gauge symmetry breaking and doublet-triplet splitting
in agreement with the proton decay bounds, improving on the MSSM
prediction for $\alpha_3(m_Z)$, and reproducing the measured quark
and lepton masses.
}

\title{
\normalsize
\hspace*{\fill}
\begin{tabular}[t]{l}\preprintnumber\end{tabular}
\vspace{3\baselineskip}\\\Large\bfseries \vskip 1cm \titletext\bigskip}
\author{\begin{minipage}[t]{0.95\textwidth}
\normalsize\centering\authortext
\end{minipage}}
\date{}

\begin{document}

\bigskip
\maketitle
\begin{abstract}\normalsize\noindent
\abstracttext
\end{abstract}\normalsize\vspace{\baselineskip}

 \clearpage

\section{Introduction}                 %
\label{sec:introduction}              %

Neutrino masses presumably arise at a scale much larger than the
electroweak scale. If this is the case, a model-independent effective
description of neutrino masses is possible at lower scales in terms of
the dimension 5 operator $l_i l_j h_u h_u/\Lambda$~\cite{Weinberg79},
where $l_i$ is the $i$-th family lepton doublet, $h_u$ is the $Y=+1/2$
Higgs doublet and $\Lambda$ is the scale at which the operator is generated,
which can be as large as $10^{15}\GeV$. While this effective description
is essentially unique, the high-energy mechanism leading to the above
dimension 5 operator is not. Generally, it is assumed to arise from
the tree-level exchange of SM singlet fermions (type I seesaw
mechanism~\cite{seesawI}).
In this case, the low-energy information from lepton masses and mixing
only determines 9 out of the 18 high-energy seesaw parameters.
Due to this arbitrariness, it is not possible to make definite predictions
for the lepton asymmetry generated in the decays of the heavy singlet
neutrinos~\cite{FY86}, nor, in supersymmetric models, for the
lepton flavour violating (LFV) effects induced by their Yukawa
interactions~\cite{BM86}.

On the other hand, the exchange of singlet fermions is not the only
possible origin of the $l_i l_j h_u h_u/\Lambda$ operator. In this paper,
we consider
the exchange of an SU(2)$_{\rm L}$ triplet scalar with $Y=\pm1$ (type II
or triplet seesaw mechanism~\cite{seesawII}). More precisely, we consider
the SO(10)~\cite{SO10} implementation of the triplet seesaw mechanism
proposed in Ref.~\cite{FHLR08}, in which all flavour parameters
contributing to low-energy observables are determined in terms of
the SM fermion masses and mixings\footnote{Actually, the higher-dimensional
operators needed to account for the measured masses of down quarks
and charged leptons may affect this relation between high-energy and
low-energy flavour parameters. We argue in Section~\ref{subsec:MdMe}
that the impact of this on physical observables is generally small,
and neglect it in the following.}.
This allows to make testable predictions for these observables,
up to a few unknown flavour-blind parameters. Particularly interesting
is the possibility~\cite{FHLR08} of accounting for the baryon asymmetry
of the universe via triplet leptogenesis.
In usual type II models, it is necessary to introduce additional triplets
or singlets in order to provide a rich enough flavour structure to induce
a non-vanishing CP asymmetry in triplet decays~\cite{triplet_lepto}.
This brings back into the game a number of flavour unknowns.
On the contrary, in the scenario that we consider, the additional states
are heavy quarks and leptons whose masses and couplings are
determined in terms of low-energy parameters through SO(10) relations.
The generated baryon asymmetry then directly depends on the light
neutrino parameters.

In this paper, we pursue the exploration of this scenario by studying
the flavour- and CP-violating effects induced by the couplings of the heavy
states to the MSSM squarks and sleptons. Assuming flavour-universal soft
supersymmetry breaking terms at the GUT scale, flavour-violating
observables are predicted up to a few unknown
scale parameters and to a mild model-dependent
uncertainty, thus allowing to test the scenario. Besides the contributions
of the type II seesaw triplet and of its SU(5) partners already studied
in Ref.~\cite{Rossi02}, the presence of heavy quarks and leptons
gives rise to additional
contributions to the slepton and squark soft terms. In particular, they induce
flavour and CP violation in the slepton singlet and squark doublet sectors,
which were absent in the SU(5) type II seesaw model.

The paper is organized as follows. In Section~\ref{sec:model}, we present
the main features of the SO(10) scenario we consider. Section~\ref{sec:lfv} contains
a qualitative discussion of its predictions for flavour and CP violation.
Model-building aspects including gauge coupling unification, doublet-triplet
splitting and proton decay are briefly discussed
in Section~\ref{sec:model-building}, and addressed in greater detail in
Appendices~\ref{app:SO10_breaking}, \ref{app:DT_splitting}
and~\ref{app:GCU}, where the main ingredients of a realistic model are given.
Section~\ref{sec:num} presents our numerical results. Finally, we give our
conclusions in Section~\ref{sec:conclusion}. The superpotential of the model,
the boundary conditions for Yukawa couplings and soft terms and a subset of
the renormalization group equations are displayed in Appendix~\ref{app:rges}.

\section{SO(10) unification with type II seesaw mechanism}   %
\label{sec:model}                                                                              %

We consider a supersymmetric SO(10) scenario with matter fields
in ${\bf 16}$ and ${\bf 10}$ representations and type II realization of the
seesaw mechanism, in which soft supersymmetry breaking terms arise
at the GUT scale or higher. Since we are interested in the
flavour-violating effects induced by the physics responsible for neutrino
masses, we assume that the soft terms are flavour universal at the GUT scale. 

Let us first describe the field content. In terms of SU(5) multiplets,
the three families of SM fermions  are described by
${\bf \bar 5}_i^{\scriptscriptstyle {\rm SU(5)}}
\oplus {\bf 10}_i^{\scriptscriptstyle {\rm SU(5)}}$, $i=1,2,3$. In conventional
SO(10) unification~\cite{SO10}, ${\bf \bar 5}_i^{\scriptscriptstyle {\rm SU(5)}}$
and ${\bf 10}_i^{\scriptscriptstyle {\rm SU(5)}}$ are unified in a single
${\bf 16}_i^{\scriptscriptstyle {\rm SO(10)}}$ together with a singlet
(right-handed) neutrino participating in the generation of light neutrino
masses through the type I seesaw mechanism. In this paper, we consider
the alternative possibility that ${\bf 10}_i^{\scriptscriptstyle {\rm SU(5)}}$
is embedded in a ${\bf 16}_i^{\scriptscriptstyle {\rm SO(10)}}$, while
${\bf \bar 5}_i^{\scriptscriptstyle {\rm SU(5)}}$ belongs to a
${\bf 10}_i^{\scriptscriptstyle {\rm SO(10)}}$ (for similar or related
approaches, see Ref.~\cite{10matter}). Having done this choice, we
drop from now on the ``SO(10)'' superscript on SO(10) representations, and
indicate the embedding of SU(5) representations into SO(10) representations
by a superscript: for instance, ${\bf \bar 5}_i^{\bf 10}$ is contained
in ${\bf 10}_i$, while 
${\bf \bar 5}_i^{\bf 16}$ belongs to ${\bf 16}_i$. Besides the three ${\bf 16}_i$
and ${\bf 10}_i$ matter multiplets, the model also involves a ${\bf 10}$,
a ${\bf 16}$ and a ${\bf 54}$ Higgs multiplets. These are needed in order
to generate the quark and charged lepton masses, as well as the neutrino
masses through the type II seesaw mechanism.
The corresponding SO(10) superpotential reads:
\begin{eqnarray}
  W & = & W_\text{Yukawa}\ +\ W_\text{seesaw}\ +\ W_\text{GUT}\ +\ W_\text{nonren.}\, ,
\label{eq:W}  \\
  W_\text{Yukawa} & = & \frac{1}{2}\, y_{ij} {\bf 16}_i {\bf 16}_j {\bf 10}\,
    +\, h_{ij} {\bf 16}_i {\bf 10}_j {\bf 16}\ ,
\label{eq:W_Yuk}  \\
  W_\text{seesaw} & = & \frac{1}{2}\, f_{ij} {\bf 10}_i {\bf 10}_j {\bf 54}\,
    +\, \frac{1}{2}\, \sigma {\bf 10}\, {\bf 10}\, {\bf 54}\, +\, \frac{1}{2}\, M_{54} {\bf 54}^2\ ,
\label{eq:WY54}
\end{eqnarray}
where $W_\text{GUT}$, whose explicit form is given in
Appendices~\ref{app:SO10_breaking} and~\ref{app:DT_splitting},
contains the terms responsible for SO(10) symmetry breaking
and for the doublet-triplet splitting, and $W_\text{nonren.}$ includes
the non-renormalizable operators needed to account for the measured
ratios of down quark and charged lepton masses.
The role of the ${\bf 10}$, ${\bf 16}$ and ${\bf 54}$ representations is the
following. The ${\bf 10}$ contains the $Y=+1/2$ MSSM Higgs doublet $h_u$
responsible for up quark masses. The ${\bf 16}$ plays  a double role: it shares
the $Y=-1/2$ MSSM Higgs doublet $h_d$ with the ${\bf 10}$ and (together
with its ${\bf \overline{16}}$ companion) it reduces
the rank of the unified gauge group down to 4 through the vev of its
SU(5)-singlet component; moreover, this vev pairs up the spare
${\bf \bar 5}_i^{\bf 16}$ and ${\bf 5}_i^{\bf 10}$ and gives them a large mass,
leaving only the ${\bf \bar 5}_i^{\bf 10}$ massless. The ${\bf 54}$ contains
the SU(2)$_{\rm L}$ triplet
mediating the type II seesaw mechanism. The three singlet neutrinos
contained in the ${\bf 16}_i$, on the other hand, do not couple to the light
neutrinos at the renormalizable level.
Assuming for definiteness that they acquire GUT-scale masses (e.g. from
couplings ${\bf 1}_i {\bf 16}_j {\bf \overline{16}}$ to three SO(10) singlets
${\bf 1}_i$), we can completely neglect their contributions to the light neutrino
masses, if any, and we are left with a pure type II seesaw mechanism.

In writing Eqs.~(\ref{eq:W_Yuk}) and~(\ref{eq:WY54}), we assumed that
the superpotential is invariant under a matter parity, with the matter fields
being the ${\bf 16}_i$ and the ${\bf 10}_i$. We also required the absence
of mass terms of the form ${\bf 10}_i {\bf 10}_j$ and of additional interactions
that would induce a vev for the ${\bf 54}$, in order to prevent a mixing
between the ${\bf \bar 5}_i^{\bf 10}$ and the ${\bf \bar 5}_i^{\bf 16}$.
The absence of such a mixing ensures that the light neutrino masses
arise from a pure type II seesaw mechanism, and is a crucial feature
of the scenario.
The scale of light neutrino masses requires $M_{54}$ to be smaller than
the GUT scale; it is therefore natural to assume that this mass term arises at the
non-renormalizable level, while it is forbidden or small at the renormalizable
level. Furthermore, the leptogenesis scenario proposed in Ref.~\cite{FHLR08}
requires that the ${\bf 24}^{\bf 54}$ component of the ${\bf 54}$ be heavier
than the $({\bf 15} \oplus {\bf \overline{15}})^{\bf 54}$ one.
We refer the reader to Appendix~\ref{app:GCU} for the discussion
of the splitting of the $\bf 54$.

The degrees of freedom surviving below the GUT scale are the MSSM states,
which are massless before electroweak symmetry breaking, three heavy pairs
of vector-like matter ${\bf \bar 5}_i^{\bf 16} \oplus {\bf 5}_i^{\bf 10}$, and the
components of the ${\bf 54}$, which are also heavy.
The light MSSM fields and the heavy ones are embedded in
the SO(10) representations of the model as follows:
\globallabel{eq:embedding}
\begin{align}
{\bf 16}_i\ & =\ {\bf 10}^{\bf 16}_i \oplus {\bf \bar 5}_i^{\bf 16} \oplus {\bf 1}_i^{\bf 16}
  & & =\ (q_i, u^c_i, e^c_i) \oplus (D^c_i,L_i) \oplus n^c_i\ ,  \notag  \\
{\bf 10}_i\ & =\ {\bf \bar 5}_i^{\bf 10} \oplus {\bf 5}_i^{\bf 10}
  & & =\ (d^c_i,l_i) \oplus (\bar D^c_i, \bar L_i)\ ,  \notag  \\
\hskip 1.5cm {\bf 54}\ & =\ {\bf 24}^{\bf 54} \oplus {\bf 15}^{\bf 54} \oplus {\bf \overline{15}}^{\bf 54}
  & & =\ (S, T, O, V, \bar V) \oplus (\Delta, \Sigma, Z)
  \oplus (\bar \Delta, \bar \Sigma, \bar Z)\ ,  \hskip 1cm  \\
{\bf 10}\ & =\ {\bf \bar 5}^{\bf 10} \oplus {\bf 5}^{\bf 10}
  & & =\ (\cdots, \cos \theta_H h_d) \oplus (\cdots, h_u)\ ,  \notag  \\
{\bf 16}\ & =\ {\bf 10}^{\bf 16} \oplus {\bf \bar 5}^{\bf 16} \oplus {\bf 1}^{\bf 16}
  & & =\ \cdots \oplus (\cdots , \sin \theta_H h_d) \oplus \cdots\ ,  \notag
\end{align}
where dots stands for heavy fields which have been integrated out at
the GUT scale (this includes in particular the coloured Higgs triplets
mediating $D=5$ proton decay). We have introduced an angle $\theta_H$
parametrizing the mixing among $Y=-1/2$ Higgs doublets
($0 < \theta_H < \pi/2$ with no loss of generality)
and assumed that the $Y=+1/2$ MSSM Higgs
doublet entirely resides in the $\bf 10$, so that all components of the
$\bf \overline{16}$ are heavy (see Appendix~\ref{app:DT_splitting} for an
explicit realization of this). 
The \SM\ quantum numbers of the ${\bf 54}$ components are
$\Delta = (1, 3)_{+1}$, $\Sigma = (6,1)_{-2/3}$, $Z = (3,2)_{+1/6}$,
$S = (1,1)_0$, $T = (1,3)_0$, $O = (8,1)_0$ and $V \sim (3,2)_{-5/6}$,
while the heavy $L_i$ and $D^c_i$ obviously carry the same quantum
numbers as the light $l_i$ and $d^c_i$. After breaking of the SO(10)
gauge symmetry, the heavy quark and lepton fields acquire
Dirac masses $(M_{D^c})_{ij} D^c_i \bar D^c_j + (M_L)_{ij} L_i \bar L_j$,
where, at the renormalizable level and neglecting the renormalization
group running below the GUT scale:
\begin{equation}
\globallabel{eq:Diracmasses}
  (M_{D^c})_{ij}\, =\, (M_L)_{ij}\, =\, h_{ij} V_1\ , 
\end{equation}
in which $V_1$ is the vev of the SU(5)-singlet component of the ${\bf 16}$.
As for the SM fermion masses, they are given by (again neglecting
corrections from non-renormalizable operators and the RG running of
the Yukawa couplings):
\begin{equation}
\begin{aligned}
\label{eq:fermionmasses}
  (M_u)_{ij}\, &=\, y_{ij} v_u\ , \\
  (M_e)_{ij}\, &=\, \sin \theta_H h_{ij} v_d\ , \\
  (M_d)_{ij}\, &=\, \sin \theta_H h^T_{ij} v_d\ ,
\end{aligned}
\end{equation}
and the neutrino masses are given by the type II seesaw formula:
\begin{equation}
\label{eq:seesaw}
  (M_\nu)_{ij}\, =\, - \frac{\sigma v^2_u}{2 M_\Delta}\, f_{ij}\ .
\end{equation}
Note that the down quark and charged lepton masses, which satisfy the
SU(5) relation $M_d = M^T_e$ at the renormalizable level, are proportional
to $\sin \theta_H$, the $Y=-1/2$ Higgs mixing parameter\footnote{This
offers the possibility of explaining the smallness of the bottom/top mass
hierarchy for moderate values of $\tan \beta$ in terms of a small Higgs mixing
angle $\theta_H$. For reasons explained in Section~\ref{subsec:DTS},
however, we shall not use this possibility here.}.
Due to the way the SM fermions are embedded into SO(10)
representations, the up quark mass matrix is not correlated with the down
quark and charged lepton mass matrices as in conventional SO(10) models,
which accounts in a natural way for the stronger mass hierarchy
observed in the up quark sector.

The role of the SO(10) symmetry is to relate the masses and couplings
of the heavy matter fields $(L_i, \bar L_i)$ and $(D^c_i, \bar D^c_i)$
to the ones of the light fermions, thus allowing to predict their contributions
to observables such as the baryon asymmetry of the universe and
the rates of flavour-violating processes.
Indeed, Eqs.~(\ref{eq:W_Yuk})--(\ref{eq:WY54})
and~(\ref{eq:Diracmasses})--(\ref{eq:seesaw}) show that
all flavour parameters involved in the superpotential,
as well as the masses of the heavy matter fields, are determined
by the SM fermion masses and mixings, up to flavour-blind factors
and to the non-renormalizable contributions
needed to make $M_d \neq M^T_e$. On the contrary, in conventional
SO(10) models where neutrino masses arise from the type I seesaw
mechanism, the Lagrangian below the GUT scale also depends on the
flavour parameters encoded in the so-called $R$ matrix~\cite{CI01}.
The SO(10) scenario studied in this paper therefore has a higher predictive
power. As for the unknowns associated with the non-renormalizable
operators in $W_\text{nonren.}$, we will argue in Section~\ref{subsec:MdMe}
that they are unlikely to affect our results in a sizable way.

\section{Flavour and CP violation}        %
\label{sec:lfv}                                             %

\subsection{General structure of radiative corrections to the MSSM soft terms}

The $f_{ij} {\bf 10}_i {\bf 10}_j {\bf 54}$ interactions introduce a new flavour
structure, directly related to the light neutrino masses, on top of the MSSM one.
As noted in Ref.~\cite{JR06}, this can be considered as a truly minimal extension
to the lepton sector of the minimal flavour violation hypothesis in the quark
sector, which does not rely on flavour basis dependent assumptions.
In any case, the $f_{ij} {\bf 10}_i {\bf 10}_j {\bf 54}$ interactions give rise to
new flavour- and CP-violating effects at low energy because of their well-known
impact on the MSSM soft terms through radiative corrections~\cite{Rossi02}.
In addition, the $y_{ij} {\bf 16}_i {\bf 16}_j {\bf 10}$ interactions induce
new flavour and CP violation in the slepton singlet sector due to the presence
of heavy lepton doublets in the ${\bf 16}_i$, as we are going to see.

In order to compute these effects, we must write the renormalisation group
equations (RGEs) for the MSSM parameters and integrate them. Before
doing so, let us note that the $l_i$ and $L_i$ fields
can mix, since they have the same quantum numbers
after breaking of the SO(10) symmetry (and similarly for the $d^c_i$ and $D^c_i$ fields).
Hence, the superpotential mass term for the lepton doublets reads,
in full generality, $L_i (M_L)_{ij} \bar L_j + l_i (M_{lL})_{ij} \bar L_j$.
The second term was omitted previously since $M_{lL}$ vanishes
at the tree level, while $M_L$ is given by Eq.~(\ref{eq:Diracmasses}).
At one loop, however, wave function renormalization induces a small
$l_i$-$L_j$ mixing:
\begin{equation}
\label{eq:leadinglog}
  (M_{lL})_{ij}\ \approx\ -\frac{\sin 2 \theta_H}{32 \pi^2}\ (M_L y^\dagger h)_{ij}
    \ln \left( \frac{M_{\rm GUT}}{\mu} \right) .
\end{equation}
This in turn necessitates a redefinition of the heavy and light lepton
doublets, which affects all couplings involving lepton doublets. However,
the size of the effect is small (especially for small or moderate $\tan \beta$),
and we shall neglect it in the following.

Before solving numerically the full 1-loop RGEs,
let us illustrate the main features of the results by using the leading-log
approximation for the exchange of the heavy degrees of freedom.
In order to be able to identify each contribution, we assume that all
components of the $\bf 54$, namely $S$, $T$, $O$ and the vector-like pairs
($V, \bar V$) ($\Delta, \bar \Delta$), ($\Sigma, \bar \Sigma$) and ($Z, \bar Z$)
have different masses (the mass $M_Z$ of the ($Z$, $\bar Z$) vector-like pair
should not be confused with the $Z$ boson mass $m_Z$). We also consider
the possibility that $M_{D^c} \neq M_L$, but ignore the effect of corrections
to the mass relation $M_d = M^T_e$. Let us denote by $m^2_{16}$ and
$m^2_{10}$ the universal soft terms for the three families of ${\bf 16}_i$
and ${\bf 10}_i$ matter fields at the GUT scale; by $m_{54}$, $m^2_{16_H}$
and $m^2_{10_H}$ the soft terms for the $\bf 54$, $\bf 16$ and $\bf 10$
SO(10) multiplets; and by $a_0$ the universal $A$-term (defined by
$A_x = a_0 x$, where $x$ is any  superpotential trilinear coupling $x$).
Using the RGEs given in Appendix~\ref{app:rges}, we obtain (in matrix form):
\globallabel{eq:LL}
\begin{align}
m^2_l\ &=\ (m^2_l)_\text{MSSM}\, -\, \frac{1}{(4\pi)^2}
    \left( 2 m^2_{10} + m^2_{54} + a_0^2 \right)
    f^\dagger\, \bigg[ \frac{3}{2} \ln \frac{M^2_\text{GUT}}{M_\Delta^2}
    + \frac{3}{2} \ln \frac{M^2_\text{GUT}}{M_Z^2}  \notag \\ 
  & \hspace{0.5cm} + \frac{3}{4} \ln \frac{M^2_\text{GUT}}{M_T^2 + M^T_L M^*_L}
    + \frac{3}{20} \ln \frac{M^2_\text{GUT}}{M_S^2 + M^T_L M^*_L}
    + \frac{3}{2} \ln \frac{M^2_\text{GUT}}{M_V^2 + M^T_{D^c} M^*_{D^c}}
    \bigg]\, f\ ,  \mytag \\
m^2_{d^c}\ &=\ (m^2_{d^c})_\text{MSSM}\, -\, \frac{1}{(4\pi)^2}
    \left( 2 m^2_{10} + m^2_{54} + a_0^2 \right)
    f\, \bigg[ 2 \ln \frac{M^2_\text{GUT}}
    {M_\Sigma^2} + \ln \frac{M^2_\text{GUT}}{M_Z^2}  \notag \\
  & \hspace{0.5cm}
    + \ln \frac{M^2_\text{GUT}}{M_V^2 + M^\dagger_L M^{\phantom{\dagger}}_L}
    + \frac{4}{3} \ln \frac{M^2_\text{GUT}}
    {M_O^2 + M^\dagger_{D^c} M^{\phantom{\dagger}}_{D^c}}
    + \frac{1}{15} \ln \frac{M^2_\text{GUT}}
    {M_S^2 + M^\dagger_{D^c} M^{\phantom{\dagger}}_{D^c}}
    \bigg]\, f^\dagger\ .  \mytag
\end{align}
In contrast to the well-known type I seesaw case, the flavour-violating
corrections to $m^2_l$ are determined
by the light neutrino mass matrix, with no ambiguity due to high-energy
flavour parameters. Therefore, while the absolute rate of a given LFV process
is not known, the correlations between different LFV channels are predicted
with little uncertainty (at least if the contributions of the slepton singlet sector,
to be discussed below, are subdominant). The first term in the squared 
brackets of Eq.~(\ref{eq:LL}a) is induced by the seesaw triplet interactions
and is present in all type II seesaw models. The second term, as well as
the first two terms in the squared  brackets of Eq.~(\ref{eq:LL}b), corresponds
to the contribution of the SU(5) partners of the triplet~\cite{Rossi02}.
The additional terms in $m^2_l$ and $m^2_{d^c}$ are due to the presence
of heavy leptons and quarks in the ${\bf 16}_i$, and are characteristic
of the model studied in this paper.
In the limit $M_L = M_{D^c}$, $M_\Delta = M_\Sigma = M_Z$,
$M_S = M_T = M_O = M_V$, the correction to the MSSM evolution
is the same for $m^2_l$ and $(m^2_{d^c})^T$, as dictated by the SU(5)
invariance of the interactions from which they arise. 

Another difference from other type II seesaw models comes from
the corrections to the MSSM running of $m^2_{e^c}$ and $m^2_q$:
\globallabel{eq:LL10}
\begin{align}
m^2_{e^c}\ & =\ (m^2_{e^c})_\text{MSSM}\, -\, \frac{\cos^2 \theta_H}{(4\pi)^2}
    \left( 2m^2_{16} + m^2_{h_d} + a_0^2 \right) y\, \bigg[
    2 \ln \frac{M^2_\text{GUT}}{M^*_{L} M^T_{L}} \bigg]\, y^\dagger\ , \mytag \\
m^2_{q}\ &=\ (m^2_{q})_\text{MSSM}\, -\, \frac{\cos^2 \theta_H}{(4\pi)^2}
    \left( 2m^2_{16} + m^2_{h_d} + a_0^2 \right) y^\dagger \bigg[
    \ln \frac{M^2_\text{GUT}}{M^{\phantom{\dagger}}_{D^c} M^\dagger_{D^c}}
    \bigg]\, y\ , \mytag
\end{align}
where $m^2_{h_d} = \cos^2\! \theta_H\, m^2_{10_H}\!
+ \sin^2\! \theta_H\, m^2_{16_H}$.
These are controlled by the up quark Yukawa couplings (evolved at the high
scale). While the corrections to $m^2_q$ do not represent a deviation from
the minimal flavour violation structure of the MSSM radiative corrections,
the corrections to $m^2_{e^c}$ are similar to the ones induced above the GUT
scale by the top quark Yukawa coupling in SU(5) models~\cite{barbieri-hall},
although their origin is different.

\subsection{Analytic approximations for the mass insertions and CP violation}
\label{subsec:analytic}

In order to be able to quickly estimate the size of various flavour- and
CP-violating observables, it is useful to provide analytic expressions
for the mass insertions parameters ($i \neq j$)~\cite{HKR86}:
\begin{equation}
  (\delta^e_{LL})_{ij}\, \equiv\, \frac{(m^2_l)_{ij}}{\overline m^2_{\tilde e_L}}\ , \quad
  (\delta^e_{RR})_{ij}\, \equiv\, \frac{(m^2_{e^c})_{ij}}{\overline m^2_{\tilde e_R}}\ , \quad
  (\delta^e_{RL})_{ij}\, \equiv\, \frac{(A_e)_{ij} v_d}
    {\overline m_{\tilde e_L} \overline m_{\tilde e_R}}\ ,
\label{eq:MIs}
\end{equation}
where $\overline m_{\tilde e_L}$ ($\overline m_{\tilde e_R}$) is an average
doublet (singlet) slepton mass, and analogous quantities are defined in the up
and down squark sectors.
These parameters can be straightforwardly derived from Eqs.~(\ref{eq:LL})
and~(\ref{eq:LL10}), but some care is needed regarding CP-violating
phases.
Even if we take real boundary conditions for the soft supersymmetry
breaking parameters at $M_{\rm GUT}$, we will end up with complex
soft terms at the weak scale because their RGEs involve
complex couplings. These couplings contain, in addition to the CKM
and PMNS phases, extra CP-violating phases
inherited from the SO(10) structure. In order to identify the latter, it is
useful to write the SO(10) Yukawa couplings in an appropriate
basis for the SO(10) matter multiplets ${\bf 16}_i$ and ${\bf 10}_i$,
namely\footnote{In the following discussion, we neglect the effect of radiative
corrections in the Yukawa sector. This allows us to disentangle the CKM
and PMNS phases from the extra SO(10) phases, which would otherwise
be mixed by the renormalization group running.}:
\begin{equation}
  y\ =\ U^T_q D_y U_q\ , \qquad h\ =\ D_h\ , \qquad f\ =\ U^* D_f U^\dagger\ ,
\label{eq:Y_SO10}
\end{equation}
where
\begin{equation}
  U_q\ =\ \mbox{Diag} (e^{i\Phi^u_1}, e^{i\Phi^u_2}, e^{i\Phi^u_3})\,
    V\, \mbox{Diag} (e^{i\Phi^d_1}, e^{i\Phi^d_2}, e^{i\Phi^d_3})\ .
\label{eq:phases_SO10}
\end{equation}
In Eqs.~(\ref{eq:Y_SO10}) and~(\ref{eq:phases_SO10}), $D_y$, $D_h$,
$D_f$ are real diagonal matrices, $U$ and $V$ are the PMNS and CKM
matrices in the standard PDG parametrization (extended to include two
``Majorana'' phases $\rho$ and $\sigma$ in the PMNS case), and
$\Phi^u_i$, $\Phi^d_i$ ($i = 1,2,3$) are extra SO(10) phases,
five of which are independent (one can impose e.g. $\Phi^u_3 = 0$).
For simplicity, we neglect the effects of the non-renormalizable operators
needed to correct the mass relation $M_d = M^T_e$, and we therefore
also assume $M_{D^c} = M_L = D_h V_1$.
Below the GUT scale, we are free to rephase independently the MSSM fields
so that the Yukawa couplings $\lambda_{u,d,e}$ and $f_\Delta$ only contain
CKM and PMNS phases; however, the extra SO(10) phases will reappear
in other couplings, such as $\hat \lambda_d$ and $f_Z$, 
which enter the RGEs for the MSSM soft terms (we refer
to Appendix~\ref{app:rges} for the definition of the superpotential couplings
below the GUT scale). The outcome of this
is that the mass insertion parameters depend on the phase differences
$\Phi^u_i - \Phi^u_j$ and $\Phi^d_i - \Phi^d_j$ in addition to the 4
low-energy phases $\delta_{CKM}$, $\delta_{PMNS}$, $\rho$ and $\sigma$.

We are now ready to write the mass insertion parameters
$(\delta^{e,d,u}_{MN})_{ij}$ ($M, N = L, R$) in the leading-log approximation,
assuming for simplicity $M_\Delta = M_\Sigma = M_Z \equiv M_{15}$,
$M_S = M_T = M_O = M_V \equiv M_{24}$, a common soft supersymmetry
breaking mass $m_0$ for all SO(10) chiral multiplets and, as before,
a common A-terms $a_0$ for all superpotential trilinear couplings. We obtain:
\begin{eqnarray}
  16\pi^2 (\delta^e_{LL})_{ij} & \approx &
    -\, \frac{3 m^2_0 + |a_0|^2}{\overline m^2_{\tilde e_L}}\ C^f_{ij}\ ,  \\
  16\pi^2 (\delta^e_{RR})_{ij} & \approx & 
    -\, \frac{2 (3m^2_0 + |a_0|^2)}{\overline m^2_{\tilde e_R}}\
    \mbox{e}^{i(\Phi^d_i - \Phi^d_j)}\, (C^{\lambda_u}_{ij})^*\ ,  \\
  16\pi^2 (\delta^e_{RL})_{ij} & \approx &
    -\, \frac{3 a_0}{2 \overline m_{\tilde e_L}\! \overline m_{\tilde e_R}}\, \left[\, m_{e_i} C^f_{ij}
    + 3\, \mbox{e}^{i(\Phi^d_i - \Phi^d_j)}\, (C^{\lambda_u}_{ij})^* m_{e_j}\, \right] ,
\label{eq:delta_e}
\end{eqnarray}
\begin{eqnarray}
  16\pi^2 (\delta^d_{LL})_{ij} & \approx & -\, \frac{3 m^2_0 + |a_0|^2}{\overline m^2_{\tilde d_L}}\,
    \left[\, (C^{\lambda_u}_{\scriptscriptstyle \rm MSSM})_{ij} + C^{\lambda_u}_{ij}\, \right] ,  \\
  16\pi^2 (\delta^d_{RR})_{ij} & \approx & -\, \frac{3 m^2_0 + |a_0|^2}{\overline m^2_{\tilde d_R}}\
      \mbox{e}^{i(\Phi^d_i - \Phi^d_j)}\, (C^f_{ij})^* ,  \\ 
  16\pi^2 (\delta^d_{RL})_{ij} & \approx &
    -\, \frac{3 a_0}{2 \overline m_{\tilde d_L}\! \overline m_{\tilde d_R}}\,
    \left\{ m_{d_i} \left[\, (C^{\lambda_u}_{\scriptscriptstyle \rm MSSM})_{ij}
    + 3\, C^{\lambda_u}_{ij}\, \right]
    +\, \mbox{e}^{i(\Phi^d_i - \Phi^d_j)}\, (C^f_{ij})^* m_{d_j} \right\} ,
\label{eq:delta_d}
\end{eqnarray}
\begin{eqnarray}
  16\pi^2 (\delta^u_{LL})_{ij} & \approx & -\, \frac{3 m^2_0 + |a_0|^2}{\overline m^2_{\tilde u_L}}\,
    \left[\, (C^{\lambda_d}_{\scriptscriptstyle \rm MSSM})_{ij}
    + \mbox{e}^{-2i (\Phi^u_i - \Phi^u_j)}\, D^{\lambda_u}_{ij}\, \right] ,  \\
  16\pi^2 (\delta^u_{RR})_{ij} & \approx & 0\ ,  \\
  16\pi^2 (\delta^u_{RL})_{ij} & \approx &
    -\, \frac{3 a_0}{2 \overline m_{\tilde u_L}\! \overline m_{\tilde u_R}}\  m_{u_i}
    \left[\, (C^{\lambda_d}_{\scriptscriptstyle \rm MSSM})_{ij}
    + \mbox{e}^{-2i (\Phi^u_i - \Phi^u_j)}\, D^{\lambda_u}_{ij}\, \right] ,
\label{eq:delta_u}
\end{eqnarray}
where the dependence
on the low-energy flavour parameters and phases is encapsulated
in the coefficients $(C^{\lambda_u}_{\scriptscriptstyle \rm MSSM})_{ij}$,
$(C^{\lambda_d}_{\scriptscriptstyle \rm MSSM})_{ij}$, $C^f_{ij}$,
$C^{\lambda_u}_{ij}$ and $D^{\lambda_u}_{ij}$.
The first two correspond to the CKM-induced MSSM contributions
and are given by:
\bea
  & (C^{\lambda_u}_{\scriptscriptstyle \rm MSSM})_{ij}\, \simeq\, \lambda^2_t\, V^*_{ti} V_{tj}\,
    \ln \left( \frac{M^2_{\rm GUT}}{m^2_Z} \right) ,  \quad
  (C^{\lambda_d}_{\scriptscriptstyle \rm MSSM})_{ij}\, =\, \sum_k\, \lambda^2_{d_k} V_{ik} V^*_{jk}\,
    \ln \left( \frac{M^2_{\rm GUT}}{m^2_Z} \right) ,
\label{eq:C_MSSM}
\eea
where terms suppressed by $\lambda_c / \lambda_t$ have been neglected
in $(C^{\lambda_u}_{\scriptscriptstyle \rm MSSM})_{ij}$. The other three
coefficients correspond to the contributions of the heavy states present
below the GUT scale and also contain a dependence on their masses.
The one that involves the seesaw couplings is given by:
\bea
  & C^f_{ij}\, =\, 3\, (f^\dagger f)_{ij}\, \ln \left( \frac{M^2_{\rm GUT}}{M^2_{15}} \right)
    +\frac{12}{5}\, \sum_a\, f^*_{ai} f_{aj}\,
    \ln \left( \frac{M^2_{\rm GUT}}{M^2_{24}+M^2_{5_a}} \right)  \hskip 5cm  \nonumber  \\
  & =\, 3\, \sum_k\, f^2_k\, U_{ik} U^*_{jk}\, \ln \left( \frac{M^2_{\rm GUT}}{M^2_{15}} \right)
    +\frac{12}{5}\, \sum_{k,l}\, f_k f_l\, U_{ik} U^*_{jl}\, \sum_a\, U_{ak} U^*_{al}\,
    \ln \left( \frac{M^2_{\rm GUT}}{M^2_{24}+M^2_{5_a}} \right) ,
\label{eq:C_f}
\eea
where $f_i = (2 M_\Delta / \sigma v^2_u)\, m_{\nu_i}$ and the $M_{5_a}$ ($a=1,2,3$)
are the common masses of the heavy quarks and leptons,
and the coefficients that involve the up quark Yukawa couplings read:
\bea
  & C^{\lambda_u}_{ij}\, \simeq\, \cos^2\! \theta_H\, \lambda^2_t\, V^*_{ti} V_{tj}\,
    \ln \left( \frac{M^2_{\rm GUT}}{M^2_{5_3}} \right) ,  \quad
  D^{\lambda_u}_{ij}\, =\, \cos^2\! \theta_H\, \lambda_{u_i} \lambda_{u_j}
    \sum_a\, V^*_{ia} V_{ja}\, \ln \left( \frac{M^2_{\rm GUT}}{M^2_{5_a}} \right) , 
\label{eq:D_u}
\eea
where terms suppressed by $\lambda_c / \lambda_t$ have been neglected
in $C^{\lambda_u}_{ij}$. Due to this (very good) approximation, only the CKM
and PMNS phases appear in Eqs.~(\ref{eq:C_MSSM}) to~(\ref{eq:D_u}).

Let us have a closer look at the mass insertion parameters~(\ref{eq:delta_e})
to~(\ref{eq:delta_u}). In the up squark sector, the new contributions are
subdominant with respect to the MSSM corrections, which are enhanced
by $\tan^2\! \beta$ and by a large logarithm. The situation is more interesting
in the down squark and charged slepton sectors. In the former, the new
corrections controlled by the top quark Yukawa coupling are again
subdominant with respect to the MSSM corrections, while they lead to
$RR$ flavour violation in the latter.
The new corrections induced by the seesaw couplings, on the other hand,
are not suppressed by small CKM angles
and can give rise to large flavour violations
in the RR down squark sector and in the LL slepton sector~\cite{Rossi02}.
One observes the following correlations:
\begin{eqnarray}
  \overline m^2_{\tilde d_R} (\delta^d_{RR})_{ij} & = &
    \overline m^2_{\tilde e_L} (\delta^e_{LL})^*_{ij}\, \mbox{e}^{i(\Phi^d_i - \Phi^d_j)}\ ,
\label{eq:correlations_dRR}  \\
  \overline m^2_{\tilde e_R} (\delta^e_{RR})_{ij} & = &
    2 R\, \overline m^2_{\tilde d_L}\! (\delta^d_{LL})^*_{ij}\, \mbox{e}^{i(\Phi^d_i - \Phi^d_j)}\ ,
\label{eq:correlations_eRR}
\end{eqnarray}
where $R \equiv \cos^2\! \theta_H \ln ( M^2_{\rm GUT} / M^2_{5_3} ) /
\left[\, \ln (M^2_{\rm GUT} / m^2_Z) + \cos^2\! \theta_H \ln (M^2_{\rm GUT} / M^2_{5_3})\, \right]$.
The first one is characteristic of the SU(5) extension of the type II seesaw
mechanism, while the second one is analogous (although from a different
origin) to the one arising from the running between $M_P$ and $M_{\rm GUT}$
in the minimal SU(5) model. In the case that the seesaw-induced corrections
dominate, one further has:
\begin{equation}
 (\delta^e_{RR})_{ij}\ \ll\ (\delta^e_{LL})_{ij}\ , \qquad
    (\delta^d_{LL})_{ij}\ \ll\ (\delta^d_{RR})_{ij}\ ,
\label{eq:fij_dominance}
\end{equation}
\begin{equation}
  \overline m_{\tilde d_R} \overline m_{\tilde d_L} (\delta^d_{RL})_{ij}\ =\
    \overline m_{\tilde e_R} \overline m_{\tilde e_L} (\delta^e_{RL})_{ji}\,
    \mbox{e}^{i(\Phi^d_i - \Phi^d_j)}\ .
\label{eq:correlations_dRL}
\end{equation}
Eqs.~(\ref{eq:correlations_dRR}), (\ref{eq:correlations_eRR})
and~(\ref{eq:correlations_dRL}) correlate the size of flavour-changing
neutral currents (FCNCs) in the lepton and B/K sectors, as well as the
charged lepton and quark electric dipole moments (EDMs), as we are
going to see.

For illustration, we give in Table~\ref{tab:deltas_numerical} numerical
values for the leptonic $\delta$'s. These values were obtained by numerically
solving the RGEs
for the following choice of high-energy parameters, motivated
by the leptogenesis scenario of Ref.~\cite{FHLR08}:
$M_\Delta = M_\Sigma = M_Z \equiv M_{15} = 10^{12}\, \mbox{GeV}$,
$M_S = M_T = M_O = M_V \equiv M_{24} = 10^{13}\, \mbox{GeV}$,
$\sigma = 3.4\times 10^{-2}$,
$V_1 = M_{\rm GUT} = 2 \times 10^{16}\, \mbox{GeV}$
and $\tan \theta_H = 1$. For the heavy $({\bf 5}_i, {\bf \bar 5}_i)$ masses,
we used $M_{5_i} = m_{e_i} V_1 / \sin \theta_H v_d$,
neglecting SU(5)-breaking corrections. The supersymmetric parameters
were chosen to be $m_0 = M_{1/2} = 1.5\, \mbox{TeV}$, $a_0 = 0$,
and $\tan \beta = 10$ ($50$), yielding the following superpartner masses
(for $\tan \beta = 10$):
$M_{\tilde \chi^0_1} = 112\, \mbox{GeV}$,
$M_{\tilde \chi^+_1} = 218\, \mbox{GeV}$, $M_{\tilde g} = 723\, \mbox{GeV}$,
sleptons around $(1.5 - 1.6)\, \mbox{TeV}$ and first two generation squarks
around $1.8\, \mbox{TeV}$ (as well as $\mu = 562\, \mbox{GeV}$).
Finally, we took the best fit values of Ref.~\cite{Schwetz:2008er} for the
measured neutrino oscillation parameters, together with $m_1=0.005 \eV$
and $\sin^2\theta_{13} = 0$ ($0.05$) for the yet unknown parameters.
All phases were set to zero.
The resulting values of the leptonic mass insertions shown in
Table~\ref{tab:deltas_numerical} can be compared with the bounds
coming from the non-observation of LFV decays of charged
leptons (see e.g. Refs.~\cite{masina-savoy,paradisi,ciuchini}).
Although the sleptons are heavy in this example, the relatively
large values of the $\delta$'s lead to a branching ratio for
$\mu \to e \gamma$ just below the experimental bound for
$\tan \beta = 10$, and above it for $\tan \beta = 50$.
Using Eqs.~(\ref{eq:correlations_dRR}), (\ref{eq:correlations_eRR})
and~(\ref{eq:correlations_dRL}), one can also compare the figures in
Table~\ref{tab:deltas_numerical} with
the constraints on hadronic $\delta$'s coming from $B$ and $K$
physics (see e.g. Refs.~\cite{GGMS96,ciuchini}).

\begin{table}\begin{center}
\begin{tabular}{|c||c|c|c|c|}
\hline
$\sin^2\theta_{13}$ & 0.05 & 0 & 0.05 & 0 \\
$\tan\beta$ & 10 & 10 & 50 & 50 \\
\hline\hline
$|\delta_{e\mu}^{LL}|$
& $2.0 \times 10^{-3}$ & $1.7 \times 10^{-4}$
& $2.0 \times 10^{-3}$ & $1.5 \times 10^{-4}$ \\
$|\delta_{e\tau}^{LL}|$
& $1.8 \times 10^{-3}$ & $1.0 \times 10^{-4}$
& $1.9 \times 10^{-3}$ & $1.4 \times 10^{-4}$ \\
$|\delta_{\mu\tau}^{LL}|$
& $5.7 \times 10^{-3}$ & $5.9 \times 10^{-3}$
& $6.1 \times 10^{-3}$ & $6.3 \times 10^{-3}$ \\
\hline
$|\delta_{e\mu}^{RR}|$
& $2.1 \times 10^{-6}$ & $2.1 \times 10^{-6}$
& $2.8 \times 10^{-8}$ & $4.6 \times 10^{-7}$ \\
$|\delta_{e\tau}^{RR}|$
& $5.8 \times 10^{-5}$ & $5.4 \times 10^{-5}$
& $1.4 \times 10^{-4}$ & $1.2 \times 10^{-6}$ \\
$|\delta_{\mu\tau}^{RR}|$
& $4.3 \times 10^{-4}$ & $4.3 \times 10^{-4}$
& $2.9 \times 10^{-4}$ & $2.9 \times 10^{-4}$ \\
\hline
\end{tabular}
\caption{Numerical values of the leptonic mass insertion parameters
$\delta^e_{LL}$, $\delta^e_{RR}$ for two different values of
$\sin^2 \theta_{13}$ and $\tan \beta$ and for the choice of parameters
mentioned in the text.}
\label{tab:deltas_numerical}
\end{center}\end{table}

From Table~\ref{tab:deltas_numerical} we can see that the LL mass insertions
are the dominant source of lepton flavour violation in our numerical example,
i.e. we are in the situation described by Eq.~(\ref{eq:fij_dominance}). This is
due to the fact that the choice made for the model parameters leads to rather
large $f_{ij}$'s (namely $f_{33} \sim 0.1$): a smaller value of the ratio
$M_\Delta / \sigma$ would yield smaller $f_{ij}$ couplings, hence smaller
LL leptonic mass insertions, without affecting the RR mass insertions.
We remind the reader that, while the overall size of the $f_{ij}$'s is
controlled by $M_\Delta / \sigma$, their
flavour structure, up to small RGE effects, solely depends on the light
neutrino parameters. Hence, even though the predictions for LFV processes
span many orders of magnitude due to their strong dependence on the
unkown scale parameter $M_\Delta / \sigma$, the ratios of rates for
different flavour channels are predicted with a mild dependence on the
high-energy parameters, as long as the corrections controlled
by the seesaw couplings dominate.
This is precisely the case in our numerical example, in which the dominance
of the LL mass insertions allows to estimate the ratios of LFV decays,
as in usual type II seesaw models~\cite{Rossi02}:
\begin{equation}
  \frac{BR(\tau \rightarrow \mu\gamma)}{BR(\mu \rightarrow e\gamma)}\
  \approx\ \frac{|\delta^{LL}_{\mu \tau}|^2}{|\delta^{LL}_{e \mu}|^2}\
  \frac{BR(\tau \rightarrow \mu \overline{\nu}_\mu \nu_\tau)}
  {BR(\mu \rightarrow e \overline{\nu}_e \nu_\mu)}\ =\
    \left\{ \begin{array}{ll} \mathcal{O}(100) & (\sin^2\theta_{13}=0)  \\
    \mathcal{O}(1) & (\sin^2\theta_{13}=0.05)
  \end{array} \right.
\label{eq:tmg_meg}
\end{equation}
\begin{equation}
  \frac{BR(\tau \rightarrow e \gamma)}{BR(\mu \rightarrow e\gamma)}\
  \approx\ \frac{|\delta^{LL}_{e \tau}|^2}{|\delta^{LL}_{e \mu}|^2}\
  \frac{BR(\tau \rightarrow e \overline{\nu}_e \nu_\tau)}
  {BR(\mu \rightarrow e \overline{\nu}_e \nu_\mu)}\ =\
    \left\{ \begin{array}{ll} \mathcal{O}(0.1) & (\sin^2\theta_{13}=0)  \\
    \mathcal{O}(0.1) & (\sin^2\theta_{13}=0.05)
  \end{array} \right. ,
\label{eq:teg_meg}
\end{equation}
where we used $BR(\tau \rightarrow \mu \overline{\nu}_\mu \nu_\tau) /
BR(\mu \rightarrow e \overline{\nu}_e \nu_\mu)
\simeq BR(\tau \rightarrow e \overline{\nu}_e \nu_\tau) /
BR(\mu \rightarrow e \overline{\nu}_e \nu_\mu) \simeq 0.17$.
The present upper bound on $BR(\mu \rightarrow e \gamma)$ thus
excludes the possibility of observing $\tau \rightarrow e \gamma$
in foreseeable experiments, while $\tau \rightarrow \mu \gamma$ may be
accessible at a super-B factory~\cite{SFF} if $\theta_{13}$ is small.
These ratios may change if the SO(10) parameters
controlling the masses of the heavy states
are varied but, as long as Eq.~(\ref{eq:fij_dominance}) is
satisfied, the dependence is only logarithmic, while the branching ratios
themselves are roughly proportional to $(M_\Delta / \sigma)^4$.

Let us now turn our attention to CP violation. It is well known that, even if
the ``flavour-independent'' phases carried by the diagonal A-terms and
by the mu term (in the convention that the universal gaugino mass
parameter is real) are absent, significant contributions to the leptonic
and hadronic EDMs may arise from phases carried by flavour
non-diagonal soft terms~\cite{BS95,masina-savoy,FV_EDMs}.
In the mass insertion approach, this corresponds
to multiple insertions of $\delta$'s.
For down quark and charged lepton EDMs, the dominant ``flavour-violating''
contribution (coming respectively from gluino and bino diagrams) is proportional
to:
\begin{equation}
  \mbox{Im} \left[\, (\delta^f_{LL})_{i3}\, (\delta^f_{LR})_{33}\,
    (\delta^f_{RR})_{3i}\, \right]\ \approx\ - \mu \tan \beta\, m_{f_3}\,
    \mbox{Im} \left[\, (\delta^f_{LL})_{i3}\, (\delta^f_{RR})_{3i}\, \right]
    \qquad (f = d, e)\ ,
\label{eq:FV_EDMs}
\end{equation}
in which the suppression due to the double flavour change is
compensated for by the $\tan \beta\, m_{f_3}$ enhancement.
As for the up quark EDM, since $(\delta^u_{RR})_{ij} \approx 0$ its
flavour-violating contributions are suppressed by small Yukawa
couplings or CKM angles, making it much smaller than the down
quark EDM. Together with Eqs.~(\ref{eq:correlations_dRR})
and~(\ref{eq:correlations_eRR}),
this implies a correlation between the electron
and neutron EDMs, which depend on the same combination of phases
and flavour couplings:
\begin{equation}
  \mbox{Im} \left[\, \mbox{e}^{i(\Phi^d_3 - \Phi^d_1)}\,
    C^f_{13}\, C^{\lambda_u}_{13}\, \right] .
\end{equation}
Using formulae available in the literature (see e.g. Ref.~\cite{Hisano09}),
one can estimate the size of the charged lepton and neutron EDMs
for the choice of parameters made in Table~\ref{tab:deltas_numerical}.
With the superpartner masses given above and an average slepton
(down squark) mass $\overline m_{\tilde e} = 1.56\, \mbox{TeV}$
($\overline m_{\tilde d} = 1.8\, \mbox{TeV}$), one obtains,
for $\tan \beta = 10$:
\begin{eqnarray}
  d_{e_i} & \simeq & \left(- 1.3 \times 10^{-24}\, e\, \mbox{cm} \right)
\mbox{Im} \left[\, (\delta^e_{LL})_{i3}\, (\delta^e_{RR})_{3i}\, \right] ,  \\
  d_n & \simeq & \left( 1 \pm 0.5 \right) \left(- 2.9 \times 10^{-22}\, e\, \mbox{cm} \right)
\mbox{Im} \left[\, (\delta^d_{LL})_{13}\, (\delta^d_{RR})_{31}\, \right] .
\end{eqnarray}
For the neutron EDM, we used the formula $d_n = (1 \pm 0.5)\,
[ 1.4 (d_d - 0.25 d_u) + 1.1 e (d^c_d + 0.5 d^c_u) ]$~\cite{Pospelov05},
in which we only kept the dominant gluino contributions to down quark
electric and chromo-electric dipole moments, while we neglected
the up quark dipole moments.
The values of the $\delta$'s given in Table~\ref{tab:deltas_numerical},
together with
$|(\delta^d_{LL})_{13}\, (\delta^d_{RR})_{31}| = 2.2 \times 10^{-7}$
(for $\theta_{13} = 0$) and $3.5 \times 10^{-6}$
(for $\sin^2 \theta_{13} = 0.05$),
lead to $d_e \lesssim (7 \times 10^{-33} - 10^{-31})\, e\, \mbox{cm}$,
$d_\mu \lesssim 3 \times 10^{-30}\, e\, \mbox{cm}$ and
$d_n \lesssim (1 \pm 0.5)\, (6 \times 10^{-29} - 10^{-27})\, e\, \mbox{cm}$,
where the upper ranges correspond to varying $\theta_{13}$ from $0$ to its
experimental upper limit. These bounds are well below the present
experimental upper bounds (respectively
$1.6 \times 10^{-27}\, e\, \mbox{cm}$~\cite{d_e},
$1.9 \times 10^{-19}\, e\, \mbox{cm}$~\cite{d_mu}
and $2.9 \times 10^{-26}\, e\, \mbox{cm}$~\cite{d_n}),
while some tension with the constraints from LFV processes
is already present for the same values of the $\delta$'s. Therefore,
taking into account the FCNC constraints, we do not expect significant
flavour-violating contributions to EDMs in the scenario studied
in this paper\footnote{Of course, ``flavour-independent'' phases carried
by diagonal A-terms or by the mu term may still induce large EDMs.}.

A more promising CP-violating observable is the indirect CP violation
parameter in the kaon sector, $\varepsilon_K$. A recent evaluation
of the Standard Model prediction for $\varepsilon_K$ suggests
that its measured value still allows for a significant
supersymmetric contribution~\cite{buras-guadagnoli}.
In the mass insertion language, the relevant contributions are
$\mbox{Im}\, [ (\delta^d_{MN})_{12} (\delta^d_{PQ})_{12} ]$,
with $(M,N), (P,Q) \in \{ (L,L), (R,R) \}$ or $\{ (L,R), (R,L) \}$.
Since the LR mass insertions are suppressed by the down and strange
quark masses and $|(\delta^d_{LL})_{12}|$ is suppressed by $|V_{td} V_{ts}|$,
the leading contributions to $\varepsilon_K$ (for $f_{33} \gtrsim 0.01 - 0.03$)
are proportional to $\mbox{Im}\, [ (\delta^d_{LL})_{12} (\delta^d_{RR})_{12} ]
\propto R^{-1}\, \mbox{Im}\, [\, \mbox{e}^{i(\Phi^d_1 - \Phi^d_2)}\,
(C^f_{12})^*\, C^{\lambda_u}_{12} ]$ and to $\mbox{Im}\, [ (\delta^d_{RR})^2_{12} ]
\propto \mbox{Im}\, [\, \mbox{e}^{2i(\Phi^d_1 - \Phi^d_2)}\, (C^{f*}_{12})^2 ]$,
which depend both on CKM/PMNS phases and on the unknown extra
SO(10) phases. As discussed at the end of Section~\ref{sec:num},
this might account for a significant part of the measured value
of $\varepsilon_K$.

\subsection{Model dependence related to higher-dimensional operators}
\label{subsec:MdMe}

Throughout the above discussion, we assumed that Eqs.~(\ref{eq:Diracmasses})
and~(\ref{eq:fermionmasses}) hold at the GUT scale, yielding
the following relations between the masses of the heavy $(L_i, \bar L_i)$
and $(D^c_i, \bar D^c_i)$ fields and of the SM quarks and leptons:
\begin{equation}
  M_L\ =\ \frac{V_1}{\sin \theta_H v_d}\ M_e\ , \qquad
  M_{D^c}\ =\ \frac{V_1}{\sin \theta_H v_d}\ M^T_d\ ,
\label{eq:ML_Me}
\end{equation}
as well as $M_d = M^T_e$. The last
relation is known to be in gross contradiction with the measured masses
of the first two generation fermions, but since the corresponding Yukawa couplings are
small, this can easily be cured by higher-dimensional operators involving
fields with SU(5)-breaking vevs. These operators also affect,
in a model-dependent way, the relations~(\ref{eq:ML_Me}) between
the masses of the heavy and light matter fields. Moreover, they generally
introduce a mismatch between the SO(10) bases in which $M_d$ and
$M_e$ are diagonal, as well as between the SO(10) bases in which
$M_d$ and $M_{D^c}$ ($M_e$ and $M_L$) are diagonal.
All these effects have an impact on the running
of the MSSM soft terms and on the resulting flavour violation. However
we argue below that, under reasonable assumptions, their impact
on physical observables should remain small.

The first effect, i.e. the fact that the masses of the $(L_i, \bar L_i)$ and
$(D^c_i, \bar D^c_i)$ fields are no longer determined by the light fermion
masses, introduces only a mild model dependence. Indeed, the heavy
masses $M_{L_i}$ and $M_{D^c_i}$ enter Eqs.~(\ref{eq:LL})
and~(\ref{eq:LL10}) [or equivalently Eqs.~(\ref{eq:C_f}) and~(\ref{eq:D_u})]
only logarithmically. On top of that, in the case that
$M_{24} > M_{L_2, D^c_2}$, $m^2_l$ and $m^2_{d^c}$ only depend
on $M_{L_3}$ and $M_{D^c_3}$, which are presumably little
affected by the higher-dimensional operators, while in the case that
$M_{24} > M_{L_1, D^c_1}$ some model dependence arises from
the contributions of these operators to $M_{L_2}$ and $M_{D^c_2}$.
As for $m^2_{e^c}$, it is dominated by  $M_{L_3}$-dependent
contributions.

The second effect, i.e. the mismatch between the mass eigenstate basis
of the heavy and light quark (lepton) fields, introduces new flavour
parameters in the model in the form of unitary matrices connecting a given
SO(10) basis to another. As a result, the couplings of the components
of the ${\bf 10}_i$ (namely the $l_i$, $d^c_i$, $\bar L_i$ and $\bar D^c_i$
fields) to the components of the $\bf 54$ are no longer equal at the GUT
scale, when expressed in terms of mass eigenstates.
For instance, $f_\Delta$ and $f_Z$ (see Appendix~\ref{app:rges}
for the definition of these couplings) now differ by a (model-dependent)
unitary matrix. It follows that the radiative corrections to the MSSM soft terms
do not only depend on low-energy flavour parameters, but are also sensitive
to these high-energy flavour parameters. However, since the unitary matrices
originate from the higher-dimensional operators correcting the masses
of the first two generation down quark and charged lepton,
it is natural to expect that they are characterized
by small mixing angles, thus only mildly affecting the flavour structure
of the $f_{ij}$ couplings, which on the contrary is characterized by large
$\theta_{12}$ and $\theta_{23}$ angles. We shall therefore neglect this
model dependence in the following, and simply input the $SU(5)$
relations~(\ref{eq:ML_Me}) in our numerical study\footnote{We note
in passing that there is enough
freedom in the higher-dimensional couplings (see the example below)
to account for the measured down quark and charged lepton masses
without introducing new flavour parameters in the model, i.e. with
$M_d$, $M_e$, $M_{D^c}$ and $M_L$ all diagonal in the same SO(10)
basis.}.

For completeness, we list below the D=5 operators which can correct
the mass relation $M_d=M_e^T$, assuming the field content and pattern
of vevs of Appendix~\ref{app:SO10_breaking}:
\begin{equation}
  a^{(1,2)}_{ij}\frac{({\bf 16}_i {\bf 16})_{\bf 10}
    ({\bf 10}_j {\bf 45}_{1,2})_{\bf 10}}{\Lambda}~,~~~
  b^{(1,2)}_{ij}\frac{({\bf 16}_i {\bf 45}_{1,2})_{\bf 16}
    ({\bf 10}_j {\bf 16})_{\bf \overline{16}}}{\Lambda}~,~~~ 
  c_{ij}\frac{({\bf 16}_i {\bf 16})_{\bf 10}
    ({\bf 10}_j {\bf 54}')_{\bf 10}}{\Lambda}~,
\label{eq:D=5_operators}
\end{equation}
where the subscripts specify the contraction of SO(10) indices and
$\Lambda$ is the cutoff. In Eq.~(\ref{eq:D=5_operators}),
the non-vanishing vevs of
${\bf 45}_1$ and ${\bf 45}_2$ are along the $T_{3R}$ and $B-L$ directions,
respectively, and ${\bf 54}'$ has a vev in the Pati-Salam singlet direction.
The operator $b^{(2)}_{ij}$, for instance, yields the following corrections
to the mass matrices of the heavy and light matter fields:
\begin{equation}
\begin{aligned}
  (M_e)_{ij}\ & =\ \sin \theta_H v_d \left( h_{ij} - 3 c\, b^{(2)}_{ij} \right) , \qquad
    & (M_L)_{ij}\ & =\ V_1 \left( h_{ij} + 3 c\, b^{(2)}_{ij} \right) ,  \\
  (M^T_d)_{ij}\ & =\ \sin \theta_H v_d \left( h_{ij} - c\, b^{(2)}_{ij} \right) , \qquad
    & (M_{D^c})_{ij}\ & =\ V_1 \left( h_{ij} + c\, b^{(2)}_{ij} \right) ,
\end{aligned}
\end{equation}
where $c = 2\sqrt{2/3}\, V_{B-L} / \Lambda$. Even in the simple case where
a single operator ($b^{(2)}$ in this example) is present, the experimental
values of the down quark and charged lepton masses are not sufficient
to fully determine  the couplings $h_{ij}$ and $b^{(2)}_{ij}$. Hence,
the heavy field masses and the mixing patterns in the four mass matrices
are model dependent. As argued above, however, this is unlikely to affect
the radiative corrections to the MSSM soft terms in a sizable way if the
couplings $h_{ij}$ and $b^{(2)}_{ij}$ have a hierarchical flavour structure.

\section{Model-building aspects}
\label{sec:model-building}

Before presenting our numerical results for flavour violation in the next
section, let us briefly present the main ingredients necessary to promote
the SO(10) scenario studied in this paper to a realistic model.
This includes the dynamics of gauge symmetry breaking,
a viable doublet-triplet splitting mechanism and
the generation of the intermediate scales associated with the components
of the $\bf 54$ multiplet consistently with gauge coupling unification. 
More details can be found in Appendices~\ref{app:SO10_breaking},
\ref{app:DT_splitting} and~\ref{app:GCU}.

\subsection{Gauge symmetry breaking and doublet-triplet splitting}
\label{subsec:DTS}

In order to break the SO(10) gauge group down to \SM, we introduce
two adjoint Higgs representations ${\bf 45}_1$ and ${\bf 45}_2$ with
non-vanishing vevs in the $T_{3R}$ and $B-L$ directions, respectively,
and a $\bf 54$' which breaks SO(10) down to its Pati-Salam
subgroup (we also introduce an SO(10) singlet). The rank of the gauge
group is broken by the ($\bf 16$, $\bf \overline{16}$) pair.
A superpotential that enforces this pattern of vevs is given in
Appendix~\ref{app:SO10_breaking}. In this example,
all vevs are proportional to a single mass parameter, $M_{16}$.
Hence, if all superpotential couplings are of order one,
SO(10) is broken down to \SM\ in one step.
Note that the $\bf 54$' cannot be identified with
the $\bf 54$ multiplet involved in the seesaw mechanism, since a vev
of the latter would induce an unwanted ${\bf \bar 5}^{\bf 10}_i /\,
{\bf \bar 5}^{\bf 16}_i$ mixing (see Section~\ref{sec:model}), and also
for reasons related to the doublet-triplet splitting mechanism.

In order to avoid proton decay at an unacceptable rate, we must split
the electroweak doublets from the colour triplets in the $\bf 10$ and
$\bf 16$ Higgs multiplets. This can be achieved by a generalization
of the missing vev mechanism~\cite{DW} involving an additional $\bf 10'$
Higgs multiplet and an adjoint Higgs representation with a vev aligned
along the $B-L$ direction, which is precisely the role of the ${\bf 45}_2$
(see Appendix~\ref{app:DT_splitting} for details). After SO(10) symmetry
breaking, all colour triplets acquire GUT-scale masses, and a single
pair of Higgs doublets ($h_u$, $h_d$) remains massless, the $Y = +1/2$
electroweak doublet in $\bf 10$ and a combination of the $Y=-1/2$
electroweak doublets in $\bf 10$ and $\bf 16$:
\beq
  h_u\, =\, H^{10}_u\ , \qquad h_d\, =\, \cos \theta_H H^{10}_d
    + \sin \theta_H H^{16}_d\ ,
\eeq
where the Higgs mixing angle
is given by $\tan \theta_H = \bar \eta \overline V_{\! 1} /  M_{16}\, $, with
$\bar \eta$ a superpotential coupling.
As discussed in Section~\ref{sec:model}, $\sin \theta_H \neq 0$ is necessary
to give masses to the down quarks and charged leptons, which live in the
${\bf 10}_i$'s. Furthermore, the leptogenesis scenario of Ref.~\cite{FHLR08}
assumes that $h_d$ contains a non-negligible $H^{10}_d$ component
($\cos \theta_H \neq 0$).

Since the down quark and charged lepton  masses are proportional
to $\sin \theta_H$ at the renormalizable level,
the top/bottom mass hierarchy could in principle be
explained by a small Higgs mixing, rather than by a hierarchy in the Yukawa
couplings or a large value of $\tan \beta$. This, however, would imply
rather heavy vector-like matter fields, since according to Eq.~(\ref{eq:ML_Me})
their masses are inversely proportional to $\sin \theta_H$.
On the contrary, leptogenesis requires a relatively light $(L_1, \bar L_1)$
pair in order to satisfy the condition for a non-vanishing CP asymmetry
($M_\Delta > 2 M_{L_1}$).
We shall therefore stick to the case $\sin \theta_H \sim 1$.

\subsection{Proton decay}
\label{subsec:pdecay}

Although all colour Higgs triplets acquire GUT-scale masses through
the doublet-triplet splitting mechanism, their contribution to the proton decay
mode $p \rightarrow K^+ \bar \nu$ can still exceed the experimental limit.
To suppress it further, one must impose additional constraints on the
combination of doublet-triplet splitting parameters which enters the
proton decay rate. More precisely, one must either arrange some
cancellation among these parameters, or allow a pair of Higgs doublets
to lie at an intermediate scale $M_H \lesssim 10^{14}\, \mbox{GeV}$
(see Appendix~\ref{subapp:D=5} for details).
In this paper, we choose the latter option, which in  turn affects the running
of the gauge couplings and leads us to split the
$({\bf 15} \oplus {\bf \overline{15}})^{\bf 54}$ and ${\bf 24}^{\bf 54}$ multiplets
in order to restore successful gauge coupling unification, as we discuss
below.

\subsection{Intermediate scales and gauge coupling unification}
\label{subsec:unification}

The leptogenesis scenario of Ref.~\cite{FHLR08} requires the
${\bf 15} \oplus {\bf \overline{15}}$ component of the $\bf 54$
to be lighter than its ${\bf 24}$ component. Actually the full $SU(5)$
multiplets are not needed for leptogenesis: only the $(\Delta, \bar \Delta)$
pair in ${\bf 15} \oplus {\bf \overline{15}}$, which is responsible for
the type II seesaw mechanism, and the $S$ and $T$ fields
in $\bf 24$ (or just one of them) are really necessary. The main reason
for using complete $SU(5)$ representations in Ref.~\cite{FHLR08}
was gauge coupling unification. However, we have seen that
the experimental constraint on the $p \rightarrow K^+ \bar \nu$ rate
can be satisfied by allowing a pair of Higgs doublets to lie
at an intermediate scale, which in turn spoils unification. As shown in
Appendix~\ref{app:GCU}, this can be cured by splitting the masses 
of the components of $({\bf 15} \oplus {\bf \overline{15}})^{\bf 54}$
and ${\bf 24}^{\bf 54}$ in an appropriate manner. This also has the
advantage of maintaining perturbativity above the GUT scale,
and of avoiding too strong flavour-violating effects in the region
of the parameter space relevant for leptogenesis, where the
$f_{ij}$ couplings are large and induce important radiative corrections
to the MSSM soft terms.

\begin{figure}[t]
 \centering
\subfigure[]{\label{fig:heavy-spectrum} \includegraphics[width=.45\textwidth]{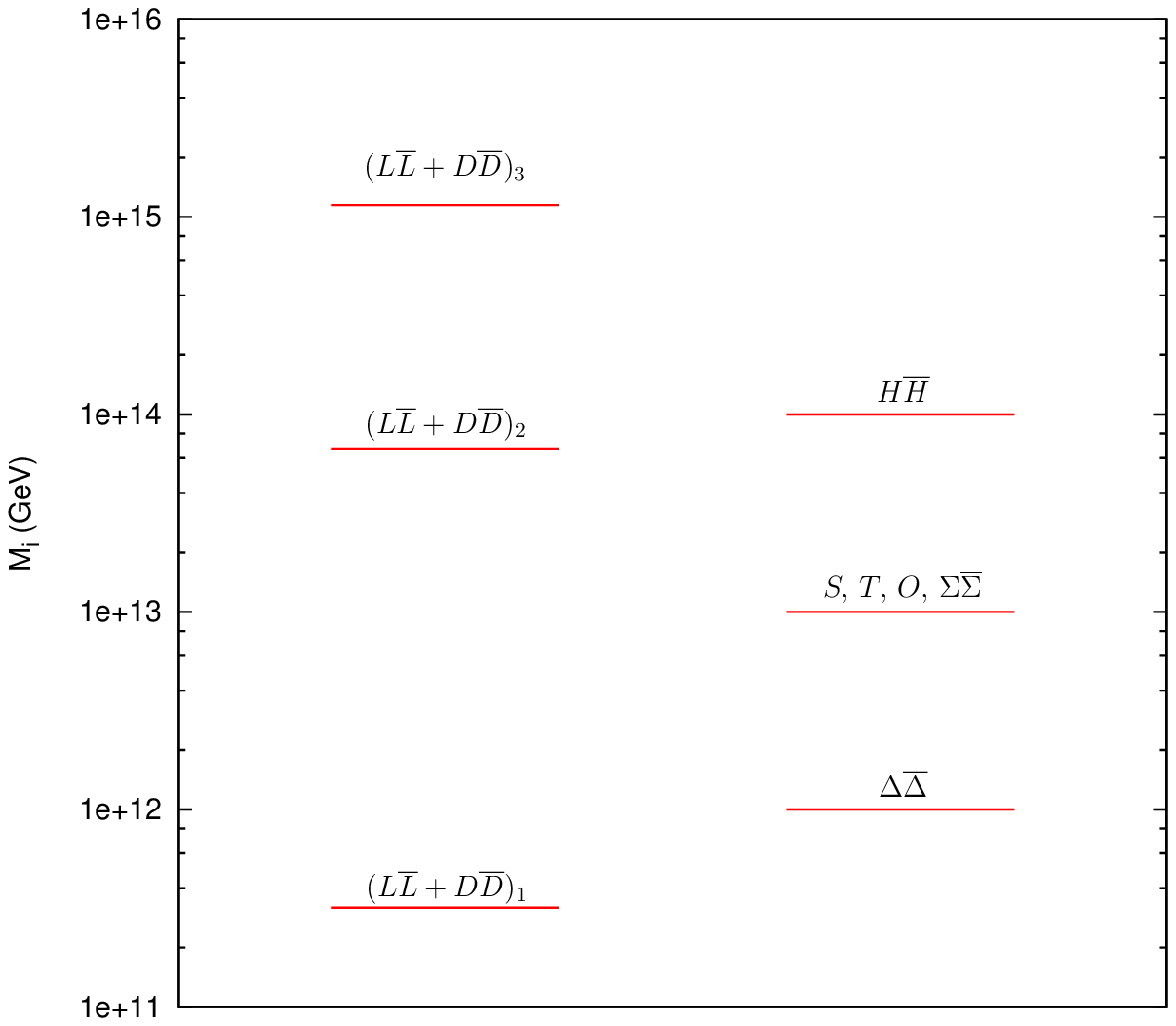}}
\hspace{0.5cm}
\subfigure[]{\label{fig:GCU} \includegraphics[width=.45\textwidth]{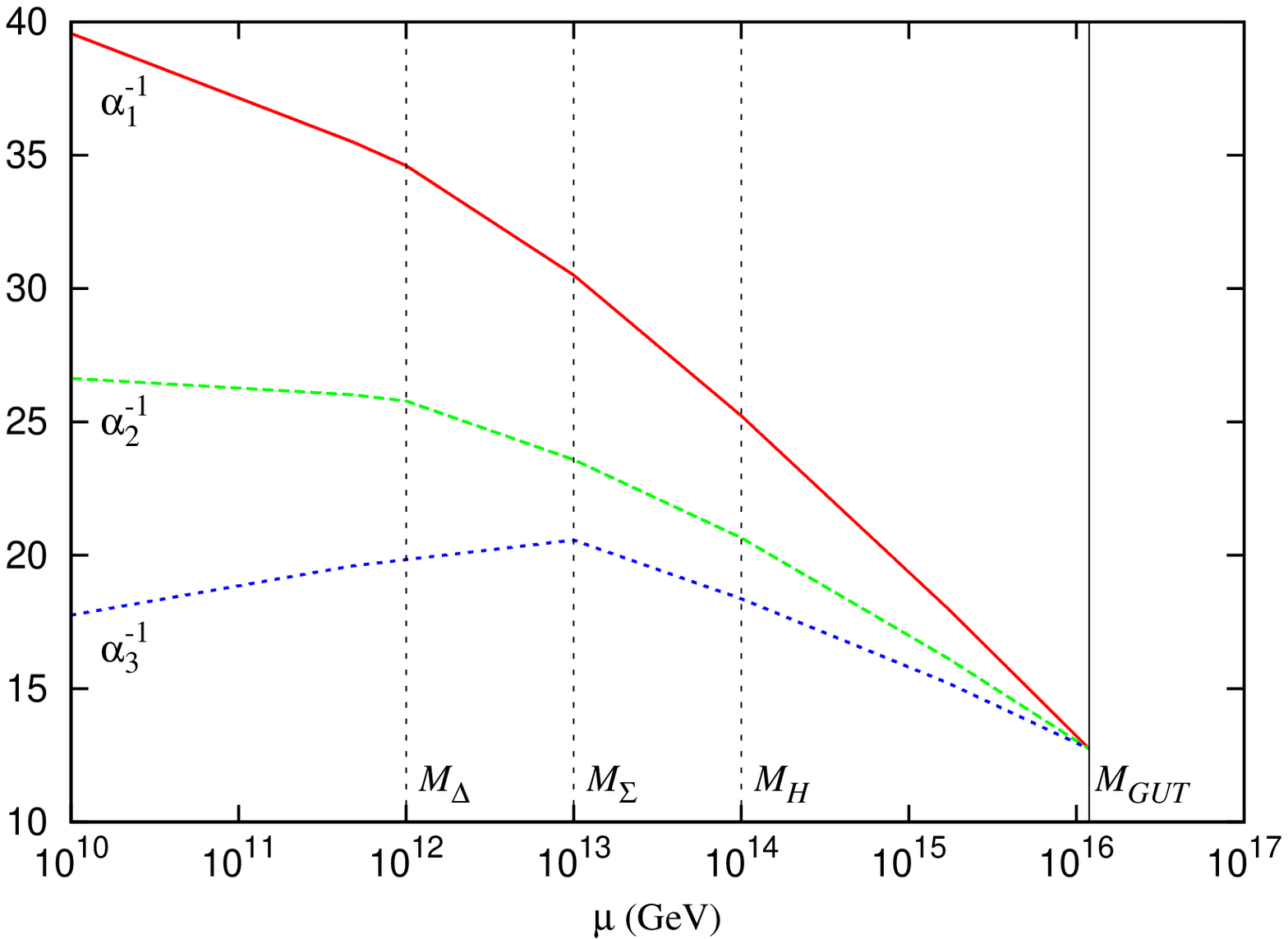}}
\caption{(a) Spectrum of heavy states below the GUT scale in model (ii)
for the following choice of parameters: $M_\Delta = 10^{12}\, \mbox{GeV}$,
$M_\Sigma = M_S = M_T = M_O = 10^{13}\, \mbox{GeV}$,
$M_H = 10^{14}\, \mbox{GeV}$, $V_1 = M_{\rm GUT}$,
$\lambda_H \equiv |\sigma_{\bar \Delta} (\mu = M_\Delta)| = 0.045$,
$\tan \beta = 10$ and $\tan \theta_H = 1$. (b) Renormalization group
running of $\alpha^{-1}_i (\mu)$ ($i=1,2,3$) between
$\mu = 10^{10} \GeV$ and $\mu = M_{\rm GUT}$.}
\end{figure}

Two simple possibilities emerge from the analysis of gauge coupling
unification done in Appendix~\ref{app:GCU} (the ratio $M_T / M_\Delta = 10$
is motivated by leptogenesis):
\begin{itemize}
  \item {\bf model (i):} intermediate $(\Delta, \bar \Delta)$, $(\Sigma, \bar \Sigma)$
and $T$, with
\begin{equation}
  M_\Delta = M_\Sigma\ , \qquad M_T = 10 M_\Delta\ ,
    \qquad M_H = 10^{14}\, \mbox{GeV}\ ,
\label{eq:model1}
\end{equation}
  \item {\bf model (ii):} intermediate $(\Delta, \bar \Delta)$, $(\Sigma, \bar \Sigma)$,
$S$, $T$ and $O$, with ($S$ is actually irrelevant for gauge coupling unification)
\begin{equation}
  M^3_\Sigma / M^2_\Delta \approx 10^{15}\, \mbox{GeV}\ ,
    \qquad M_S = M_T = M_O = 10 M_\Delta\ ,
    \qquad M_H = 10^{14}\, \mbox{GeV}\ ,
\label{eq:model2}
\end{equation}
\end{itemize}
all other components of the $\bf 54$ having GUT-scale masses.
The desired splittings can be achieved by appropriate higher-dimensional
operators (see Appendix~\ref{app:GCU}). In model (i), the unification scale
lies almost one order of magnitude lower than in the MSSM,
which is at odds with the experimental limit on $p \rightarrow \pi^0 e^+$.
While this might be cured by 2-loop running and
GUT threshold corrections, we prefer to adopt model (ii) for the numerical
study of the next section. We note in passing that both models give similar
(although quantitatively different) results for flavour violation. For illustration,
let us consider model (ii) with $M_\Delta = 10^{12}\, \mbox{GeV}$,
$M_\Sigma = M_S = M_T = M_O = 10^{13}\, \mbox{GeV}$ and
$M_H = 10^{14}\, \mbox{GeV}$ (the other parameters are chosen to be
$V_1 = M_{\rm GUT}$,
$\lambda_H \equiv |\sigma_{\bar \Delta} (\mu = M_\Delta)| = 0.045$,
$\tan \beta = 10$ and $\tan \theta_H = 1$). The spectrum of heavy states
below the GUT scale is shown in Fig.~\ref{fig:heavy-spectrum}, and
the renormalization group running of the gauge couplings
in Fig.~\ref{fig:GCU}. At the 1-loop level, one finds:
\begin{equation}
  M_{\rm GUT} = 1.2 \times 10^{16}\, \mbox{GeV}\ , \qquad
  \alpha_{\rm GUT} = 1/12.5\ .
\end{equation}
The prediction for $\alpha_3 (m_Z)$, including supersymmetric
thresholds and the 2-loop MSSM running, agrees with the measured
value within $1 \sigma$. Finally, the Landau pole lies one order
of magnitude above $M_{\rm GUT}$.

%

\section{Numerical results}
\label{sec:num}

%
\begin{table}[t]
\begin{center}
\begin{tabular}{|c|c|c|}
\hline
Observable & Bound & Ref.\\
\hline \hline
$\text{BR}(\mu\to e\gamma)$ & $<1.2\times10^{-11}$ & \cite{Brooks:1999pu} \\
$\text{BR}(\mu\to eee)$ & $< 1.0 \times10^{-12}$ & \cite{Bellgardt:1987du} \\
$\text{CR}(\mu\to e ~{\rm in ~ Ti})$ & $< 4.3 \times10^{-12}$ & \cite{sindrum} \\
$\text{BR}(\tau\to\mu\gamma)$ & $<4.4\times10^{-8}$ & \cite{Aubert:2009tk} \\
$\text{BR}(\tau\to e\gamma)$ & $<3.3\times10^{-8}$ & \cite{Aubert:2009tk}   \\
\hline
 $\varepsilon_K$ &  $(2.229\pm0.012)\times10^{-3}$  & \cite{pdg}\\
 $\Delta m_K$ &  $(3.483\pm 0.006)\times10^{-12}$ MeV & \cite{pdg}\\
$\Delta m_{B_s}$ & $(17.77\pm 0.12)$ ${\rm ps^{-1}} $ & \cite{pdg} \\
$\text{BR}(b\to s\gamma)$ & $2.77\times10^{-4}-4.33\times10^{-4}$ (3$\sigma$) & \cite{Barberio:2007cr} \\
$\text{BR}(B_s\to\mu\mu)$ & $<4.7\times10^{-8}$ & \cite{bsmumu} \\
$\Delta m_{B}$ &  $(0.507\pm 0.005)$ ${\rm ps^{-1}} $   & \cite{pdg} \\
$\Delta m_{D}$  & $(0.0237^{+0.0066}_{-0.0071})$ ${\rm ps^{-1}} $& \cite{pdg} \\
\hline
\end{tabular}
\end{center}
\caption{Flavour-violating observables studied in the numerical analysis
and their experimental values or upper bounds.}
\label{tab:FCNCs}
\end{table}

In this section, we perform a numerical analysis of flavour violation
in the SO(10) model defined above.
Since the flavour-violating effects we want to study arise from radidative
corrections to the
sfermion soft terms, we numerically solve the full 1-loop RGEs
from the unification scale $M_{\rm GUT}$ down to the low-energy scale
$M_{\rm EWSB} \equiv \sqrt{m_{\tilde{t}_1}m_{\tilde{t}_2}}\,$, taking into
account the presence of additional states and interactions below the GUT
scale, which modify the running of the MSSM soft terms as outlined in
Section~\ref{sec:lfv}. A subset of these RGEs is given in
Appendix~\ref{app:rges}, where the definitions of the various superpotential
couplings and soft terms can also be found.

In order to set the initial conditions at the GUT scale for the superpotential
couplings and to compute the masses
of the heavy (${\bf 5}_i,~{\bf \bar 5}_i$) pairs (which depend on the
SO(10) couplings $h_{ij}$), we first evolve the Yukawa matrices and
the low-energy neutrino mass operator from low energy up to the seesaw
scale $M_\Delta$, where the couplings $f_{ij}$ are computed according
to the seesaw formula
$(M_\nu)_{ij} = - \lambda_H f_{ij} v^2_u / 2 M_\Delta$, in which
$\lambda_H \equiv |\sigma_{\bar \Delta} (\mu = M_\Delta)|$. Then we evolve
the $f_{ij}$ together with the other Yukawa couplings up to the GUT scale.
This procedure is iterated until convergence of the heavy
(${\bf 5}_i,~{\bf \bar 5}_i$)
masses $M_{5_i}$ is reached. The other mass scales are taken as inputs
and varied in order to study their impact on leptogenesis and on the
flavour-violating observables listed in Table~\ref{tab:FCNCs}.
Moreover, to satisfy the proton decay constraint consistently with gauge
coupling unification, we split the
components of the $\bf 54$ as indicated in Eq.~(\ref{eq:model2}).
Thus, specifying the triplet mass $M_\Delta$ is enough to fix the masses
of all $\bf 54$ components. The value of $M_H$ has a weaker impact
on gauge coupling unification, and we keep it to be $10^{14}\, \mbox{GeV}$.
The other inputs in the procedure, apart from the SM and
supersymmetric parameters, are the seesaw coupling $\lambda_H$,
$V_1$ and $\tan \theta_H$, which are needed to compute the  heavy
(${\bf 5}_i,~{\bf \bar 5}_i$) masses.

In order to single out the flavour-violating effects arising from radiative
corrections, we assume universal boundary conditions for the soft
supersymmetry breaking terms at the GUT scale, with common gaugino
mass parameter $M_{1/2}$, scalar soft mass $m_0$ and $A$-term $a_0$.
For definiteness, we set $a_0 = 0$ in the following.
In a later stage, we shall also comment about the effect of relaxing
the equality of the soft mass parameters for different SO(10) multiplets,
still assuming flavour-blind supersymmetry breaking.
After having performed the running from the GUT scale to low energy,
we check that
electroweak symmetry breaking does take place and that no tachyonic
states are present in the spectrum. We then compute the masses of all
Higgs bosons and superpartners and impose the mass limits
coming from direct searches at LEP and at the Tevatron.
For the rates of LFV processes, we use the expressions
of Ref.~\cite{Hisano:1995cp}. The supersymmetric contributions to
$\text{BR}(B^0_{d,s} \to \mu^+\mu^-)$ are estimated using the formulae
of Ref.~\cite{Isidori:2002qe}, while $\text{BR}(b\to s\,\gamma)$ is computed
with the help of the routine {\tt SusyBSG}~\cite{Degrassi:2007kj}.
The supersymmetric contributions to the meson mass splittings
$\Delta m_K$, $\Delta m_D$, $\Delta m_B$, $\Delta m_{B_s}$ and to the
indirect CP violation parameter $\varepsilon_K$ are
computed in the mass insertion approximation, using the formulae of
Ref.~\cite{Ciuchini:1998ix}. We recall in Table~~\ref{tab:FCNCs}
the experimental values and upper limits for these observables.

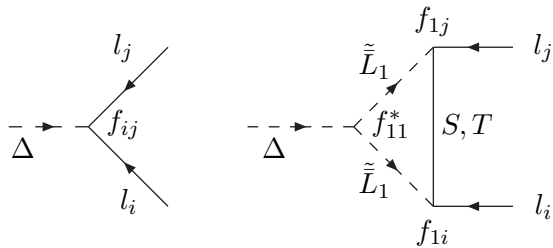
\begin{figure}
\vskip .3cm
\begin{center}
\begin{picture}(140,80)(-50,10)
\DashArrowLine(-100,50)(-70,50){5}
\ArrowLine(-40,80)(-70,50)
\ArrowLine(-40,20)(-70,50)
\Text(-95,42)[]{$\Delta$}
\Text(-55,22)[]{$l_i$}
\Text(-57,80)[]{$l_j$}
\Text(-57,51)[]{$f_{ij}$}
\DashArrowLine(-10,50)(30,50){5}
\ArrowLine(90,80)(60,80)
\ArrowLine(90,20)(60,20)
\Line(60,80)(60,20)
\DashArrowLine(30,50)(60,80){5}
\DashArrowLine(30,50)(60,20){5}
\Text(0,42)[]{$\Delta$}
\Text(38,30)[]{$\tilde{\bar L}_1$}
\Text(38,75)[]{$\tilde{\bar L}_1$}
\Text(73,50)[]{$S,T$}
\Text(102,80)[]{$l_j$}
\Text(102,20)[]{$l_i$}
\Text(44,51)[]{$f^*_{11}$}
\Text(60,90)[]{$f_{1j}$}
\Text(60,10)[]{$f_{1i}$}
\end{picture}
\end{center}
\vskip -0.3cm
\caption{Feynman diagrams responsible (together with the CP-conjugated
diagrams) for the CP asymmetry in the decays of the scalar triplet
into Standard Model lepton doublets.}
\label{fig:epsilon_Delta}
\end{figure}

%
\begin{figure}[t]
 \centering
\includegraphics[width=.55\textwidth,angle=-90]{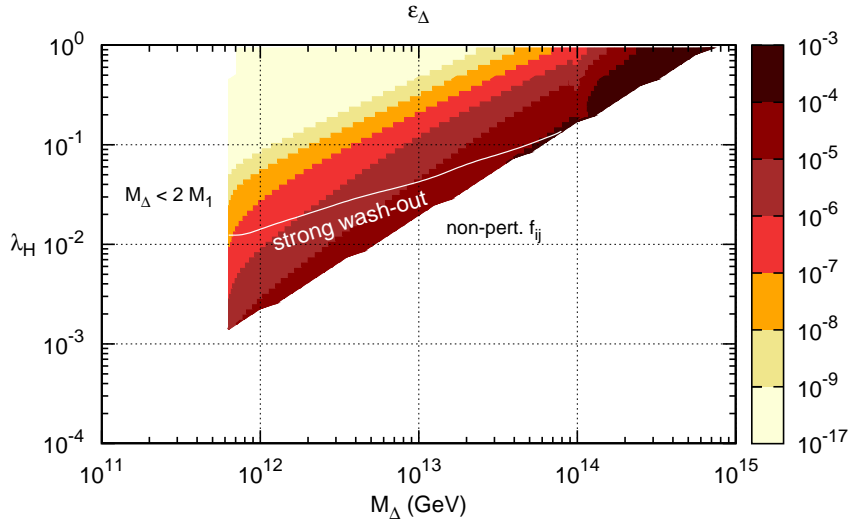}
\caption{Dependence of the CP asymmetry $\varepsilon_\Delta$ on
$M_\Delta$ and $\lambda_H$ for $\tan\beta=10$, $V_1 = M_{\rm GUT}$
and maximal Higgs mixing ($\tan \theta_H = 1$). The unmeasured
neutrino parameters are chosen to be $m_1=0.005\, \mbox{eV}$,
$\sin^2\theta_{13} = 0.05$, $\rho = \pi / 4$, $\sigma = \pi / 2$ and
$\delta = 0$. The area below the white line does not satisfy the conditions
for a large efficiency.}
\label{fig:epsD}
\end{figure}

\begin{figure}[t]
 \centering
\includegraphics[width=.55\textwidth, angle=-90]{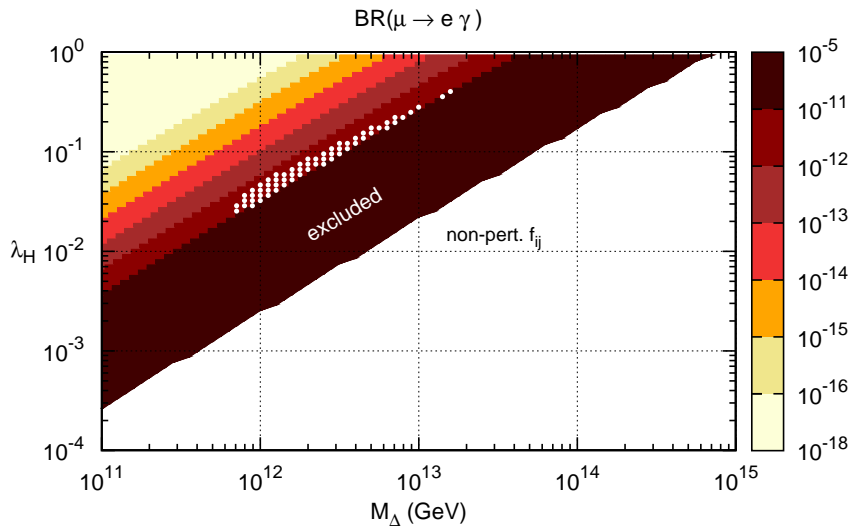}
\caption{Dependence of  ${\rm BR}(\mu\to e\gamma)$ on $M_\Delta$ and
$\lambda_H$, for the following choice of mSUGRA parameters:
$m_0=M_{1/2}=700$~GeV, $a_0=0$, $\tan \beta = 10$ and $\mu>0$.
The other parameters are chosen as in Fig.~\ref{fig:epsD}.
The white dots indicate the area where both
${\rm BR}(\mu\to e\gamma) < 1.2\times 10^{-11}$ and
$\varepsilon_\Delta > 2\times 10^{-8}$.}
\label{fig:meg-epsD}
\end{figure}

As observed in Section~\ref{subsec:analytic} and illustrated in
Fig.~\ref{fig:meg-epsD}, the predictions of the model for LFV processes
span many orders of magnitude, due to their strong dependence
on the ratio $M_\Delta / \lambda_H$. In order to obtain definite
predictions, we shall restrict the seesaw parameter space to the region
favoured by the leptogenesis scenario of Ref.~\cite{FHLR08},
which is built in the SO(10) model considered here.
The source of the lepton asymmetry is the CP asymmetry
in triplet decays, which arises from the interference
between the tree-level and one-loop diagrams shown in
Fig.~\ref{fig:epsilon_Delta} and is given by (in the limit
$M_{L_1} \ll M_\Delta < M_{L_1} + M_{L_2}$):
$\varepsilon_\Delta \simeq (0.1 / 10 \pi)\,
{\rm Im} [f_{11} (f^*ff^*)_{11}]\, / \left[ {\rm Tr} (f f^*) + |f_{11}|^2
+ |\lambda_H|^2 (1 + \cos^4 \theta_H) \right]$, where
$M_S = M_T = 10 M_\Delta$ has been used. The presence of heavy
lepton fields with hierarchical masses is crucial for generating
a non-vanishing CP asymmetry. In particular, the condition
$M_\Delta > 2 M_{L_1}$ must be satisfied for the loop integral
to have an imaginary part.

Let us identify the region of the $(M_\Delta, \lambda_H)$ parameter
space in which successful leptogenesis is possible.
One can distinguish between two regimes~\cite{FHLR08}: the first one,
characterized by a large value of the CP asymmetry in triplet decays and
by a strong washout, requires order one values of the $f_{ij}$ couplings.
This however is in conflict with the experimental upper bounds on lepton
flavour violation, unless the supersymmetric spectrum is very heavy.
In the second regime, the CP asymmetry is smaller, but the efficiency factor
accounting for the dilution of the generated lepton asymmetry
by washout processes can be of order one. Given that the observed baryon
asymmetry is reproduced for $\eta \varepsilon_\Delta \simeq 10^{-8}$,
where $\eta$ is the efficiency factor, successful leptogenesis is possible for
$\varepsilon_\Delta \gtrsim 2 \times 10^{-8}$. In Fig.~\ref{fig:epsD}, we plot
$\varepsilon_\Delta$ as a function of the seesaw parameters $M_\Delta$
and $\lambda_H$.
The light neutrino parameters are chosen as in Ref.~\cite{FHLR08},
with the best fit values for the measured oscillation parameters taken
from Ref.~\cite{Schwetz:2008er} and $m_1=0.005\, \mbox{eV}$,
$\sin^2\theta_{13} = 0.05$, $\rho = \pi / 4$,  $\sigma = \pi / 2$ and
$\delta = 0$.
In the region $M_\Delta \lesssim 6 \times 10^{11}\, \mbox{GeV}$,
the condition for a non-vanishing CP asymmetry, $M_\Delta > 2 M_{L_1}$,
is not satisfied, while in the small $\lambda_H$ region
some of the $f_{ij}$ couplings become non-perturbative below
the GUT scale, making the computation of $\varepsilon_\Delta$ non reliable.
The white line separates the weak and strong washout
regimes: below this line, the conditions for a large efficiency,
$\Gamma (\Delta \rightarrow \tilde{\bar L}_1 \tilde{\bar L}_1) < H (M_\Delta)$
and $\Gamma (\Delta \rightarrow H_u H_u) > H(M_\Delta)$, are not satisfied
(see Ref.~\cite{FHLR08} for details). As shown by Fig.~\ref{fig:epsD},
$\varepsilon_\Delta$ can reach sizable values in the large efficiency region.

Let us now see how the LFV decay $\mu \rightarrow e \gamma$
constrains the parameter space of Fig.~\ref{fig:epsD}. This is shown in
Fig.~\ref{fig:meg-epsD} for a point of the MSSM (mSUGRA) parameter space
giving a rather light superpartner spectrum: $m_0=M_{1/2}=700$~GeV,
$a_0=0$, $\tan \beta = 10$ and $\mu>0$.
The brown (dark) area is excluded by the experimental upper limit
${\rm BR}(\mu\to e\gamma) < 1.2 \times 10^{-11}$. Comparing
Fig.~\ref{fig:meg-epsD} with Fig.~\ref{fig:epsD}, one observes a tension
between the requirement of successful leptogenesis and the experimental
constraint on $\mu \to e \gamma$, due to the fact that both
$\varepsilon_\Delta$ and ${\rm BR}(\mu\to e\gamma)$ grow with the ratio
$M_\Delta / \lambda_H$ (to which the $f_{ij}$ couplings are proportional).
Indeed, the region of the $(M_\Delta, \lambda_H)$ parameter space
consistent with both ${\rm BR}(\mu\to e\gamma) <1.2\times 10^{-11}$
and $\varepsilon_\Delta > 2 \times 10^{-8}$, marked with white dots
in Fig.~\ref{fig:meg-epsD}, is rather small for the chosen MSSM
parameters\footnote{In passing, this shows that the supersymmetric version
of the leptogenesis scenario proposed in Ref.~\cite{FHLR08} can be
excluded on the basis of low-energy flavour physics measurements,
if the supersymmetric partners are accessible at the LHC (and barring
cancellation with other sources of flavour violation in the soft terms).}.
An heavier superpartner spectrum and/or a smaller value of $\tan \beta$
would increase the size of this region by relaxing the $\mu \to e \gamma$
constraint, without affecting leptogenesis. A smaller value of
$\theta_{13}$ would also relax the $\mu \to e \gamma$
constraint, but it would simultaneously reduce $\varepsilon_\Delta$.
Nevertheless, the requirement
of successful leptogenesis (together with the non-observation of
$\mu \to e \gamma$) considerably restricts the seesaw
parameter space, thus allowing us to make testable predictions for
flavour-violating observables.

\begin{figure}[t]
 \centering
\includegraphics[width=.5\textwidth, angle=-90]{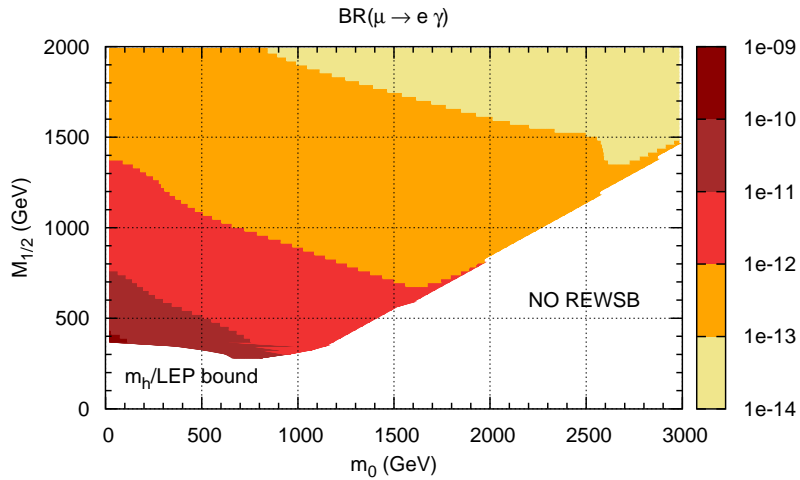}
\vskip -.3cm
\caption{Predictions for $\text{BR}(\mu\to e\gamma)$ in the ($m_0$, $M_{1/2}$)
plane, assuming $a_0=0$, $\tan\beta=10$ and $\mu>0$. The parameters
of the SO(10) model are: $M_\Delta = 10^{12}\, \mbox{GeV}$,
$\lambda_H = 0.045$, $V_1 = M_{\rm GUT}$ and $\tan \theta_H = 1$.
The neutrino parameters are chosen as in Fig.~\ref{fig:epsD}.}
\label{fig:meg}
\end{figure}
\begin{figure}[t]
 \centering
\subfigure[]{\label{fig:spectrum-contours}\includegraphics[width=.55\textwidth]{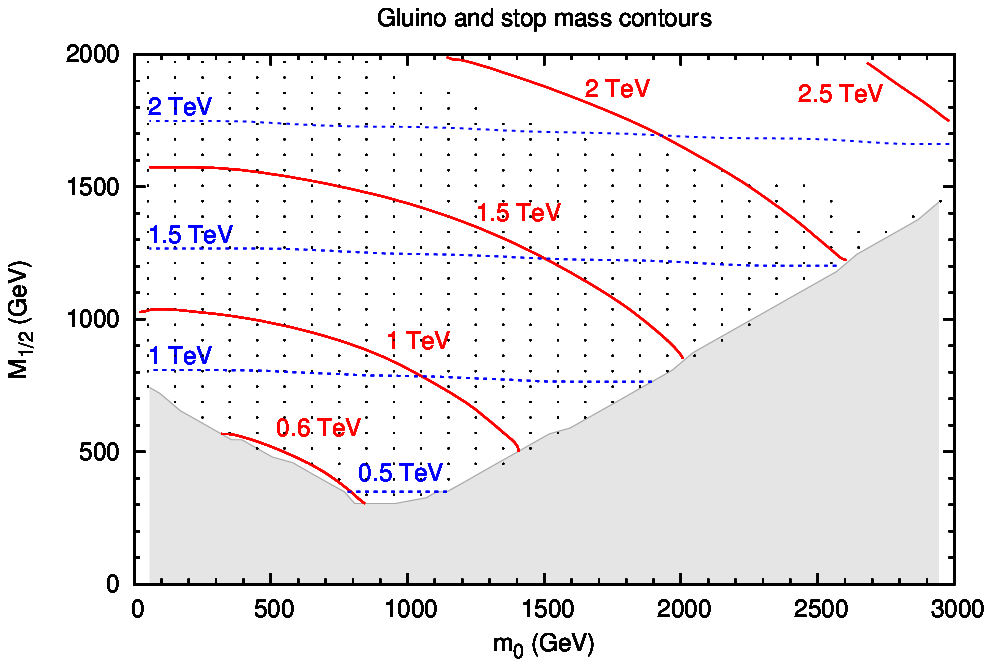}}
\hspace{0.3cm}
\subfigure[]{\label{fig:spectrum}\includegraphics[width=.40\textwidth]{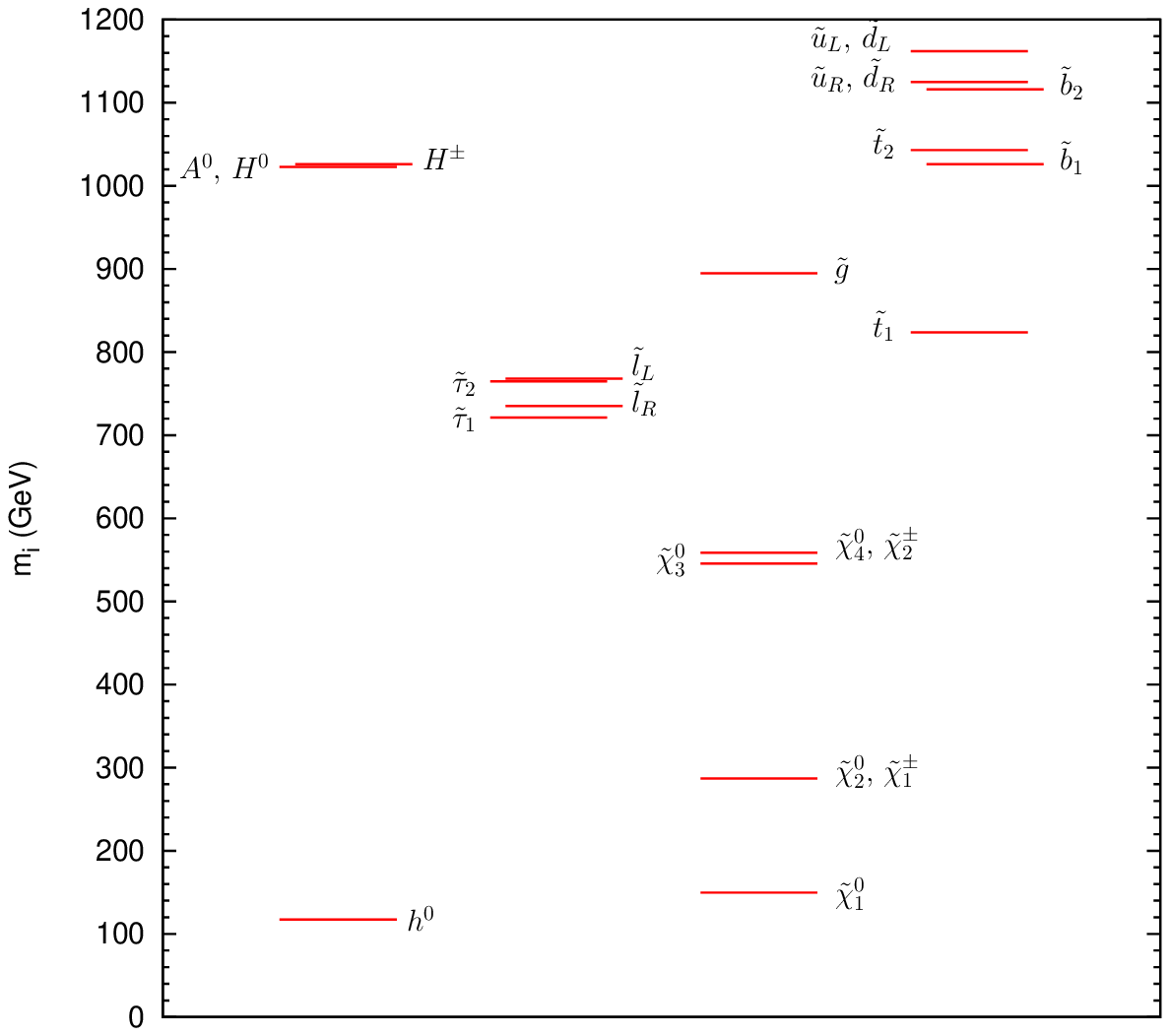}}
\caption{(a) Contours of the gluino mass (blue dashed lines) and of the
lightest stop mass (red solid lines) in the $(m_0, M_{1/2})$ plane, for the
same choice of parameters as in Fig.~\ref{fig:meg}. The grey region is
excluded by the constraints discussed in the text, including the experimental
upper limit on $\text{BR}(\mu\to e\gamma)$. The dotted area will be probed
by MEG. (b) Supersymmetric spectrum corresponding to the mSUGRA point
$m_0 = M_{1/2} = 700\, \mbox{GeV}$, $a_0 = 0$, $\tan \beta = 10$
and $\mu > 0$.}
\end{figure}

In the following, we present our results for a set of seesaw
parameters belonging to the region where successful leptogenesis is
possible, namely we take $M_\Delta = 10^{12}\, \mbox{GeV}$
and $\lambda_H = 0.045$, together with $V_1 = M_{\rm GUT}$ and
$\tan \theta_H = 1$ (which gives
$\varepsilon_\Delta \simeq 2.5\times 10^{-8}$). The corresponding spectrum
of heavy states is shown in Fig.~\ref{fig:heavy-spectrum}. As for the neutrino
parameters, we choose them as in Fig.~\ref{fig:epsD}. We are now ready to
present the predictions of the model
for various flavour-violating observables as a function of the MSSM
parameters. In Fig.~\ref{fig:meg}, the contours of ${\rm BR}(\mu\to e\gamma)$
are plotted in the ($m_0$, $M_{1/2}$) plane for $a_0 = 0$, $\tan \beta=10$
and $\mu > 0$ (the contours corresponding to different values of $\tan \beta$
can be easily deduced from Fig.~\ref{fig:meg} by noting that
${\rm BR}(\mu \to e\gamma)$ approximately scales as $\tan^2\beta$).
The parameter space is bounded by the LEP limits on superpartner and
Higgs boson masses for low values of $M_{1/2}$, and by the absence of
radiative electroweak symmetry breaking for large values of $m_0$ and
moderate $M_{1/2}$. ${\rm BR}(b \to s\gamma)$ gives a constraint similar
to the Higgs mass bound, while the other hadronic observables of
Table~\ref{tab:FCNCs} do not significantly restrict the parameter space.
This implies that the flavour physics signatures of the model are expected
to show up in the lepton sector rather than in the hadronic sector (with the
possible exception of $\varepsilon_K$ discussed at the end of this section).
One can see from Fig.~\ref{fig:meg} that the on-going experiment
MEG~\cite{meg}, which aims at a sensitivity of ${\cal O} (10^{-13})$
on ${\rm BR}(\mu\to e\gamma)$, will probe the model over a large portion
of the MSSM parameter space. Interestingly, this region approximately
corresponds to the one that is accessible at the LHC. This can be seen
in Fig.~\ref{fig:spectrum-contours}, where the contours of the gluino and of
the lightest stop masses are plotted  in the $(m_0, M_{1/2})$ plane, and
the area that will be probed by MEG is marked with dots. For completeness,
we show in Fig.~\ref{fig:spectrum} the full supersymmetric spectrum
corresponding to the mSUGRA point of Fig.~\ref{fig:meg-epsD}, i.e.
$m_0 = M_{1/2} = 700\, \mbox{GeV}$, $a_0 = 0$,
$\tan \beta=10$ and $\mu > 0$.
Note that this spectrum is characterized by rather heavy sfermions compared
to neutralinos and charginos. This is due to the relatively large value of
the unified gauge coupling, which enhances the gauge
contributions to the running of sfermion masses at high energy.
As a result, the lightest neutralino is found to be the LSP over the whole
($m_0$, $M_{1/2}$) parameter
space. For the spectrum showed in Fig.~\ref{fig:spectrum}, the sleptons are
heavier than all neutralinos and charginos and cannot be produced in
cascade decays of squarks at the LHC. However, this possibility is
recovered for $M_{1/2} \gg m_0$.

\begin{figure}[t]
 \centering
\subfigure[]{\label{fig:lfv} \includegraphics[width=.45\textwidth, angle=-90]{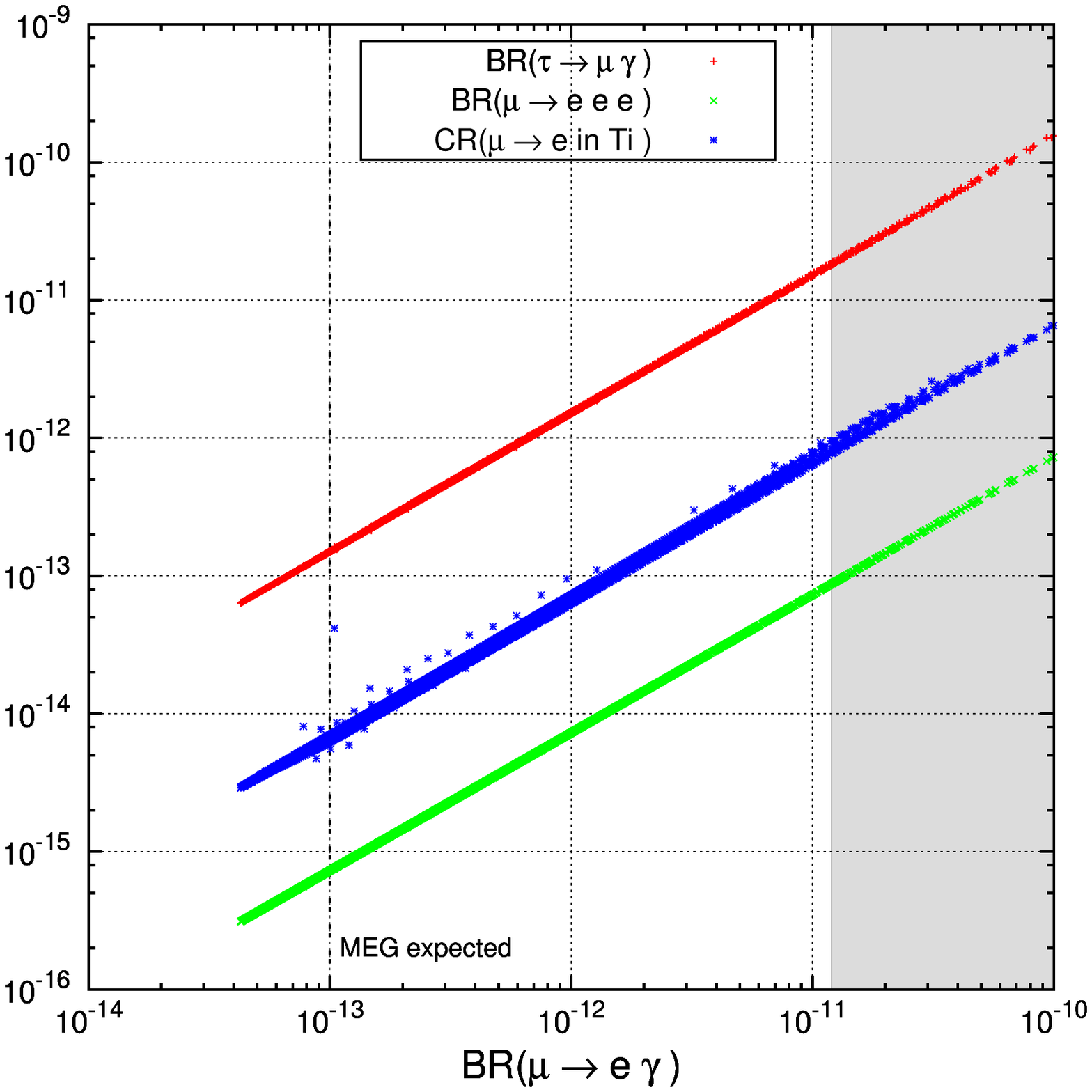} }
\hspace{0.5cm}
\subfigure[]{\label{fig:scatter} \includegraphics[width=.45\textwidth, angle=-90]{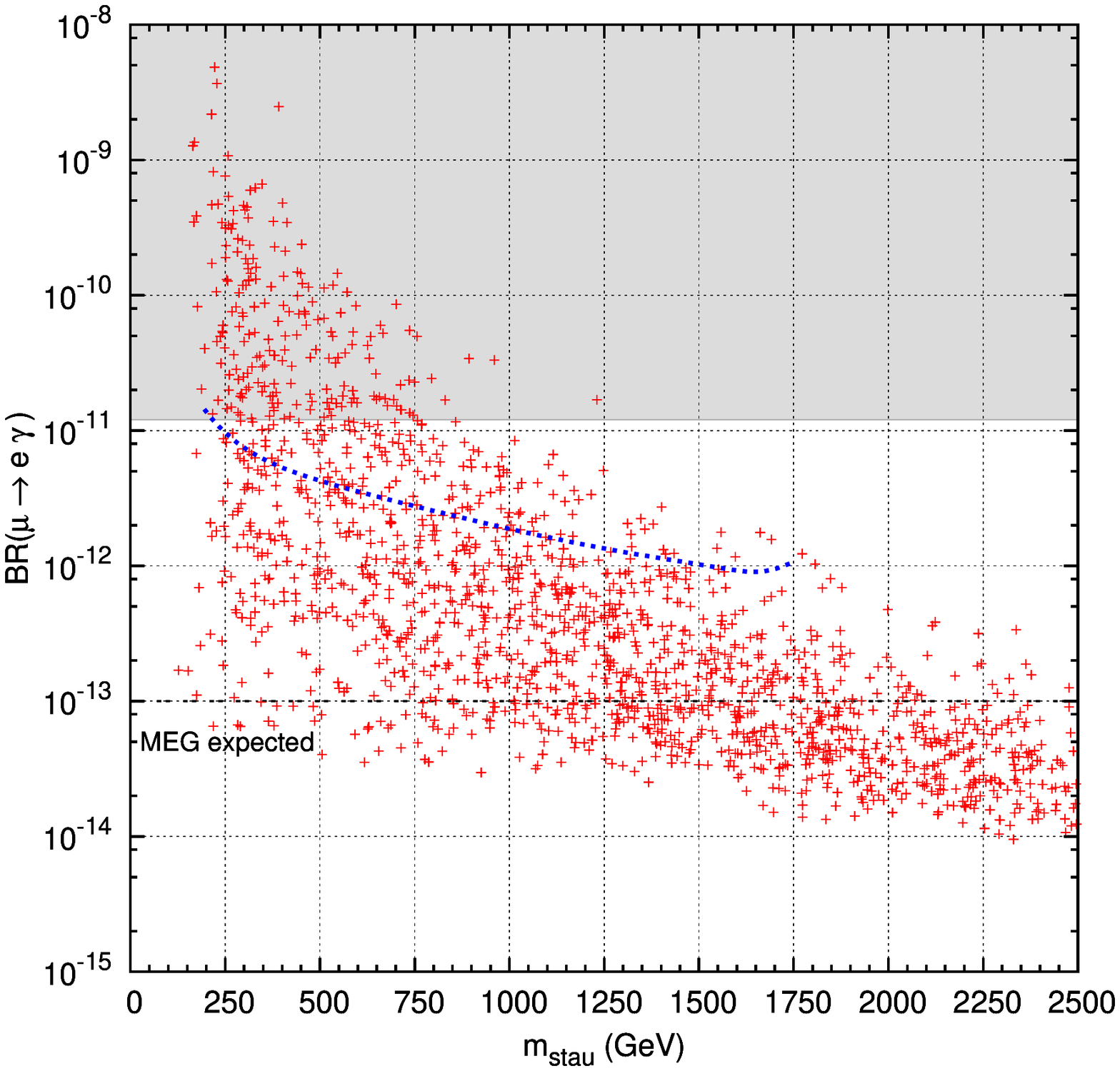}}
\caption{(a) ${\rm BR}(\tau\to \mu\gamma)$, ${\rm BR}(\mu\to eee)$ and
${\rm CR}(\mu\to e~{\rm in~ Ti})$ versus ${\rm BR}(\mu\to e\gamma)$
for the same choice of parameters as in Fig.~\ref{fig:meg}, scanning over
$m_0$ and $M_{1/2}$ in the ranges $0 < m_0 < 3\, \mbox{TeV}$ and
$0 < M_{1/2} < 2\, \mbox{TeV}$. The grey area is excluded by the
experimental upper limit on ${\rm BR}(\mu\to e\gamma)$. 
(b) ${\rm BR}(\mu\to e\gamma)$ as a function of the lightest slepton mass
for non-universal soft scalar masses $m_{16}$, $m_{10}$, $m_{16_H}$,
$m_{10_H}$ and $m_{54}$ in the $\left[ 0, 3\, \mbox{TeV} \right]$ range,
and $M_{1/2} = 700\, \mbox{GeV}$. The blue dotted line corresponds
to the universal case $m_{16} = m_{10} = m_{16_H} = m_{10_H} = m_{54}
\equiv m_0$.}
\end{figure}

Let us now present the predictions of the model for other LFV processes.
In Fig.~\ref{fig:lfv}, ${\rm BR}(\tau\to \mu\gamma)$, ${\rm BR}(\mu\to eee)$
and the $\mu-e$ conversion rate in the Titanium nucleus are plotted against
${\rm BR}(\mu\to e\gamma)$ for the same parameter choice as in
Fig.~\ref{fig:meg}.
${\rm BR}(\tau \to e \gamma)$ is not shown since, as discussed in
Section~\ref{subsec:analytic}, it is generally smaller than
${\rm BR}(\tau\to \mu\gamma)$.
Given the experimental upper limits shown in Table~\ref{tab:FCNCs},
${\rm BR} (\mu \to e\gamma)$ is at present the most constraining LFV
observable.
Since ${\rm BR}(\tau\to \mu\gamma)/{\rm BR}(\mu\to e\gamma)
= \mathcal{O}(1)$, in agreement with Eq.~(\ref{eq:tmg_meg}) and with
the correlation observed in type II seesaw models for large
$\theta_{13}$~\cite{Rossi02}, $\tau \rightarrow \mu \gamma$
is out of reach of super B factories, which are expected to achieve
a sensivity of $10^{-9}$ on its branching ratio~\cite{SFF}.
$\mu - e$ conversion looks more promising,
given that proposed experiments at Fermilab~\cite{Carey:2008zz}
and at J-PARK~\cite{prism} aim at respective sensitivities of $10^{-16}$
and $10^{-18}$ on ${\rm CR}(\mu\to e~{\rm in~ Ti})$. If approved,
these experiments would test the model well beyond the MEG reach.

In order to estimate the impact of non-universal (but flavour-blind) boundary
conditions on the predictions for LFV observables, we performed
a random scan of the soft scalar masses for different SO(10) multiplets
between $0$ and $3\, \mbox{TeV}$, assuming a fixed value
$M_{1/2} = 700\, \mbox{GeV}$ of the common gaugino mass parameter.
The results for ${\rm BR}(\mu\to e\gamma)$ are shown in Fig.~\ref{fig:scatter},
where the blue dotted line\footnote{In the universal case, radiative
electroweak symmetry breaking does
not take place for large $m_0$ values, which explains why the blue dotted
line stops at $m_{\tilde{\tau}_1} \approx 1750\, \mbox{GeV}$.} corresponds
to the universal case
$m_{16} = m_{10} = m_{16_H} = m_{10_H} = m_{54} \equiv m_0$.
Relaxing the universality of soft scalar masses can enhance or suppress
${\rm BR}(\mu\to e\gamma)$ by up to 2 orders of magnitude,
but most of the points still remain within the reach of MEG
(unless the lightest slepton is very heavy).

\begin{figure}[t]
 \centering
\includegraphics[width=.45\textwidth, angle=-90]{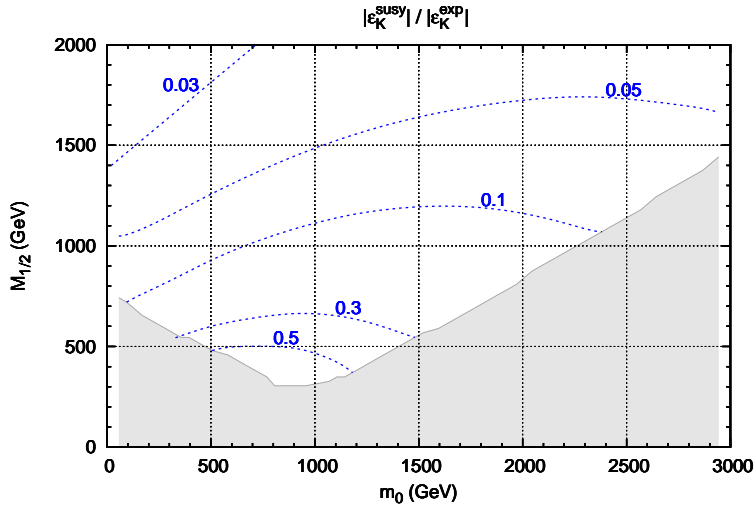}
\caption{Contours of $|\varepsilon^{\rm susy}_K| / |\varepsilon^{\rm exp}_K|$
in the $(m_0, M_{1/2})$ plane (where $\varepsilon^{\rm susy}_K$ is the
supersymmetric contribution to $\varepsilon_K$ and
$\varepsilon^{\rm exp}_K$ its central experimental value), for the same
choice of parameters as in  Fig.~\ref{fig:meg} and assuming
$\arg \left[ (\delta^d_{RR})_{12} \right] = 0.5$.}
\label{fig:epsK}
\end{figure}

As mentioned earlier, LFV observables provide much stronger constraints
on the model studied in this paper than hadronic observables.
A possible exception is represented by the indirect CP violation parameter
in the kaon sector, $\varepsilon_K$. It is well known that $\varepsilon_K$
is very sensitive to new sources of flavour and CP violation in the 1-2 down
squark sector. In the present model, radiative corrections generate large
off-diagonal entries both in the LL slepton mass matrix (which leads to large
LFV rates) and in the RR down squark mass matrix. Moreover, order one
phases in the PMNS matrix, hence in the $f_{ij}$ couplings, are needed
to account for the baryon asymmetry of the universe. Unfortunately,
unknown extra SO(10) phases spoil the link between CP violation in the
neutrino sector and the RG-induced CP violation in the sfermion mass
matrices (see the discussion in Section~\ref{subsec:analytic}). Nevertheless,
barring cancellations between contributions carrying different phases, it is
reasonable to expect a large imaginary part of $(\delta^d_{RR})_{12}$.
In Fig.~\ref{fig:epsK}, the contours of the supersymmetric contribution
to $\varepsilon_K$ are plotted in the $(m_0, M_{1/2})$ plane for the same
choice of parameters as in Fig.~\ref{fig:meg}, assuming
$\arg \left[ (\delta^d_{RR})_{12} \right] = 0.5$. While low values of $M_{1/2}$
would give too large a contribution to $\varepsilon_K$, a rather light
superpartner spectrum can account for up to $(10 - 20) \%$ of its
experimental value. According to Ref.~\cite{buras-guadagnoli}, this is
precisely what is needed in order to reconcile the SM prediction
for $\varepsilon_K$ with experiment.

\section{Conclusions}
\label{sec:conclusion}

We have studied flavour violation in a supersymmetric SO(10)
implementation of the type II seesaw mechanism, which provides
a predictive realization of triplet leptogenesis.
In this scenario, the high-energy flavour parameters
involved in the computation of
the bayon asymmetry of the universe and of
flavour-violating observables
are determined in terms of the Standard Model fermion
masses and mixing, up to mild model-dependent uncertainties. The overall
size of the FCNC effects is then controlled by a few unknown
flavour-blind parameters, while the ratios of FCNC rates for different
flavour channels, such as $BR(\tau \rightarrow \mu\gamma) /
BR(\mu \rightarrow e\gamma)$, mainly depend on low-energy parameters.

The features of flavour violation in the SO(10) scenario studied in this paper
present some interesting differences with the SU(5) implementation of the
type II seesaw mechanism~\cite{Rossi02}, because of additional
contributions coming from the heavy matter fields crucial for leptogenesis.
These give rise, for example, to radiative corrections to the soft mass
matrices $m^2_{e^c}$ and $m^2_q$ controlled by the top quark Yukawa
coupling. Moreover, the presence of a built-in leptogenesis
mechanism provides a criterion for fixing the values of the unknown
flavour-blind parameters,
thus yielding testable predictions for the rates of flavour-violating processes. 
Imposing the conditions for successful leptogenesis together with
the experimental constraints on FCNCs, we found that the predicted
branching ratio for $\mu \to e \gamma$ lies within the sensitivity of the
MEG experiment if the superpartner spectrum is accessible at the LHC,
while $\tau \to \mu \gamma$ is out of reach of future super B factories.
$\mu-e$ conversion on Titanium is also a promising
process, with a predicted rate within the reach of proposed
experiments at Fermilab and J-PARK even in regions of the 
parameter space where
$BR(\mu \rightarrow e \gamma)$ lies below the MEG sensitivity.
Hadronic observables only receive small contributions
once the experimental bounds on LFV processes are imposed, with
the possible exception of $\varepsilon_K$.

We have also studied flavour-violating contributions to leptonic and
hadronic EDMs, as well as to the $\varepsilon_K$ parameter.
The CKM and PMNS phases, together with
new CP-violating phases associated with the SO(10) structure,
can in principle induce sizable contributions to these observables,
even if the soft terms are real at the high scale.
The experimental bounds on LFV processes,
however, prevent significant contributions to the EDMs, while a sizable
contribution to $\varepsilon_K$ (which might be necessary to account
for its measured value) is still possible. 

The predictivity of the scenario studied in this paper
relies on the SO(10) relations between the flavour structures
of the SM and heavy matter fields. These relations, however, only hold
at the tree level and may be affected by model-dependent corrections
from the non-renormalizable operators necessary to account for the
measured quark and lepton masses. 
Nevertheless, the impact of these corrections on our results can be
estimated to be mild.

The presence of heavy states below the GUT scale, besides inducing
low-energy flavour and CP violation through radiative corrections, also
has an impact on the superpartner spectrum. Indeed, their contribution
to the beta functions leads to a relatively large value of the unified gauge
coupling, which enhances the gauge contribution to the running of sfermion
masses. One of the consequences of this is that the lightest neutralino
turns out to be the LSP in an unusually large portion of the parameter space. 

Finally, we have shown that the SO(10) implementation of the type II seesaw
mechanism studied in this paper can be promoted to a consistent model
including the dynamics of gauge symmetry breaking, doublet-triplet splitting
consistent with the present experimental bounds on proton decay, and
gauge coupling unification. The doublet-triplet splitting is achieved through
a generalization of the missing vev mechanism. The proton decay rate is
suppressed by arranging, without fine-tuning, a pair of Higgs doublets to lie
at an intermediate scale.
Gauge coupling unification is then restored by an appropriate splitting of the
SU(5) components of the {\bf 54} containing the type II seesaw
triplet, which in turn brings $\alpha_3(m_Z)$  within $1\sigma$ from its
experimental value, thus improving on the MSSM prediction.
In passing, we provided an updated, comprehensive analysis of proton
decay from D=5 and D=6 operators.

\vskip .3cm


\section*{Acknowledgments}
We thank F. Joaquim and A. Rossi for useful comments.
The work of MF was supported in part by the
Marie-Curie Intra-European Fellowship MEIF-CT-2007-039968.
AR acknowledges partial support from the RTN European Program
``UniverseNet'' (MRTN-CT-2006-035863).

\vfill
\eject


\renewcommand{\theequation}{A.\arabic{equation}}
\setcounter{equation}{0}  

\begin{appendix}

\section{SO(10) gauge symmetry breaking}
\label{app:SO10_breaking}

The purpose of this appendix is to provide an explicit sector
breaking the SO(10) gauge symmetry down to the SM gauge group. 
It is well known that this breaking can be realized by the most general
renormalizable superpotential involving the following Higgs representations:
one ${\bf 45}$, one ${\bf 54}$ and one $({\bf 16},{\bf \overline{16}})$ pair.
In this case, both SM singlets in the ${\bf 45}$ acquire a nonzero vev.
However, we need a ${\bf 45}$ multiplet with a vev aligned in the $B-L$
direction to implement the missing vev mechanism
for doublet-triplet splitting (see Appendix~\ref{app:DT_splitting}).
In order to obtain such a vacuum alignment, we must introduce
additional fields and consider a non-generic superpotential.
A simple realization of this is provided by the following superpotential:
\beq
W_{\rm SB}\, =\, \frac 12 f_1{\bf 54}' {\bf 45}_1{\bf 45}_1
+ (\lambda_{12} {\bf S} + f_{12} {\bf 54}'){\bf 45}_1 {\bf 45}_2 
+ \frac 13 \lambda\, {\bf 54}'{\bf 54}'{\bf 54}'
+ {\bf\overline{16}}(M_{16}+g {\bf 45}_1) {\bf 16}  ~,
\label{eq:WG}
\eeq
where ${\bf S}$ is an SO(10) singlet and the contractions of SO(10) indices
are understood. As shown below, the ${\bf 54}'$ acquires a GUT-scale vev
and therefore cannot be identified with the ${\bf 54}$ multiplet involved
in the type II seesaw mechanism. 

Altogether, eight SM singlets can acquire a nonzero vev at the GUT scale.
We normalize their vevs as follows:
$\langle{\rm Tr}({\bf 54}'^\dag {\bf 54}')\rangle = |V_{PS}|^2$,
which breaks SO(10) down to its Pati-Salam subgroup, 
$\langle{\rm Tr}({\bf 45}_i^\dag {\bf 45}_i)\rangle = |V_{B-L}^{(i)}|^2+|V_{R}^{(i)}|^2$
($i=1,2$), $\langle {\bf 16}^\dag {\bf 16} \rangle = |V_1|^2$,
$\langle {\bf \overline{16}}^\dag {\bf \overline{16}} \rangle = |\overline V_{\! 1}|^2$
and $\langle {\bf S} \rangle = S$.
In order to preserve supersymmetry, these vevs must satisfy F-flatness
and D-flatness conditions. 
The solution to these constraints reads\footnote{Let us mention for
completeness that the D-flatness and F-flatness conditions admit
another solution characterized
by $V_{B-L}^{(2)} = V_{R}^{(1)} = 0$ and all other vevs nonzero.
This solution would also be satisfactory for our purposes, but we stick
to the first one for definiteness.}:
\beq
\ba{lll}
V_{B-L}^{(1)}\, =\, V_{R}^{(2)}\, =\, 0 ~, & V_{R}^{(1)}\, =\, \dfrac{1}{2g}M_{16} ~, &
V_{B-L}^{(2)}\, =\, -\dfrac{3\sqrt{3}f_1}{10\sqrt{2}f_{12}g} M_{16}  ~,\\
V_{PS}^2\, =\, -\dfrac{3f_1}{8\lambda g^2} M_{16}^2 ~,\quad &
S\, =\, -\dfrac{\sqrt{3}f_{12}}{2\sqrt{5}\lambda_{12}}V_{PS} ~,\quad &
\overline V_{\! 1} V_1\, =\, \dfrac{\sqrt{3}f_1}{8\sqrt{5}g^2} V_{PS}M_{16} ~,
\ea
\label{eq:vacuum}
\eeq
together with $|V_1|=|\overline V_{\! 1}|$.
If the dimensionless couplings are of order one, then all nonzero vevs
are of order $M_{16}$, the unique mass parameter in $W_{\rm SB}$.
Therefore SO(10) is broken down to the SM gauge group in one step.
The two ${\bf 45}$ vevs are aligned along the  $T_{3R}$ and $B-L$ directions,
respectively; in the rest of the paper we will rename them
$V_R \equiv V_{R}^{(1)}$ and $V_{B-L} \equiv V_{B-L}^{(2)}$.

The SM vacuum defined by Eq.~(\ref{eq:vacuum}) provides all uneaten chiral
superfields with a GUT-scale mass,
except for a pair of charged states
with \SM\ quantum numbers $(1,1,\pm 1)$, which remains
massless\footnote{In the alternative vacuum characterized by
$V_{B-L}^{(2)} = V_{R}^{(1)} = 0$,
the massless states are $(\overline 3,1,-2/3) \oplus (3,1,2/3)$.}.
This problem can be cured by adding the following terms to $W_{\rm SB}$:
\beq
\frac 12\, \lambda_3\, {\bf S}\, {\bf 45}_3{\bf 45}_3\,
+\, \lambda_{123}\, {\bf 45}_1 {\bf 45}_2 {\bf 45}_3 ~.
\eeq
This does not modify the vacuum alignment discussed above
and does not introduce extra mass parameters either. The F-flatness
conditions imply that the ${\bf 45}_3$ does not acquires a vev.
The coupling $\lambda_{123}$ is sufficient to make the unwanted
massless states heavy, while the coupling $\lambda_3$ guarantees
that all ${\bf 45}_3$ components are also massive.

For completeness, we give the masses of the $X$ and $Y$ gauge
bosons, which can mediate proton decay through $D=6$ operators
(see Appendix~\ref{subapp:D=6}):
\beq
M^2_X\, =\, M^2_Y\, =\, g_{\rm GUT}^2
  \left( \frac 56\, |V_{PS}|^2 + \frac 13\, |V_{B-L}|^2 +  \frac 12\, |V_R|^2 \right)~,
\eeq
where $g_{\rm GUT}$ is the SO(10) gauge coupling.

We remark that the desired vev alignment is obtained because several
couplings allowed  by the SO(10) gauge symmetry are absent
in Eq.~(\ref{eq:WG}).
In fact, one can add mass terms for ${\bf 45}_1$ and ${\bf 54}'$ without
upsetting the vev alignment. On the contrary, it is crucial to forbid
all ${\bf 45}_2$ couplings except $\lambda_{12}$ and $f_{12}$.
We do not try to justify the non-generic form of $W_{\rm SB}$
by global symmetries, which would require a more complicated
set of fields and interactions.

Note finally that it is possible to align the vev of one of the two ${\bf 45}$'s
without introducing the SO(10) singlet ${\bf S}$. In fact, if one replaces
$\lambda_{12}\, {\bf S}$ in Eq.~(\ref{eq:WG}) by a bare mass $M_{12}$,
the F-flatness equations still have a solution with $V_{R}^{(1)}=0$
and a second one with $V_{B-L}^{(1)}=0$, all other vevs being nonzero. 
This is sufficient to realize the doublet-triplet splitting, but this requires
the presence of two mass parameters ($M_{16}$ and $M_{12}$)
in $W_{\rm SB}$.


\renewcommand{\theequation}{B.\arabic{equation}}
\setcounter{equation}{0}  

\section{Doublet-triplet splitting and proton decay}
\label{app:DT_splitting}

In this appendix, we provide an explicit doublet-triplet splitting
mechanism for the SO(10) scenario studied in this paper, and derive
the conditions imposed by the non-observation of proton decay
on the heavy spectrum.

\subsection{Doublet-triplet splitting}
\label{subapp:DT}

In the SO(10) implementation of the type II seesaw mechanism
considered in this paper, up quark masses
arise from the superpotential couplings ${\bf 16}_i {\bf 16}_j {\bf 10}$,
while down quark and charged lepton masses arise from the couplings
${\bf 16}_i {\bf 10}_j {\bf 16}$.
The $Y = + 1/2$ MSSM Higgs doublet $h_u$ should therefore reside
mainly in the $\bf 10$ in order to account for the large top quark mass,
while the $Y = - 1/2$ Higgs doublet $h_d$ should contain a significant
$H^{16}_d$ component, and we require that it also has a non-vanishing
$H^{10}_d$ component\footnote{Indeed, the leptogenesis scenario
of Ref.~\cite{FHLR08} assumes that the dominant decay modes of
the heavy matter fields preserve $B-L$, which is guaranteed
if $h_d$ contains a non-negligible $H^{10}_d$ component.}.
The doublet-triplet splitting mechanism, besides rendering all colour
triplets coupling to light matter fields heavy, should therefore leave
$H^{10}_u$ and an admixture of $H^{10}_d$ and $H^{16}_d$ massless. 

This can be achieved by a generalization of the missing vev
mechanism~\cite{DW} involving an additional $\bf 10'$ Higgs multiplet
and an adjoint Higgs field with a non-vanishing vev in the $B-L$ direction
(which motivates the choice made for ${\bf 45}_2$ in
Appendix~\ref{app:SO10_breaking}).
The superpotential that accomplishes the desired doublet-triplet splitting
reads:
\beq
W_\text{DT}\, =\, \frac 12\, M_{10'} {\bf 10}' {\bf 10}' + h {\bf 10}' {\bf 45}_2 {\bf 10}
+ {\bf\overline{16}} (M_{16} + g {\bf 45}_1) {\bf 16}
+\frac 12\, {\bf\overline{16}}\, {\bf\overline{16}}\, (\bar{\eta} {\bf 10} + \bar{\rho} {\bf 10}')~,
\label{eq:W_DT}
\eeq
where the third term
also plays a role in SO(10) symmetry breaking (and therefore appears
in Eq.~(\ref{eq:WG})), while the coupling
$\bar \rho\, {\bf\overline{16}}\, {\bf\overline{16}}\, {\bf 10}'$ is optional.
As in the case of Eq.~(\ref{eq:WG}), this is not  the most general
superpotential for the fields involved. After SO(10) symmetry breaking, 
one obtains the following doublet and triplet mass matrices, written in
the $({\bf 10}, {\bf 10}', {\bf 16})$ (left) and $({\bf 10}, {\bf 10}', {\bf \overline{16}})$
(right) bases:
\beq
M_D\, =\,
    \left( \begin{array}{ccc}
  0 & 0 & - \bar{\eta} \overline V_{\! 1}  \\
  0 & M_{10'} & - \bar{\rho} \overline V_{\! 1}  \\
  0 & 0 & M_{16}
    \end{array} \right) , \quad
M_T\, =\, 
    \left( \begin{array}{ccc}
  0 & \frac{h}{\sqrt{6}} V_{B-L} & - \bar{\eta} \overline V_{\! 1}  \\
  -\frac{h}{\sqrt{6}} V_{B-L} & M_{10'} & - \bar{\rho} \overline V_{\! 1}  \\
  0 & 0 & M_{16} + 2 g V_R
     \end{array} \right) .
\label{eq:M_D_T}
\eeq
From Eq.~(\ref{eq:M_D_T}) we can see that all colour triplets and
two pairs of Higgs doublets acquire GUT-scale masses, while $H^{10}_u$
and a combination of $H^{10}_d$ and $H^{16}_d$ remain massless:
\beq
  h_u\, =\, H^{10}_u\ , \qquad h_d\, =\, \cos \theta_H H^{10}_d
    + \sin \theta_H H^{16}_d\ ,
\label{eq:light_higgses}
\eeq
where the Higgs mixing angle $\theta_H$ is given by (with no loss of generality,
we assume $\bar\eta \overline V_{\! 1}$ and $M_{16}$ to be real):
\begin{equation}
  \tan \theta_H\ =\ \frac{\bar \eta \overline V_{\! 1}}{M_{16}}\ .
\end{equation}
Note that the alignment of the ${\bf 45}_2$ vev along the $B-L$ direction
is crucial for the splitting of the doublet and triplet components.

A few comments are in order about the above doublet-triplet
mechanism. First, the SO(10)-breaking sector contains a ${\bf 54}'$
representation with a vev in the Pati-Salam direction, whose couplings
to $\bf 10\, 10$ and ${\bf 10\, 10}'$ must be forbidden as they would make
all Higgs doublets heavy.
On the contrary,  the coupling ${\bf 10}\, {\bf 10}\, {\bf 54}$ involved in the
seesaw mechanism is harmless since $\bf 54$ does not acquire a vev.
Second, the mass term $M_{16} {\bf \overline{16}} {\bf 16}$ in
Eqs.~(\ref{eq:WG}) and~(\ref{eq:W_DT})
is crucial both for GUT symmetry breaking and for the doublet-triplet
splitting: if it were absent, F-flatness would imply either $V_1=0$,
leaving the SO(10) rank unbroken, or $V_R=0$, leaving a pair of colour
triplets massless.

\subsection{Higgs-mediated proton decay}
\label{subapp:D=5}

In general, giving
GUT-scale masses to the colour Higgs triplets is not enough to suppress
the proton decay rate below its experimental upper limit. To achieve this,
one must impose additional constraints on the triplet mass matrix.
We show below that this can be done in a simple way in the doublet-triplet
splitting scenario presented above.

Let us first adapt the standard computation of the Higgs-mediated proton
decay rate~\cite{ENR82,ACN85,HMY93,Nath06} to our case.
Upon integrating out the heavy Higgs triplet superfields, one obtains
the following D=5 operators:
\beq
  \left. \frac 1 2\, \kappa_{ijkl}\, (Q_i Q_j) (Q_k L_l)\ \right|_{\theta^2}\,
    +\ \left. \frac 1 2\, \kappa'_{ijkl}\,
    u^c_i u^c_j d^c_k e^c_l\ \right|_{\theta^2}\, +\ \mbox{h.c.}\, ,
\label{eq:D=5_ops}
\eeq
where the contraction of gauge indices is understood,
and the dimensionful coefficients $\kappa_{jikl}$ and $\kappa'_{jikl}$
generated by the superpotential~(\ref{eq:W}) are given by\footnote{Note
that the flavour structure of $\kappa_{ijkl}$ and $\kappa'_{ijkl}$ is the same
as in the minimal SU(5) model. This is an obvious consequence of the
way the SM matter fields are embedded into SO(10) representations.}:
\begin{equation} 
  \kappa_{ijkl}\, =\, -\, y_{ij} h_{kl}
    \left( M^{-1}_T \right)_{T^{10} \overline T^{16}}\, , \qquad
  \kappa'_{ijkl}\, =\, -\, (y_{il} h_{jk} - y_{jl} h_{ik})
    \left( M^{-1}_T \right)_{T^{10} \overline T^{16}}\, ,
\label{eq:kappa}
\end{equation}
where $\left( M^{-1}_T \right)_{T^{10} \overline T^{16}}$ is the
$(T^{10}, \overline T^{16})$ entry of the inverse triplet mass matrix.
Since colour invariance implies $\kappa_{iiil} = 0$ and $\kappa'_{iikl} = 0$,
the dominant proton decay modes arising from the above operators
involve a kaon, and in practice $p \rightarrow K^+ \bar \nu$ dominates.
The corresponding amplitude is obtained by ``dressing'' the D=5 operators
of Eq.~(\ref{eq:D=5_ops}) with gaugino/higgsino loops. If the first two
generation squarks are close in mass, as the observed weakness
of hadronic flavour-violating processes may suggest, the dominant
contributions come from the wino dressing of the $QQQL$ operator
and from the charged higgsino dressing of the $u^c u^c d^c e^c$ operator.
Here we consider only the former contribution; the latter (which is significant
for the $K^+ \bar \nu_\tau$ channel only) would give a stronger constraint
on $M_T$ only for large values of $\tan \beta$. The proton partial decay
width then reads:
\begin{eqnarray}
  \Gamma (p \rightarrow K^+ \bar \nu)\
    =\ \frac{(m^2_p - m^2_{K^+})^2}{32 \pi f^2_\pi m^3_p}\ |\beta_H|^2
    \hskip 8cm  \nonumber \\
    \times\, \sum_l\, \left| \left[ 1 + \frac{m_p}{3 m_{\Lambda}}\, (D+3F) \right] C_{112l}\,
    +\, \left[ \frac{m_p}{2 m_{\Sigma^0}}\, (D-F)
    + \frac{m_p}{6 m_{\Lambda}}\, (D+3F) \right] C_{121l}\, \right|^2 , \hskip 0.1cm
\label{eq:pdecay_D=5}
\end{eqnarray}
where $f_\pi = 130$~MeV is the pion decay constant;
$\beta_H$ is the hadronic parameter defined by $\beta_H P_L u_p
= \langle 0| (u_L d_L) u_L |p \rangle$, where $u_p$ is the proton spinor
and the parenthesis indicates the contraction of Lorentz indices;
D and F are chiral Lagrangian parameters;
and $C_{ijkl}$ is the Wilson coefficient of the four-fermion operator 
$(u_{Li} d_{Lj}) (d_{Lk} \nu_l)$.
The most recent lattice determination of $\beta_H$
is $|\beta_H| = 0.0120 (26)\, \mbox{GeV}^3$~\cite{UKQCD08}, while an analysis
of hyperon decay measurements gives $F+D = 1.2670 \pm 0.0030$
and $F-D = - 0.341 \pm 0.016$~\cite{CSW03}. The Wilson coefficients
$C_{ijkl}$ read:
\begin{equation}
  C_{ijkl}\ =\ \frac{\alpha_2}{4 \pi}\, A_R\, \sum_{m,n}\, V_{mj} V_{nk}
    (\kappa_{imnl} - \kappa_{nmil})
    \left[ f ({\tilde d}^{\, \prime}_{Li},\tilde u_{Lm}) + f (\tilde u_{Ln}, \tilde e_{Ll}) \right] ,
\label{eq:Cijkl}
\end{equation}
where $A_R$ is a renormalization factor, $V$ is the CKM matrix,
the $\kappa$ coefficients are expressed in the basis $q = (u,d') \equiv (u, V d)$,
and the loop function $f$ is given by ($M_2$ is the wino mass):
\begin{equation}
  f(a,b)\ =\ \frac{M_2}{m^2_a - m^2_b}
    \left[\, \frac{m^2_a}{m^2_a - M^2_2} \ln \left( \frac{m^2_a}{M^2_2} \right)
    - \frac{m^2_b}{m^2_b - M^2_2} \ln \left( \frac{m^2_b}{M^2_2} \right) \right] .
\end{equation}
For degenerate sfermion masses ($m_a = m_b \equiv \tilde m$) and
a hierarchy $M^2_2 \ll \tilde m^2$, $f(a,b)$ reduces to $M_2 / \tilde m^2$
to a good approximation. In Eq.~(\ref{eq:Cijkl}), the quantity
$(\kappa_{imnl} - \kappa_{nmil})$ is evaluated at the GUT scale,
where the operator $QQQL$ is generated. The renormalization factor
$A_R = A_{SD} A_{LD}$ contains a short-distance piece $A_{SD}$
which accounts for the renormalization of the superpotential operator
$QQQL$ from $M_{\rm GUT}$ to the supersymmetry breaking scale
(here identified with $m_Z$), and a long-distance piece $A_{LD}$
which encodes the renormalization of the four-fermion operator
$(u_{Li} d_{Lj}) (d_{Lk} \nu_l)$ from $m_Z$ to $\mu_{had} = 1$~GeV.
The latter is given by:
\begin{equation}
  A_{LD}\ =\ \left( \frac{\alpha_3 (\mu_{had})}{\alpha_3 (m_c)} \right)^{2/9}
     \left( \frac{\alpha_3 (m_c)}{\alpha_3 (m_b)} \right)^{6/25}
     \left( \frac{\alpha_3 (m_b)}{\alpha_3 (m_Z)} \right)^{6/23}\ \simeq\ 1.4\ ,
\label{eq:A_LD}
\end{equation}
and the former by (neglecting the Yukawa contributions):
\begin{equation}
  A_{SD}\ =\ \prod_I\ \left( \frac{\alpha_1 (M_I)}{\alpha_1 (M_{I+1})} \right)^{\! -1/5 b^I_1}
     \left( \frac{\alpha_2 (M_I)}{\alpha_2 (M_{I+1})} \right)^{\! -3/ b^I_2}
     \left( \frac{\alpha_3 (M_I)}{\alpha_3 (M_{I+1})} \right)^{\! -4/ b^I_3} ,
\label{eq:A_SD_D=5}
\end{equation}
where $I$ runs over all mass thresholds $M_I$ between $m_Z$ and
$M_{\rm GUT}$, and $b^I_{1,2,3}$ are the beta function coefficients between
$M_I$ and $M_{I+1}$. Given the hierarchy among the quark Yukawa
couplings at the GUT scale, one has $C_{112l} \simeq C_{121l}$, with:
\begin{eqnarray}
  C_{112l}\, \simeq\, \frac{\alpha_2}{4 \pi}\, A_R
    \left( M^{-1}_T \right)_{T^{10} \overline T^{16}}
    \frac{\lambda_{d_l}\! V^*_{ul}\, e^{-i \Phi^d_l}}{\sin \theta_H}
    \sum_n\, e^{2i \Phi^u_n} \lambda_{u_n}\! V_{nd} V_{ns}
    \left[ f (\tilde d^{\, \prime}_L,\tilde u_{Ln}) + f (\tilde u_{Ln}, \tilde e_{Ll}) \right] , \hskip 0.5cm
\label{eq:C112l}
\end{eqnarray}
where $\Phi^u_n$ and $\Phi^d_n$ ($n=1,2,3$, $\Phi^u_1 + \Phi^u_2
+ \Phi^u_3 = 0$) are high-energy phases, and the sum is dominated
by the $n=2$ term (i.e. by the $\tilde c_L$ loops).

We are now in a position to derive a lower bound on
$\left( M^{-1}_T \right)_{T^{10} \overline T^{16}}$ from the experimental
constraint $\tau (p \rightarrow K^+ \bar \nu) > 2.3 \times 10^{33}$~yrs
($90 \%$~C.L.)~\cite{p_K+nubar}.
For definiteness, we consider the model~(ii) of
Section~\ref{subsec:unification} with $M_\Delta = 10^{12}$~GeV,
$\tan \beta = 10$ and a spectrum close to the one of Fig.~\ref{fig:spectrum},
with squark masses in the TeV range (except for the lightest stop),
slepton masses around $750$~GeV and $M_2 \approx 280$~GeV.
This spectrum gives $f (\tilde d^{\, \prime}_L,\tilde c_L)
+ f (\tilde c_L, \tilde e_{Ll}) \approx 0.45\, \mbox{TeV}^{-1}$.
Evaluating the Yukawa couplings at the GUT scale, we find
$\lambda_u V_{ud} V_{us} = 5.2 \times 10^{-7}$,
$\lambda_c V_{cd} V_{cs} = - 2.6 \times 10^{-4}$,
$\lambda_t |V_{td} V_{ts}| = 7.5 \times 10^{-5}$, and
$(\lambda_d V_{ud}, \lambda_s V_{us}, \lambda_b |V_{ub}|)
=  (0.43, 1.9, 2.0) \times 10^{-4}$.
Depending on the values of the high-energy phases, a partial cancellation
between the $\tilde t_L$ and the $\tilde c_L$ contributions is possible.
Here we do not consider this possibility and keep only the dominant
$\tilde c_L$ contribution, which gives:
\begin{eqnarray}
  \sqrt{\sum_l |C_{112l}|^2}\ \simeq\ 9.9 \times 10^{-13}\, \mbox{GeV}^{-1}
    \left( \frac{A_{SD}}{8.1} \right) \left( \frac{1}{\sin \theta_H} \right)
    \left( \frac{2 \bar f}{0.45\, \mbox{TeV}^{-1}} \right)
    \left( M^{-1}_T \right)_{T^{10} \overline T^{16}}\, , \hskip .5cm
\label{eq:C112l_bis}
\end{eqnarray}
where we have replaced $f (\tilde d^{\, \prime}_L,\tilde c_L)
+ f (\tilde c_L, \tilde e_{Ll})$ by an average value $2 \bar f$,
and $A_{SD} \simeq 8.1$ (to be compared with $7.8$ in the MSSM
with $\tan \beta = 10$) takes into account the various thresholds
shown in Fig.~\ref{fig:heavy-spectrum}.
From Eqs.~(\ref{eq:pdecay_D=5}) and~(\ref{eq:C112l_bis}), we finally obtain:
\begin{eqnarray}
  \frac{\tau (p \rightarrow K^+ \bar \nu)}{2.3 \times 10^{33}\, \mbox{yrs}}\
    =\ \left( \frac{0.012\, \mbox{GeV}^3}{|\beta_H|} \right)^{\! 2}\!
    \left( \frac{\sin \theta_H}{1} \right)^{\! 2}\!
    \left( \frac{0.45\, \mbox{TeV}^{-1}}{2 \bar f} \right)^{\! 2}\!
    \left( \frac{(4.4 \times 10^{18}\, \mbox{GeV})^{-1}}
    { \left( M^{-1}_T \right)_{T^{10} \overline T^{16}}} \right)^{\! 2} ,  \hskip .5cm
\end{eqnarray}
which yields the upper bound $\left( M^{-1}_T \right)_{T^{10} \overline T^{16}}
\lesssim \left[\, 4.4 \times 10^{18}\, \mbox{GeV} \left( 1 / \sin \theta_H \right)
\left(\tan \beta / 10 \right)\, \right]^{-1}$,
where we have restored the approximate $\tan \beta$ dependence of the
proton decay amplitude (one can indeed see from Eq.~(\ref{eq:C112l})
that $C_{112l}$ and $C_{121l}$ scale as $1 / \sin 2 \beta$).
This bound is a conservative one, and several effects could relax it:
there are still large uncertainties on the hadronic parameter $\beta_H$;
$A_{SD}$ overestimates the running of the $\kappa$ coefficients, since
it does not include the Yukawa contribution; a superpartner spectrum
characterized by heavier sfermions and/or a stronger sfermion/gaugino
mass hierarchy would reduce the proton decay amplitude; corrections
to the mass relation $M_d = M^T_e$ could affect the couplings of the Higgs
triplets to quarks and leptons; and finally, some sfermion mixing patterns
compatible with the observed level of flavour violation
can significantly reduce the proton decay amplitude~\cite{BFS02}.

Let us now translate this bound into constraints on the doublet-triplet
splitting parameters. From Eq.~(\ref{eq:M_D_T}) we have:
\begin{equation}
  \left( M^{-1}_T \right)_{T^{10} \overline T^{16}}\
    =\ \frac{\sqrt{3}\, \overline V_{\! 1} \left( \sqrt{6}\, \bar \eta M_{10'}
    - \bar \rho h V_{B-L} \right)}{\sqrt{2}\, M_{16}\, (h V_{B-L})^2}\
    =\ \frac{3 \tan \theta_H M_{10'}}{(h V_{B-L})^2}
    - \frac{\sqrt{3}\, \bar \rho \overline V_{\! 1}}{\sqrt{2}\, M_{16}\, h V_{B-L}}\ ,
\end{equation}
where we made use of $V_R = M_{16} / 2 g$ and $\tan \theta_H
= \bar \eta \overline V_{\! 1} / M_{16}$.
Since the SO(10) gauge symmetry is broken in one step
(see Appendix~\ref{app:SO10_breaking}),
$\overline V_{\! 1} \sim V_{B-L} \sim M_{16} \sim M_{\rm GUT}$ and 
the desired suppression of $\left( M^{-1}_T \right)_{T^{10} \overline T^{16}}$
with respect to its natural value $M^{-1}_{\rm GUT}$ can be achieved
by a partial cancellation in the combination
$(\sqrt{6}\, \bar \eta M_{10'} - \bar \rho h V_{B-L})$, or by taking
$\bar \rho \ll 1$ and $M_{10'}$ in the $(10^{13}-10^{14})$~GeV range,
where the lower value corresponds to the conservative bound
on $\left( M^{-1}_T \right)_{T^{10} \overline T^{16}}$. Note that
$\tan \theta_H \ll 1$ would not help, since the upper bound
on $\left( M^{-1}_T \right)_{T^{10} \overline T^{16}}$ becomes
stricter for smaller values of $\theta_H$. In this paper
we take\footnote{Note that the mass scale $M_{10'} \sim 10^{14}$ GeV
can be generated by an interaction term $\lambda\, {\bf S} {\bf 10}' {\bf 10}'$
with $\lambda \sim 10^{-2}$, where $\bf S$ is the SO(10) singlet
introduced in Appendix~\ref{app:SO10_breaking}.}
$M_{10'} = 10^{14}$~GeV in order to avoid a strong cancellation among
unrelated superpotential parameters. One can check that all colour Higgs
triplets acquire GUT-scale masses in this case, while one pair of Higgs
doublet sits at the intermediate scale $M_{10'}$. The consequences
for gauge coupling unification are discussed in Appendix~\ref{app:GCU},
and summarized in Section~\ref{subsec:unification}.

\subsection{Gauge-mediated proton decay}
\label{subapp:D=6}

Let us now discuss the contribution of SO(10) gauge interactions to proton
decay~\cite{Nath06}. Due to the way the SM matter fields are embedded
into SO(10) representations, there are no new gauge contributions with
respect to the SU(5) case, i.e. only the $X$ and $Y$ gauge bosons mediate
proton decay. The dominant decay mode from $(X,Y)$ exchange is
$p \rightarrow \pi^0 e^+$, and its rate is given by (neglecting the non-gauge
contributions, which are subdominant):
\begin{equation}
  \Gamma (p \rightarrow \pi^0 e^+)\
    =\ \frac{(m^2_p - m^2_{\pi^0})^2}{64 \pi f^2_\pi m^3_p}\ |\alpha_H|^2\,
    (1+D+F)^2 \left( |C_{RL}|^2 + |C_{LR}|^2 \right) ,
\label{eq:pdecay_D=6}
\end{equation}
where $\alpha_H$ is the hadronic parameter defined by $\alpha_H P_L u_p
= \langle 0| (u_R d_R) u_L |p \rangle$,
D and F are the same chiral Lagrangian parameters as above,
and $C_{RL}$ and $C_{LR}$ are the Wilson coefficients of the four-fermion
operators $(u_R d_R) (u_L e_L)$ and $(u_L d_L) (u_R e_R)$, respectively.
The most recent lattice determination of $\alpha_H$
is $|\alpha_H| = 0.0112 (25)\, \mbox{GeV}^3$~\cite{UKQCD08}.
The Wilson coefficients take the same form as in minimal SU(5):
\begin{equation}
  C_{RL}\ =\ A_R\ \frac{g^2_{\rm GUT}}{M^2_V}\ , \qquad
    C_{LR}\ =\ A_R\ \frac{g^2_{\rm GUT}}{M^2_V} \left( 1 + |V_{ud}|^2 \right) ,
\label{eq:Wilson_D=6}
\end{equation}
where $M_V \equiv M_X = M_Y$ is the mass of the heavy $(X, Y)$
gauge bosons. The long-distance piece of the renormalization factor
$A_R = A_{SD} A_{LD}$ is given by Eq.~(\ref{eq:A_LD}),
while its short-distance piece reads:
\begin{equation}
  A_{SD}\ =\ \prod_I\
    \left( \frac{\alpha_1 (M_I)}{\alpha_1 (M_{I+1})} \right)^{\! - \frac{23}{30 b^I_1}}
    \left( \frac{\alpha_2 (M_I)}{\alpha_2 (M_{I+1})} \right)^{\! - \frac{3}{2 b^I_2}}
    \left( \frac{\alpha_3 (M_I)}{\alpha_3 (M_{I+1})} \right)^{\! - \frac{4}{3 b^I_3}}\, ,
\end{equation}
where, as in Eq.~(\ref{eq:A_SD_D=5}), the Yukawa contributions have been
neglected, and $I$ runs over all mass thresholds $M_I$ between $m_Z$
and $M_{\rm GUT}$. Together with Eq.~(\ref{eq:Wilson_D=6}),
Eq.~(\ref{eq:pdecay_D=6}) gives:
\begin{eqnarray}
  \tau (p \rightarrow \pi^0 e^+)\ =\ \left(8.2 \times 10^{34}\, \mbox{yrs} \right)
    \left( \frac{0.0112\, \mbox{GeV}^3}{|\alpha_H|} \right)^{\! 2}
    \left( \frac{2.4}{A_{SD}} \right)^{\! 2}
    \left( \frac{1/24}{\alpha_{\rm GUT}} \right)^{\! 2}
    \left( \frac{M_V}{10^{16}\, \mbox{GeV}} \right)^{\! 4} ,  \hskip .5cm
\label{eq:plifetime_D=6}
\end{eqnarray}
to be compared with the experimental upper bound
$\tau (p \rightarrow \pi^0 e^+) > 8.2 \times 10^{33}$~yrs
($90 \%$~C.L.)~\cite{p_e+pi0}.
In Eq.~(\ref{eq:plifetime_D=6}), $A_{SD} = 2.4$
and $\alpha_{\rm GUT} = 1/24$ are the reference MSSM values.
Consider now the model~(i) of Section~\ref{subsec:unification}
with $M_\Delta = M_\Sigma = 10^{12}$~GeV,
$M_T = 10^{13}$~GeV, $M_H = 10^{14}$~GeV and the same heavy fermion
spectrum as in Fig.~\ref{fig:heavy-spectrum}. At the one-loop level
and omitting both low-energy and GUT thresholds, one obtains
$\alpha_{\rm GUT} = 1/14$, $M_{\rm GUT} = 3.5 \times 10^{15}$~GeV and
$A_{SD} = 2.4$. These values are at odds with the experimental
limit on the proton lifetime (assuming $M_V = M_{\rm GUT}$ leads to
$\tau (p \rightarrow \pi^0 e^+) = 4.2 \times 10^{32}$~yrs). On the contrary,
for model~(ii) with the spectrum of Fig.~\ref{fig:heavy-spectrum} (with
$M_\Delta = 10^{12}$~GeV, $M_\Sigma = M_S = M_T = M_O = 10^{13}$~GeV
and $M_H = 10^{14}$~GeV), one finds $\alpha_{\rm GUT} = 1/12$,
$M_{\rm GUT} = 1.2 \times 10^{16}$~GeV and $A_{SD} = 2.5$, leading to
$\tau (p \rightarrow \pi^0 e^+) = 3.9 \times 10^{34}$~yrs for $M_V = M_{\rm GUT}$.
We shall therefore adopt model~(ii) as our reference model, although
model~(i) may be viable if 2-loop running and threshold effects
conspire to increase the GUT scale. Also, the corrections needed to depart
from the minimal SU(5) relation $M_d = M^T_e$ will in general modify
Eq.~(\ref{eq:Wilson_D=6}) by introducing non-trivial fermion mixing angles
in the heavy gauge boson couplings, and this could significantly
reduce the proton decay rate~\cite{DF04}.

\subsection{Proton decay from non-renormalizable operators}
\label{subapp:nonrenormalizable}

For completeness, we mention that proton decay could also be
induced by non-renormalizable superpotential operators of the form:
\begin{equation}
  \frac{d_{ijkl}}{\Lambda^2}\
  {\bf 16}_i {\bf 16}_j {\bf 16}_k {\bf 10}_l {\bf \overline{16}}\, ,
\label{eq:nonren_operators}
\end{equation}
where $\Lambda$ is the cutoff, and the SU(5)-singlet component
of the ${\bf \overline{16}}$ acquires a GUT-scale vev. These operators,
if present, will generate the $D=5$ operators of Eq.~(\ref{eq:D=5_ops})
after SO(10) symmetry breaking. In order to avoid a confict with the
experimental bound on proton lifetime, they must either be forbidden
by some symmetry, or the ones involving light generation fields must be
suppressed by small coefficients $d_{ijkl}$. This is actually what one would
expect in a theory of flavour capable of explaining the hierarchy of fermion
masses. Note that the required suppression of the coefficients is less severe
than in conventional SO(10) models, where the dangerous operators,
of the form ${\bf 16}_i {\bf 16}_j {\bf 16}_k {\bf 16}_l$, have dimension 5.

\renewcommand{\theequation}{C.\arabic{equation}}
\setcounter{equation}{0}  

\section{Intermediate scales and gauge coupling unification}
\label{app:GCU}

In this appendix, we study the constraints imposed by the requirement
of successful gauge coupling unification on the extra heavy states present
below the GUT scale, using 1-loop renormalization group equations.

As shown in Appendix~\ref{subapp:D=5}, the experimental constraint on
the $p \rightarrow K^+ \bar \nu$ rate can easily be satisfied by allowing
a pair of Higgs doublets to lie at an intermediate
scale. This tends to spoil unification, but we will show that the problem
can be cured by splitting the components of the $\bf 54$. 
In fact, it is also desirable to give GUT-scale masses to some components
of the  SU(5) multiplets $({\bf 15}, {\bf \overline{15}})$ and $\bf 24$
contained in the $\bf 54$ in order to preserve perturbative unification.
Indeed, the presence of additional chiral superfields at intermediate 
scales increases the value of the unified gauge coupling $\alpha_{\rm GUT}$
with respect to the MSSM. One has to check that a Landau pole is not
reached soon above (or even below) the GUT scale, because this would
make the predictions of the model very sensitive to the effects of
higher-dimensional operators.
In the SO(10) scenario studied in this paper, keeping the full $\bf 54$
close to the seesaw scale would give 
$\alpha_{\rm GUT} \simeq 1/4$, and the Landau pole would be reached
at a scale $\Lambda \approx 2 M_{\rm GUT}$.
Thus, motivated both by perturbativity and by proton decay,
we are led to split the $({\bf 15}, {\bf \overline{15}})$ and $\bf 24$ multiplets,
keeping below the GUT scale only the components that are necessary for
the seesaw mechanism and for leptogenesis, together with possible
additional components needed to achieve unification.

As is customary, we define $M_{\rm GUT}$ as the scale where
$\alpha_1$ and $\alpha_2$ meet. Denoting by $M_n$ the masses of the
intermediate-scale states and by $b^{(n)}_i$ ($i=1,2,3$) their contributions
to the beta-function coefficients, the 1-loop predictions for $\alpha_3 (m_Z)$,
$M_{\rm GUT}$ and $\alpha_{\rm GUT}$ read:
\begin{align}
\frac{1}{\alpha_3(m_Z)}-\frac{1}{\alpha^{0}_3(m_Z)}\ &=\ \frac{1}{2\pi}
  \sum_n\, \frac{b_{32}^{(n)}b_{21}-b_{21}^{(n)}b_{32}}{b_{21}}\,
  \ln \frac{M^0_{\rm GUT}}{M_n}\ ,
  \label{eq:alpha3}  \\
\ln \frac{M_{\rm GUT}}{M^0_{\rm GUT}}\ &=\
  - \sum_n\, \frac{b_{21}^{(n)}}{b_{21}}\, \ln \frac{M^0_{\rm GUT}}{M_n}\ ,
  \label{eq:MGUT}  \\
\frac{1}{\alpha_{\rm GUT}} - \frac{1}{\alpha^0_{\rm GUT}}\ &=\
  \frac{1}{2\pi} \sum_n\, \frac{b_{2} b_{21}^{(n)}-b_{2}^{(n)}b_{21}}{b_{21}}\,
  \ln \frac{M^0_{\rm GUT}}{M_n}\ ,
  \label{eq:alphaGUT}
\end{align}
where the superscript ``$0$'' refers to MSSM quantities, with
$(b^0_1, b^0_2, b^0_3) = (\frac{33}{5}, 1, -3)$,
$M^0_{\rm GUT} \simeq 2 \times 10^{16}$~ GeV,
$\alpha^0_{\rm GUT} \simeq 1/24$, and
\begin{equation}
  b_{ij}\, \equiv\, b_i - b_j\ ,  \qquad  b_i\, \equiv\, b^0_i + \sum_n b^{(n)}_i\ .
\label{bis}
\end{equation}
Given the fact that the MSSM prediction for $\alpha_3 (m_Z)$ significantly
deviates from the measured value (using 2-loop RGEs and including
low-energy supersymmetric thresholds, one finds
$[1/\alpha_3^{\rm exp}(m_Z)-1/\alpha_3^0(m_Z)]\simeq +4.3/(2\pi)$,
which corresponds to a $5 \sigma$ deviation),
a positive contribution of the extra states to Eq.~(\ref{eq:alpha3})
would be welcome.

Replacing Eq.~(\ref{eq:MGUT}) into Eq.~(\ref{eq:alphaGUT}) one finds:
\begin{equation}
  \frac{1}{\alpha_{\rm GUT}}\ =\ \frac{1}{\alpha^0_{\rm GUT}}\,
    -\, \frac{b_2}{2\pi}\, \ln \frac{M_{\rm GUT}}{M^0_{\rm GUT}}\,
    -\, \sum_n\, \frac{b_{2}^{(n)}}{2\pi}\, \ln \frac{M^0_{\rm GUT}}{M_n}\ .
\end{equation}
The third term always increases $\alpha_{\rm GUT}$, while the second one
can decrease or increase it, depending on whether $M_{\rm GUT}$ is
smaller or larger than $M^0_{\rm GUT}$. 
However, the contribution of the second term is bounded by 
the requirement $M_{\rm GUT} \gtrsim 5\times 10^{15}$~GeV coming from
proton decay (see Apprendix~\ref{subapp:D=6}). 
Therefore, in order to avoid a too large value of $\alpha_{\rm GUT}$,
the additional intermediate-scale fields besides the ones needed to realize
leptogenesis and to suppress proton decay [namely $(\Delta, \overline{\Delta})$,
$S$ and/or $T$, $({\bf 5}_i, {\bf \overline{5}}_i)$ and $(H, \overline H)$]
should better be SU(2)$_{\rm L}$ singlets.
There are two such $\bf 54$ components: $(\Sigma, \overline{\Sigma})$
and $O$.
Since $O$ is an electroweak singlet, it does not affect $M_{\rm GUT}$
nor $\alpha_{\rm GUT}$ (but it corrects the prediction for $\alpha_3$). 
Adding only $O$
would push $M_{\rm GUT}$ above the Planck scale.  We are thus left with
two possibilities: (i) intermediate $(\Sigma, \overline{\Sigma})$;
(ii) intermediate $(\Sigma, \overline{\Sigma})$ and $O$.
The right-hand side of Eq.~(\ref{eq:alpha3}) reads, for each of the two cases:
\globallabel{eq:alpha3_1}
\begin{align}
  ({\rm i}) ~~ 
  &\frac{1}{10\pi}\, \bigg[
  41 \ln \frac{M^0_{\rm GUT}}{M_\Sigma}
  - 22 \ln \frac{M^0_{\rm GUT}}{M_\Delta}
  - 20 \ln \frac{M^0_{\rm GUT}}{M_T}
  - 7 \ln \frac{M^0_{\rm GUT}}{M_{H}} \bigg]\ , \mytag \\
({\rm ii}) ~~
&\frac{1}{10\pi}\, \bigg[
  33 \ln \frac{M^0_{\rm GUT}}{M_\Sigma}
  - 21 \ln \frac{M^0_{\rm GUT}}{M_\Delta}
 - 15 \ln \frac{M^0_{\rm GUT}}{M_T}
  - 6 \ln \frac{M^0_{\rm GUT}}{M_{H}} 
  + 15 \ln \frac{M^0_{\rm GUT}}{M_O}\bigg]\ , \mytag
\end{align}
while $M_{\rm GUT}$ and $\alpha_{\rm GUT}$
are given by the same expression in both cases:
\begin{align}
\label{eq:MGUT_1}
  \ln \frac{M_{\rm GUT}}{M^0_{\rm GUT}}\ & =\
   \frac{1}{15} \bigg[ -8\ln \frac{M^0_{\rm GUT}}{M_\Sigma}
  + \ln \frac{M^0_{\rm GUT}}{M_\Delta}
  + 5 \ln \frac{M^0_{\rm GUT}}{M_T}
  + \ln \frac{M^0_{\rm GUT}}{M_H}\bigg]\ ,     \\
  \frac{1}{\alpha_{\rm GUT}}\, -\, \frac{1}{\alpha^0_{\rm GUT}}\ &=\
  \frac{1}{30\pi} \bigg[88\ln \frac{M^0_{\rm GUT}}{M_\Sigma}
  - 71 \ln \frac{M^0_{\rm GUT}}{M_\Delta}
  - 85 \ln \frac{M^0_{\rm GUT}}{M_T}
  - 26 \ln \frac{M^0_{\rm GUT}}{M_{H}}  \nonumber  \\
  & \hspace{7.3cm} 
   - 15 \ln \frac{(M^0_{\rm GUT})^3}{M_{5_1} M_{5_2} M_{5_3}}  \bigg]\ ,
\label{eq:alphaGUT_1}
\end{align}
where we assumed $M_{L_i}=M_{D^c_i}\equiv M_{5_i}$.
From the above equations, we can see that lowering $M_\Sigma$ decreases
$\alpha_{\rm GUT}$ as desired (and improves the prediction for $\alpha_3$),
but decreases $M_{\rm GUT}$.

Let us first consider case (i). Assuming
$M_\Sigma = M_\Delta \equiv M_{15} = 10^{12}$~GeV,
$M_T \equiv M_{24} = 10^{13}$~GeV and $M_H = 10^{14}$~GeV (as well
as $\lambda_H = 0.045$, $V_1 = M_{\rm GUT}$, $\tan \beta = 10$ and
$\tan \theta_H = 1$ to fix the masses of the heavy $({\bf 5}_i, {\bf \bar 5}_i)$
pairs), we obtain, at the 1-loop level:
\beq
  \frac{1}{\alpha_3(m_Z)}-\frac{1}{\alpha^{0}_3(m_Z)}\, =\, - \frac{0.19}{2\pi}\ ,\
  \quad  M_{\rm GUT}\, =\, 3.5 \times 10^{15} ~\rm GeV\ ,
  \quad  \frac{1}{\alpha_{\rm GUT}}\, =\, 14.1\ .
\eeq
In this case, the prediction for $\alpha_3(m_Z)$ is only slightly larger
than in the MSSM, and the value of the unified coupling remains
reasonable. Unfortunately, the unification scale lies almost one order
of magnitude below the MSSM prediction,
which leads to a too fast proton decay rate (see Appendix~\ref{subapp:D=6}).
Gauge coupling unification is approximately preserved if $M_{15}$
and $M_{24}$ are varied while keeping the ratio $M_{24}/M_{15}$ fixed.
In particular, increasing $M_{15}$ while keeping
$M_{24} / M_{15} = 10$ slightly increases $M_{\rm GUT}$ 
(and decreases $\alpha_3$ and $\alpha_{\rm GUT}$).
For instance, for $M_{15} = 10^{13}$~GeV, one obtains
$1/\alpha_3 (m_Z) - 1/\alpha^0_3 (m_Z) = + 0.27/(2\pi)$,
$M_{\rm GUT} = 4.8 \times 10^{15}$~GeV and $\alpha_{\rm GUT} = 1/15.9$.
Still the unification scale is dangerously close to the proton decay bound. 

Let us now consider case (ii). As anticipated, only
$\alpha_3 (m_Z)$ is affected by the presence of $O$ below $M_{\rm GUT}$.
This gives the possibility of increasing $M_{\rm GUT}$ with respect to case (i)
by increasing $M_\Sigma$, while correcting the prediction for
$\alpha_3 (m_Z)$ by adjusting $M_O$. For instance, the choice
$M_\Delta = 10^{12}$~GeV, $M_T = M_O = M_\Sigma = 10^{13}$~GeV
and $M_H = 10^{14}$~GeV gives, at the 1-loop level:
\beq
  \frac{1}{\alpha_3(m_Z)}-\frac{1}{\alpha^{0}_3(m_Z)}\, =\, + \frac{2.2}{2\pi}\ ,\
  \quad  M_{\rm GUT}\, =\, 1.2 \times 10^{16} ~\rm GeV\ ,
  \quad  \frac{1}{\alpha_{\rm GUT}}\, =\, 12.5\ .
\eeq
In this case, unification works better than in the MSSM. Indeed, we checked
that the prediction for $\alpha_3(m_Z)$, including low-energy thresholds
and the 2-loop MSSM running, matches the measured value within $1 \sigma$.
Moreover, there is no conflict between the value of $M_{\rm GUT}$
and proton decay, and the Landau pole lies one order of magnitude above
$M_{\rm GUT}$. From Eq.~(\ref{eq:alpha3_1}b), we can see that the
contributions of $T$ and $O$ cancel if $M_T=M_O$. 
Therefore, unification is still preserved if the masses of the various
states are varied while keeping $M_T=M_O$ and 
$M^3_\Sigma / M^2_\Delta \approx 10^{15}$~GeV.

%
\begin{table}[t]
\begin{center}
\begin{tabular}{|c|c|c|}
\hline
Operator & Massive $\bf 54$ components & Mass \\
\hline 
    \raisebox{0pt}[15pt][5pt]{${\bf 54}.{\bf 45}_3.{\bf 54}'$}
    &  $(Z, \overline Z), (V, \overline V)$
    &  $V_{PS}$ \\
     \raisebox{0pt}[10pt][10pt]{$\dfrac{1}{\Lambda} ({\bf 54}.{\bf 45}_3)_{\bf 45}
     ({\bf 16}.{\bf \overline{16}})_{\bf 45}$}
    & $S, T, O$
    &  $\dfrac{1}{\Lambda}V_1 \overline{V}_{\! 1}$  \\
    \raisebox{0pt}[10pt][10pt]{$\dfrac{1}{\Lambda}\, ({\bf 54}.{\bf 45}_2)_{\bf 54}
    ({\bf 54}'.{\bf 54}')_{\bf 54}$}
    &  $(\Sigma, \overline \Sigma)$
    &  $\dfrac{1}{\Lambda}V_{B-L} V_{PS}$ \\
    \raisebox{0pt}[10pt][10pt]{$\dfrac{1}{\Lambda^2}\, ({\bf 54}.{\bf 54})_{\bf 1}
    ({\bf\overline{16}}.{\bf 45}_2.{\bf 16})_{\bf 1}$}
    & $(\Delta,\overline{\Delta})$
    & $\dfrac{1}{\Lambda^2} V_1 \overline{V}_{\! 1} V_{B-L}$ \\
\hline
\end{tabular}
\end{center}
\caption{The operators listed in the first column generate masses
for the $\bf 54$ components given in the second column. 
The order of magnitude of these masses is reported in the third column.
\label{tab:case2}}
\end{table}
The splitting of the masses of the $\bf 54$ components needed to realize
case (ii) can be achieved with the set of operators shown in
Table~\ref{tab:case2}, where $V_{PS}$ is the Pati-Salam invariant vev
of the $\bf 54'$, $V_{B-L}$ is the vev of the ${\bf 45}_2$, which is aligned
along the $B-L$ direction,
and ${\bf 45}_3$ has no vev (see Appendix~\ref{app:SO10_breaking}).
At the renormalizable level, only $(Z, \bar Z)$ and $(V, \bar V)$ acquire
a (GUT-scale) mass. The dimension-5 operators provide
$M_S=M_T=M_O \sim M_{\rm GUT}^2/\Lambda$ as well as
$M_\Sigma \sim M_{\rm GUT}^2/\Lambda$
(more precisely, when the GUT-scale masses of the ${\bf 45}_3$ and
${\bf 54}'$ components are taken into account, $M_{S,T,O}$ and $M_\Sigma$
are further suppressed by a mild seesaw-like mechanism).
Finally, the dimension-6 operator generates $M_\Delta\sim M_{\rm GUT}^3/\Lambda^2$.


\renewcommand{\theequation}{D.\arabic{equation}}
\setcounter{equation}{0}  

\section{Renormalization group equations}
\label{app:rges}

Below $M_{\rm GUT}$, the superpotential terms~(\ref{eq:W_Yuk})
and~(\ref{eq:WY54}) read, in terms of
the SM components~(\ref{eq:embedding}):
\begin{align}
W\ &=\  (\lambda_u)_{ij}\, u^c_i q_j h_u + (\lambda_d)_{ij}\, d^c_i q_j h_d
+ (\lambda_e)_{ij}\, e^c_i l_j h_d  \nonumber  \\
&+ (\hat{\lambda}_d)_{ij}\, D^c_i q_j h_d + (\hat{\lambda}_e)_{ij}\, e^c_i L_j h_d
+ (M_L)_{ij} L_i\bar L_j + (M_{D^c})_{ij} D^c_i\bar D^c_j  \nonumber  \\
&+ \frac{1}{2}\, (f_\Delta)_{ij}\, l_i\Delta l_j
+ \frac{1}{2}\, (f_{\bar{\Delta}})_{ij}\, \bar{L}_i\bar{\Delta}\bar{L}_j
+ \frac{1}{\sqrt{2}}\, (f_Z)_{ij}\, d^c_i Z l_j
+ \frac{1}{\sqrt{2}}\, (f_{\bar{Z}})_{ij}\, \bar{D}^c_i \bar{Z} \bar{L}_j  \nonumber  \\
&+ \frac{1}{2}\, (f_\Sigma)_{ij}\, d^c_i \Sigma d^c_j
+ \frac{1}{2}\, (f_{\bar{\Sigma}})_{ij}\, \bar{D}^c_i \bar{\Sigma} \bar{D}^c_j
+ \frac{1}{\sqrt{2}}\, (f_V)_{ij}\, d^c_i V \bar{L}_j
+ \frac{1}{\sqrt{2}}\, (f_{\bar{V}})_{ij}\, \bar{D}^c_i \bar{V} l_j  \nonumber  \\ 
&+ \frac{1}{\sqrt{2}}\, (f_O)_{ij}\, d^c_i O \bar{D}^c_j
+ \frac{1}{\sqrt{2}}\, (f_T)_{ij}\, \bar{L}_i T l_j
+ \sqrt{\frac{3}{20}}\, (f_{S_l})_{ij}\, \bar{L}_i S l_j 
- \frac{1}{\sqrt{15}}\, (f_{S_d})_{ij}\, d^c_i S \bar{D}^c_j  \nonumber  \\
&+ \frac{1}{\sqrt{2}}\, \sigma_T\, h_u T h_d
+ \sqrt{\frac{3}{20}}\, \sigma_S\, h_u S h_d
+ \frac{1}{2}\, \sigma_\Delta\, h_d \Delta h_d
+ \frac{1}{2}\, \sigma_{\bar{\Delta}}\, h_u \bar{\Delta} h_u  \nonumber  \\
&+ M_\Delta \Delta\bar \Delta + M_Z Z \bar Z + M_\Sigma \Sigma \bar\Sigma 
+ M_V V \bar V + \frac 12 (M_S\, S^2 + M_T\, T^2 +M_O\, O^2)\ ,
\label{eq:fullsp}
\end{align}
where Clebsch-Gordan coefficients have been factorized out,
and contractions of $SU(3)_C$ and $SU(2)_L$ indices are understood.
The interactions of fields with GUT-scale masses, namely the right-handed
neutrinos and the components of $\bf 16$ and $\bf 10$ other than the light
Higgs doublets, have been omitted\footnote{Similarly, the interactions
of the $\bf 54$ components with GUT-scale masses should not appear
in Eq.~(\ref{eq:fullsp}), nor in the RGEs.}.
The boundary conditions at the GUT scale for the superpotential
couplings and mass parameters are:
\begin{align}
&\lambda_u = y~, \quad  \lambda_e =  \lambda_d^T = \sin\theta_H h   ~,
  \quad \hat{\lambda}_e = \hat{\lambda}_d  = \cos\theta_H y ~, \nonumber\\
& f_X = f~~~ {\rm for} ~X=\Delta,\bar\Delta,Z,\bar Z,\Sigma,\bar\Sigma,
  V,\bar V,O,T,S_l,S_d ~, \nonumber\\
& \sigma_\Delta = \cos^2\theta_H \sigma ~, \quad
  \sigma_{\bar{\Delta}} = \sigma ~, \quad
  \sigma_T = \sigma_S = \cos\theta_H \sigma~, \\
& M_L=M_{D^c}=h V_1\ ,  \nonumber
\label{boundY}
\end{align}
where the Higgs mixing angle $\theta_H$ is defined by
Eq.~(\ref{eq:light_higgses}), and $V_1$ is the vev of the $SU(5)$ singlet
component of the $\bf 16$. We did not write boundary conditions
for the masses of the $\bf 54$ components, since, as discussed
in Appendix~\ref{app:GCU}, they are assumed to be split by
operators not included in the superpotential~(\ref{eq:WY54}).

The soft supersymmetry breaking terms for the SO(10) fields
are defined by:
\begin{align}
- L_{soft}\ &=\ (m_{16}^2)_{ij} {\bf 16}_i^* {\bf 16}_j + (m_{10}^2)_{ij} 
{\bf 10}_i^* {\bf 10}_i + m_{16_H}^2 {\bf 16}^* {\bf 16}
+ m_{10_H}^2 {\bf 10}^*{\bf 10} + m_{54}^2 {\bf 54}^* {\bf 54} \nonumber \\
&+  \left( \frac 12 (A_y)_{ij} {\bf 16}_i {\bf 16}_j {\bf 10}
+ (A_h)_{ij} {\bf 16}_i {\bf 10}_j {\bf 16}
+ \frac 12 (A_f)_{ij} {\bf 10}_i {\bf 10}_j {\bf 54}
+ \frac 12 A_\sigma {\bf 10}\, {\bf 10}\, {\bf 54} + \mbox{h.c.} \right) \nonumber \\
&+\left(\frac 12 M_{1/2} \lambda^a \lambda^a + \mbox{h.c.} \right)~,
\end{align}
where we used the same notation for the chiral multiplets and for their
scalar components,
$\lambda^a$ are the SO(10) gauginos, and we omitted the $B$-terms.
The soft terms of the SM components are given, at the GUT scale, by the
following boundary conditions:
\begin{align}
& m^2_q = m^{2T}_{u^c} = m^{2T}_{e^c} = m^2_{L}
  = m^{2T}_{D^c}= m^2_{16}\, ,  \nonumber  \\
& m^2_l = m^{2T}_{d^c} = m^2_{\bar{L}}
  = m^{2T}_{\bar D^c} = m^2_{10}\, ,  \nonumber  \\
& m^2_{h_u} = m^2_{10_H}\, , \quad m^2_{h_d} = \cos^2\! \theta_H\,
  m^2_{10_H}\! + \sin^2\! \theta_H\, m_{16_H}^2\, ,  \\
& m^2_X=m^2_{54} ~~~{\rm for}~X=\Delta,\bar{\Delta},Z,\bar Z,
  \Sigma,\bar \Sigma,S,T,O,V, \bar V\, ,  \nonumber  \\
& M_1=M_2=M_3=M_{1/2}\, .  \nonumber
\end{align}
The boundary conditions for the $A$-terms, not included in the above list,
are analogous to the ones for the corresponding Yukawa couplings.
One has for instance:
\begin{align}
& A_u = A_y~, \quad  A_e =  A_d^T = \sin \theta_H A_h   ~,
  \quad \hat A_e = \hat A_d  = \cos \theta_H A_y ~.
\end{align}

In the numerical study of Section~\ref{sec:num}, we solved the
1-loop RGEs for all superpotential
couplings and soft terms below $M_{\rm GUT}$. For brevity, we only list
below the RGEs for the MSSM parameters, which are sufficient for
a leading-log analysis of flavour-violating effects.
Let us first recall the 1-loop RGEs for gauge couplings
and gaugino masses:
\begin{equation}
 \frac{d}{dt}\, \alpha_a^{-1}\, =\, -\frac{1}{2\pi} b_a~, \qquad
   b_a\, =\, -3\,C_2(G_a) + \sum_R T_a(R) ~, 
\end{equation}
\begin{equation}
 \frac{d}{dt} M_a\, =\, \frac{1}{2\pi} b_a \alpha_a M_a ~,
\end{equation}
where $t = \log (\mu/\mu_0)$, $\mu$ being the renormalization scale and
$\mu_0$ a reference scale, $C_2(G_a)$ is the second Casimir invariant
of the group $G_a$, $T_a(R)$ is the Dynkin index of the representation $R$,
and the sum in $b_a$ runs over all chiral superfields with mass smaller
than $\mu$.

The 1-loop RGEs for the MSSM Yukawa couplings are:
\begin{align}
 (4\pi)^2 \frac{d}{dt} \lambda_u\ &=\ \lambda_u\left(3 \lambda_u^\dag \lambda_u  +  \lambda_d^\dag \lambda_d+ \hat{\lambda}_d^\dag 
 \hat{\lambda}_d \right)+  \tr\left( 3 \lambda_u \lambda_u^\dag
 \right)\lambda_u \nonumber\\ 
&+  \left(\frac{3}{4} \left|\sigma_T\right|^2 + \frac{3}{20} \left|\sigma_S\right|^2 +\frac{3}{2} \left|\sigma_{\bar{\Delta}}\right|^2\right) \lambda_u 
-\left(\frac{13}{15}g_1^2+ 3  g_2^2 +\frac{16}{3}g_3^2 \right) \lambda_u ~,
\end{align}
\begin{align}
(4\pi)^2 \frac{d}{dt} \lambda_d\ &=\ \lambda_d\left(3 \lambda_d^\dag \lambda_d  +  \lambda_u^\dag \lambda_u+ 3 \hat{\lambda}_d^\dag 
\hat{\lambda}_d \right)+  
\tr\left(3 \lambda_d \lambda_d^\dag + \lambda_e \lambda_e^\dag+ \hat{\lambda}_e \hat{\lambda}_e^\dag+3  \hat{\lambda}_d \hat{\lambda}_d^\dag
\right)\lambda_d \nonumber \\
&+ \left(f_{Z}f_{Z}^\dag  + f_{V} f_{V}^\dag +2 f_{\Sigma} f_{\Sigma}^\dag + \frac{4}{3} f_{O} f_{O}^\dag
+ \frac{1}{15} f_{S_d} f_{S_d}^\dag \right)\lambda_d \nonumber \\
& + \left(\frac{3}{4} \left|\sigma_T\right|^2 + \frac{3}{20} \left|\sigma_S\right|^2 +\frac{3}{2} \left|\sigma_\Delta\right|^2\right) \lambda_d 
-\left(\frac{7}{15}g_1^2+ 3  g_2^2 +\frac{16}{3}g_3^2 \right) \lambda_d ~,
\end{align}
\begin{align}
 (4\pi)^2 \frac{d}{dt} \lambda_e\ &=\ \left(3 \lambda_e \lambda_e^\dag  + 3 \hat{\lambda}_e \hat{\lambda}_e^\dag \right)\lambda_e +  
 \tr\left(3 \lambda_d \lambda_d^\dag + \lambda_e \lambda_e^\dag+ \hat{\lambda}_e \hat{\lambda}_e^\dag
 +3  \hat{\lambda}_d \hat{\lambda}_d^\dag 
\right)\lambda_e
 \nonumber \\
&+  \lambda_e \left(
  \frac{3}{2} f_{\Delta}^\dag f_{\Delta}+ \frac{3}{2} f_{Z}^\dag f_{Z} +
  \frac{3}{2} f_{\bar{V}}^\dag f_{\bar{V}} + \frac{3}{4} f_{T}^\dag f_{T} + \frac{3}{20} f_{S_l}^\dag f_{S_l} \right) \nonumber \\
&+ \left(\frac{3}{4} \left|\sigma_T\right|^2 + \frac{3}{20} \left|\sigma_S\right|^2 +\frac{3}{2} \left|\sigma_\Delta\right|^2\right) \lambda_e
-\left(\frac{9}{5}g_1^2+ 3  g_2^2 \right) \lambda_e ~.
\end{align}

The 1-loop RGEs for the MSSM soft sfermion masses are:
\begin{align}
(4\pi)^2 \frac{d}{dt} m^2_q\ &=\
\left(m_q^2 \lambda_u^\dag \lambda_u + \lambda_u^\dag\lambda_u   m_q^2 + 2 \lambda_u^\dag\lambda_u  m_{h_u}^2 
+ 2 \lambda_u^\dag m^2_{u^c} \lambda_u +2 A_u^\dag A_u  \right) \nonumber \\ 
&+ \left(m_q^2 \lambda_d^\dag \lambda_d  + \lambda_d^\dag  \lambda_d  m_q^2 
+ 2\lambda_d^\dag \lambda_d  m_{h_d}^2 + 2 \lambda_d^\dag m^2_{d^c} \lambda_d +2 A_d^\dag A_d \right) \nonumber \\ 
&+ \left(m_q^2\hat{\lambda}_d^\dag \hat{\lambda}_d  +\hat{\lambda}_d^\dag  \hat{\lambda}_d  m_q^2
+ 2\hat{\lambda}_d^\dag \hat{\lambda}_d  m_{h_d}^2 + 2 \hat{\lambda}_d^\dag m^2_{D^c} \hat{\lambda}_d 
+2 \hat{A}_d^\dag \hat{A}_d \right) \nonumber \\ 
& - \left(\frac{2}{15} |M_1|^2 g_1^2 + 6|M_2|^2 g_2^2 +\frac{32}{3} |M_3|^2 g_3^2  - \frac 15 g_1^2 S \right)~,
\end{align}
\begin{align}
(4\pi)^2 \frac{d}{dt} m^2_{d^c}\ &=\
2 \left(m^2_{d^c} \lambda_d \lambda_d^\dag + \lambda_d \lambda_d^\dag  m^2_{d^c} + 2 \lambda_d \lambda_d^\dag m_{h_d}^2 
+ 2 \lambda_d m^2_q \lambda_d^\dag +2 A_d A_d^\dag \right) \nonumber \\ 
&+  \left(m^2_{d^c} f_Z f^\dag_Z + f_Z f^\dag_Z m^2_{d^c} +2 f_Z f_Z^\dag m_Z^2 
+ 2 f_Z m^2_l f_Z^\dag + 2 A_{f_Z} A_{f_Z}^\dag \right) \nonumber \\ 
&+ 2 \left(m^2_{d^c} f_\Sigma f^\dag_\Sigma + f_\Sigma f^\dag_\Sigma m^2_{d^c} +2 f_\Sigma f_\Sigma^\dag m_\Sigma^2 
+ 2 f_\Sigma m^{2T}_{d^c} f_\Sigma^\dag + 2 A_{f_\Sigma} A_{f_\Sigma}^\dag \right) \nonumber \\
&+ \frac{1}{15} \left(m^2_{d^c} f_{S_d} f^\dag_{S_d} + f_{S_d} f^\dag_{S_d} m^2_{d^c} +2 f_{S_d} f^\dag_{S_d}  m_S^2
+ 2f_{S_d} m^{2T}_{\bar{D}^c} f^\dag_{S_d} +2A_{f_{S_d}} A^\dag_{f_{S_d}} \right) \nonumber\\
& + \frac{4}{3} \left(m^2_{d^c} f_O f^\dag_O + f_O f^\dag_O m^2_{d^c} +2 f_O  f^\dag_O m_O^2
+ 2 f_O m^{2T}_{\bar{D}^c} f^\dag_O +2A_{f_O} A^\dag_{f_O} \right) \nonumber\\
&+ \left(m^2_{d^c} f_V f^\dag_V + f_V f^\dag_V m^2_{d^c} + 2 f_V  f^\dag_V m_V^2+ 2 f_V m^2_{\bar{L}} f^\dag_V 
+ 2 A_{f_V} A_{f_V}^\dag \right) \nonumber\\
&+ 2 \left(m^2_{d^cD^c} \hat{\lambda}_d \lambda_d^\dag + \lambda_d \hat{\lambda}_d^\dag m^{2\dagger}_{d^cD^c} \right) 
- \left(\frac{8}{15} |M_1|^2 g_1^2 + \frac{32}{3} |M_3|^2 g_3^2  - \frac 25 g_1^2 {S} \right)~,
\end{align}
\begin{align}
(4\pi)^2 \frac{d}{dt} m^2_{u^c}\ &=\
2 \left(m^2_{u^c} \lambda_u \lambda_u^\dag + \lambda_u \lambda_u^\dag  m^2_{u^c} + 2 \lambda_u \lambda_u^\dag m_{h_u}^2 
+ 2 \lambda_u m^2_q \lambda_u^\dag +2 A_u A_u^\dag \right) \nonumber \\ 
& - \left(\frac{32}{15} |M_1|^2 g_1^2 + \frac{32}{3} |M_3|^2 g_3^2  + \frac 45 g_1^2 {S} \right)~,
\end{align}
\begin{align}
(4\pi)^2 \frac{d}{dt} m^2_l\ &=\ \left(m_l^2 \lambda_e^\dag \lambda_e + \lambda_e^\dag \lambda_e m_l^2 
+ 2 \lambda_e^\dag  \lambda_e m_{h_d}^2 + 2 \lambda_e^\dag m^2_{e^c} \lambda_e +2 A_e^\dag A_e\right) \nonumber \\
&+ \frac{3}{2} \left(m_l^2 f^\dag_\Delta f_\Delta + f^\dag_\Delta f_\Delta m_l^2 + 2 f^\dag_\Delta  f_\Delta  m_\Delta^2 
+ 2f^\dag_\Delta m^{2 T}_l f_\Delta +2 A_{f_\Delta}^\dag A_{f_\Delta} \right)  \nonumber \\
&+ \frac{3}{2} \left(m_l^2 f^\dag_Z f_Z + f^\dag_Z f_Z m_l^2 + 2 f^\dag_Z  f_Z m_Z^2 + 2f^\dag_Z m^2_{d^c} f_Z 
+2A_{f_Z}^\dag A_{f_Z}\right)   \nonumber \\
&+ \frac{3}{2} \left(m_l^2 f^\dag_{\bar{V}} f_{\bar{V}} + f^\dag_{\bar{V}} f_{\bar{V}} m_l^2 + 2 f^\dag_{\bar{V}} f_{\bar{V}} m_{\bar{V}}^2 
+ 2f^\dag_{\bar{V}} m^{2}_{\bar{D}^c} f_{\bar{V}} +2A_{f_{\bar{V}}}^\dag A_{f_{\bar{V}}}\right)  \nonumber \\
&+ \frac{3}{4} \left(m_l^2 f^\dag_T f_T + f^\dag_T f_T m_l^2 + 2 f^\dag_T  f_T m_T^2+ 2f^\dag_T m^{2T}_{\bar{L}} f_T 
+2A_{f_T}^\dag A_{f_T}\right) \nonumber \\
&+ \frac{3}{20} \left(m_l^2 f^\dag_{S_l} f_{S_l} + f^\dag_{S_l} f_{S_l} m_l^2 + 2 f^\dag_{S_l}  f_{S_l} m_S^2
+ 2f^\dag_{S_l} m^{2T}_{\bar{L}} f_{S_L} +2A_{f_{S_l}}^\dag A_{f_{S_l}} \right)   \nonumber \\
&+ \left( \lambda_e^\dag \hat{\lambda}_e m^{2\dagger}_{lL} + m^2_{lL} \hat{\lambda}_e^\dag \lambda_e \right)
- \left(\frac{6}{5} |M_1|^2 g_1^2 + 6 |M_2|^2 g_2^2 +\frac{3}{5} g_1^2 {S} \right)~,
\end{align}
\begin{align}
(4\pi)^2 \frac{d}{dt} m^2_{e^c}\ &=\ 2\left(m^2_{e^c} \lambda_e \lambda_e^\dag + \lambda_e \lambda_e^\dag m^2_{e^c} +
2 \lambda_e  \lambda_e^\dag m_{h_d}^2 + 2 \lambda_e m^2_l \lambda_e^\dag +2A_e A_e^\dag \right) \nonumber\\ 
&+ 2\left(m^2_{e^c} \hat{\lambda}_e \hat{\lambda}_e^\dag + \hat{\lambda}_e \hat{\lambda}_e^\dag m^2_{e^c} +
2 \hat{\lambda}_e \hat{\lambda}_e^\dag m_{h_d}^2 + 2 \hat{\lambda}_e m^2_L \hat{\lambda}_e^\dag 
+2\hat{A}_e \hat{A}_e^\dag \right)_{ij} \nonumber\\ 
& - \left(\frac{24}{5} |M_1|^2 g_1^2   -\frac 65 g_1^2 {S} \right)~,
\end{align}
where the hypercharge D-term contribution $S$ is given by:
\begin{align}
{S}\ &=\ m_{h_u}^2 - m_{h_d}^2 + \tr(-m_l^2+m_{e^c}^2-2m_{u^c}^2+m_{d^c}^2+m_q^2)
+ \tr( - m_{L}^2 + m_{D^c}^2 + m_{\bar L}^2 - m_{\bar D^c}^2) \nonumber\\
&+ 3(m_\Delta^2 - m_{\bar{\Delta}}^2) + (m_Z^2-m_{\bar{Z}}^2)
+4(m_{\bar{\Sigma}}^2-m_\Sigma^2) + 5 (m_V^2 - m_{\bar{V}}^2) ~.
\end{align}
The mixing terms $m^2_{d^cD^c}d^{c*}D^c + m^2_{lL}l^*L+ \mbox{h.c.}$
appearing in the RGEs for $m_{d^c}^2$ and $m_l^2$, although absent
at $M_{\rm GUT}$, are generated by the RGE evolution and must be included
for consistency.

Finally, the 1-loop RGEs for the MSSM $A$-terms are:
\begin{align}
(4\pi)^2 \frac{d}{dt} A_u\ &=\  A_u \left( 5 \lambda_u^\dag \lambda_u  +  \lambda_d^\dag \lambda_d 
+ \hat{\lambda}_d^\dag \hat{\lambda}_d \right) + 4 \lambda_u \lambda_u^\dag A_u 
+2 \lambda_u  \left(\lambda_d^\dag A_d + \hat{\lambda}_d^\dag \hat{A}_d \right) \nonumber\\
&+ \tr\left( 3 \lambda_u \lambda_u^\dag 
\right)A_u +
  \left(\frac{3}{4} \left|\sigma_T\right|^2 + \frac{3}{20} \left|\sigma_S\right|^2 +\frac{3}{2} \left|\sigma_{\bar{\Delta}}\right|^2\right) A_u \nonumber\\
&+ \tr\left( 6 A_u \lambda_u^\dag
\right)\lambda_u +
  \left(\frac{3}{2} A_{\sigma_T}\sigma_T^* + \frac{3}{10} A_{\sigma_S}\sigma_S^* 
  + 3 A_{\sigma_{\bar{\Delta}}} \sigma_{\bar{\Delta}}^*\right) \lambda_u \nonumber \\
&+ \frac{13}{15}g_1^2 \left(2 M_1 \lambda_u - A_u \right) + 3 g_2^2 \left(2 M_2 \lambda_u - A_u \right) 
+ \frac{16}{3} g_3^2 \left(2 M_3 \lambda_u - A_u \right) ~,
\end{align}
\begin{align}
(4\pi)^2 \frac{d}{dt} A_d\ &=\  A_d \left( 5 \lambda_d^\dag \lambda_d + \lambda_u^\dag \lambda_u 
+ 5 \hat{\lambda}_d^\dag \hat{\lambda}_d \right) + 4 \lambda_d \lambda_d^\dag A_d 
+ 2\lambda_d \left(\lambda_u^\dag A_u + 2 \hat{\lambda}_d^\dag \hat{A}_d \right) \nonumber \\
&+ \left( f_{Z}f_{Z}^\dag  + f_{V} f_{V}^\dag +2 f_{\Sigma} f_{\Sigma}^\dag + \frac{4}{3} f_{O} f_{O}^\dag
+ \frac{1}{15} f_{S_d} f_{S_d}^\dag \right) A_d \nonumber \\
&+ 
\left( 2 A_{f_{Z}}f_{Z}^\dag  + 2 A_{f_{V}} f_{V}^\dag +4 A_{f_{\Sigma}} f_{\Sigma}^\dag + \frac{8}{3} A_{f_{O}} f_{O}^\dag
+ \frac{2}{15} A_{f_{S_d}} f_{S_d}^\dag \right) \lambda_d \nonumber\\
&+ \tr\left(3 \lambda_d \lambda_d^\dag + \lambda_e \lambda_e^\dag+ \hat{\lambda}_e \hat{\lambda}_e^\dag
+3  \hat{\lambda}_d \hat{\lambda}_d^\dag
\right)A_d+
 \left(\frac{3}{4} \left|\sigma_T\right|^2 + \frac{3}{20} \left|\sigma_S\right|^2 +\frac{3}{2} \left|\sigma_\Delta\right|^2\right) A_d \nonumber \\
&+ 2 \tr\left(3 A_d \lambda_d^\dag +  A_e \lambda_e^\dag+  \hat{A}_e \hat{\lambda}_e^\dag+ 3  \hat{A}_d \hat{\lambda}_d^\dag
\right)\lambda_d+
 \left(\frac{3}{2} A_{\sigma_T}\sigma_T^*  + \frac{3}{10} A_{\sigma_S}\sigma_S^* + 3 A_{\sigma_\Delta} \sigma_\Delta^* \right) \lambda_d \nonumber \\
&+ \frac{7}{15}g_1^2 \left(2 M_1 \lambda_d - A_d \right) + 3 g_2^2 \left(2 M_2 \lambda_d - A_d \right) + \frac{16}{3} g_3^2 \left(2 M_3 \lambda_d - A_d \right) ~,
\end{align}
\begin{align}
(4\pi)^2 \frac{d}{dt} A_e\ &=\ 4\left(  \lambda_e \lambda_e^\dag +  \hat{\lambda}_e \hat{\lambda}_e^\dag \right)A_e
+ 5 A_e \lambda_e^\dag \lambda_e +5 \hat{A}_e \hat{\lambda}_e^\dag \lambda_e  \nonumber\\
&+A_e \left(\frac{3}{2} f_{\Delta}^\dag f_{\Delta}+ \frac{3}{2} f_{Z}^\dag f_{Z} +
  \frac{3}{2} f_{\bar{V}}^\dag f_{\bar{V}} + \frac{3}{4} f_{T}^\dag f_{T} + \frac{3}{20} f_{S_l}^\dag f_{S_l} \right) \nonumber \\
& + \lambda_e \left( 3 f_{\Delta}^\dag A_{f_{\Delta}}+  3 f_{Z}^\dag A_{f_{Z}} +
  3 f_{\bar{V}}^\dag A_{f_{\bar{V}}} + \frac{3}{2} f_{T}^\dag A_{f_{T}} + \frac{3}{10} f_{S_l}^\dag A_{f_{S_l}}  \right) \nonumber \\
&+ \tr\left(3 \lambda_d \lambda_d^\dag + \lambda_e \lambda_e^\dag+ \hat{\lambda}_e \hat{\lambda}_e^\dag
+3  \hat{\lambda}_d \hat{\lambda}_d^\dag
 \right)A_e +
 \left(\frac{3}{4} \left|\sigma_T\right|^2 + \frac{3}{20} \left|\sigma_S\right|^2 +\frac{3}{2} \left|\sigma_\Delta\right|^2\right) A_e \nonumber \\
&+ 2\tr\left(3 A_d \lambda_d^\dag + A_e \lambda_e^\dag+ \hat{A}_e \hat{\lambda}_e^\dag+ 3  \hat{A}_d \hat{\lambda}_d^\dag
\right)\lambda_e+
 \left(\frac{3}{2} A_{\sigma_T}\sigma_T^*  + \frac{3}{10} A_{\sigma_S}\sigma_S^* + 3 A_{\sigma_\Delta} \sigma_\Delta^* \right) \lambda_e \nonumber \\
&+ \frac{9}{5}g_1^2 \left(2 M_1 \lambda_e - A_e \right) + 3 g_2^2 \left(2 M_2 \lambda_e - A_e \right)   ~.
\end{align}

\end{appendix}

\vfill
\eject



\end{document}